\documentclass[aps,prx,twocolumn,english,balance,superscriptaddress,floats,showpacs,letter, prb]{revtex4-1}
\usepackage[latin9]{inputenc}
\usepackage{amsmath}
\usepackage{amssymb}
\usepackage{graphicx}
\usepackage{esint}
\usepackage{subfigure}
\usepackage{multirow}
\usepackage{mathtools}
\usepackage{xcolor}
\usepackage{gensymb}

\makeatletter

\newcommand{\beq}{\begin{equation}}
\newcommand{\eeq}{\end{equation}}
\newcommand{\bea}{\begin{eqnarray}}
\newcommand{\eea}{\end{eqnarray}}
\newcommand{\bwt}{\begin{widetext}}
\newcommand{\ewt}{\end{widetext}}
\@ifundefined{textcolor}{}
{%
 \definecolor{BLACK}{gray}{0}
 \definecolor{WHITE}{gray}{1}
 \definecolor{RED}{rgb}{1,0,0}
 \definecolor{GREEN}{rgb}{0,1,0}
 \definecolor{BLUE}{rgb}{0,0,1}
 \definecolor{CYAN}{cmyk}{1,0,0,0}
 \definecolor{MAGENTA}{cmyk}{0,1,0,0}
 \definecolor{YELLOW}{cmyk}{0,0,1,0}
}

\newcommand{\eps}{\epsilon}
\newcommand{\bk}{\mathbf{k}}
\newcommand{\bq}{\mathbf{q}}
\newcommand{\bp}{\mathbf{p}}
\newcommand{\bg}{\mathbf{g}}
\newcommand{\br}{\mathbf{r}}
\newcommand{\bL}{\mathbf{L}}
\newcommand{\bK}{\mathbf{K}}

\newcommand{\bG}{\mathbf{G}}

\newcommand{\fvec}[1]{\boldsymbol{#1}}

\newcommand{\half}{\frac{1}{2}}

\newcommand{\rmd}{{\rm d}}

\makeatother

\begin{document}
	
\title{Non-Abelian Dirac node braiding and near-degeneracy of correlated phases at odd integer filling in magic angle twisted bilayer graphene}

\author{Jian Kang}
\email{jkang@suda.edu.cn}
\affiliation{School of Physical Science and Technology \& Institute for Advanced Study, Soochow University, Suzhou, 215006, China}
\affiliation{National High Magnetic Field Laboratory, Tallahassee, Florida, 32310, USA}

\author{Oskar Vafek}
\email{vafek@magnet.fsu.edu}
\affiliation{National High Magnetic Field Laboratory, Tallahassee, Florida, 32310, USA}
\affiliation{Department of Physics, Florida State University, Tallahassee, Florida 32306, USA}

\begin{abstract}
	We use the density matrix renormalization group (DMRG) to study the correlated electron states favored by the Coulomb interaction projected onto the narrow bands of twisted bilayer graphene within a spinless one-valley model. The Hilbert space of the narrow bands is constructed from a pair of hybrid Wannier states with opposite Chern numbers, maximally localized in one direction and Bloch extended in another direction. Depending on the parameters in the Bistritzer-Macdonald model, the DMRG in this basis determines the ground state at one particle per unit cell to be either the quantum anomalous Hall (QAH) state or a state with zero Hall conductivity which is nearly a product state. Based on this form, we then apply the variational method to study their competition, thus identifying three states: the QAH, a gapless $C_2\mathcal{T}$ symmetric nematic, and a gapped $C_2\mathcal{T}$ symmetric stripe. In the chiral limit, the energies of the two $C_2\mathcal{T}$ symmetric states are found to be significantly above the energy of the QAH.
	However, all three states are nearly degenerate at the realistic parameters of the Bistritzer-Macdonald model. The single particle spectrum of the nematic contains either a quadratic node or two close Dirac nodes near $\Gamma$. Motivated by the Landau level degeneracy found in this state, we propose it to be the state observed at the charge neutrality point once spin and valley degeneracies are restored. The optimal period for the $C_2\mathcal{T}$ stripe state is found to be $2$ unit cells. In addition, using the fact that the topological charge of the nodes in the $C_2\mathcal{T}$ nematic phase is no longer described simply by their winding numbers once the translation symmetry is broken, but rather by certain elements of a non-Abelian group that was recently pointed out, we identify the mechanism of the gap opening within the $C_2\mathcal{T}$ stripe state. Although the nodes at the Fermi energy are locally stable, they can be annihilated after braiding with other nodes connecting them to adjacent (folded) bands. Therefore, if the translation symmetry is broken, the gap at one particle per unit cell can open even if the system preserves the $C_2\mathcal{T}$ and valley $U(1)$ symmetries, and the gap to remote bands remains open.
\end{abstract}

\maketitle
\section{Introduction}
Since the discovery of correlated insulating phases and superconductivity (SC) in magic angle twisted bilayer graphene (TBG)~\cite{Pablo1,Pablo2,David,Young,Cory1,Cory2,Dmitry1,Ashoori,Dmitry2,Yazdani,Eva,Yazdani2,Shahal,Young2,Stevan} and other moire systems~\cite{Pablo3,Guanyu,Kim,Feng,Feng2,Feng3}, tremendous theoretical\cite{BMModel} effort has been devoted towards understanding the properties and the mechanisms of these correlated electron phenomena~\cite{Leon1,LiangPRX1,KangVafekPRX,Senthil1,FanYang,Louk,LiangPRX2, GuineaPNAS, Kivelson,Fernandes1,Fernandes2,Chubukov,Ma,Guo, Kuroki,Qianghua,Stauber,KangVafekPRL,Bruno,Senthil2,Ashvin1,Cenke,MacDonald,Zalatel1,Zalatel2,Senthil3Ferro,Sau,Ashvin2,Zalatel3,Dai2,YiZhang,Fengcheng}. Significant progress has been achieved in understanding the topological band properties of this material and other moire systems~\cite{Senthil1,SenthilTop,Grisha,BJYangPRX,Bernevig1,Leon2,Dai1}. Furthermore, several approaches~\cite{KangVafekPRL,Senthil3Ferro,Zalatel3,Dai2, Neupert} have revealed the similarity between the quantum hall ferromagnetism and the insulating states observed at the even integer fillings. However, two entirely different insulating phases have been observed at the filling of $\nu = 3$~\cite{Cory1,Young,Dmitry1}. While the (quantum) anomalous Hall (QAH) state has been readily identified when one of the layers of the  TBG is aligned with the hexagonal boron nitride (hBN) substrate\cite{David,Young}, the observed gapped insulating state at $\nu=3$  without the hBN alignment -- and without anomalous Hall conductance -- is much less understood.

The experiments, as well as the band calculations with lattice corrugation\cite{LiangPRX1}, have shown that the TBG near the magic angle contains four spin degenerate narrow bands separated from other remote bands by a finite band gap. The four copies of Dirac nodes at $\fvec K$ and $\fvec K'$ per each of the two spin projections of the TBG are protected by $C_2\mathcal{T}$ -- two fold rotation about the axis perpendicular to the graphene plane followed by time reversal -- and the conservation of the number of fermions within each valley i.e. $U_{v}(1)$~\cite{Senthil1, SenthilTop}. We wish to stress that this means that $C_2\mathcal{T}$ and $U_{v}(1)$ symmetries are {\em necessary} for stable nodes to exist, but they are not {\em sufficient}. As we discuss below, the spectrum may be gapped despite the presence of the $C_2\mathcal{T}$ and $U_{v}(1)$ symmetries, and despite maintaining the gap to remote bands, when the moire lattice translation symmetry is broken.

A more familiar example of Dirac nodes protected by symmetries is the monolayer graphene, with two massless Dirac fermions per spin projection and without spin-orbit coupling. In this case, the nodes are said to be protected by the time reversal ($\mathcal{T}$) and inversion ($\mathcal{I}$) symmetries. Nevertheless, strong breaking of the rotational symmetry can in principle result in an insulating state~\cite{Neto}, despite preserving $\mathcal{T}$ and $\mathcal{I}$ throughout the process of gap opening. This happens when the Dirac nodes with opposite chirality move across the Brillouin zone, meet, and annihilate.

Unlike in the monolayer graphene example, however, the two Dirac nodes in magic angle twisted bilayer graphene have the same chirality and thus cannot be annihilated by simply meeting together. Strong breaking of rotational symmetry alone will therefore not produce an insulator. Thus, in the simplest scenario with polarized spin and valley, it seems that the gap at the Dirac nodes can be opened only by breaking the $C_2 \mathcal{T}$ symmetry. The mentioned QAH, observed at the filling of $\nu =3$ with hBN alignment, is an example of such $C_2 \mathcal{T}$ symmetry breaking. However, as mentioned, without the hBN alignment experiments demonstrated that the system at $\nu = 3$ is in a gapped state without anomalous Hall effect\cite{Cory1, Dmitry1}.

One of the goals of this paper is to explore the mechanism of gap opening in a $C_2 \mathcal{T}$ and valley $U(1)$ symmetric system and how the energy of the resulting state competes with QAH at odd integer filling. Such a state would be insulating and not display the anomalous Hall effect, and thus be consistent with the experiments at $\nu=3$ without hBN alignment; the connection to the spinless one-valley model is to simply spin and valley polarize one hole per moire unit cell. We find that in the chiral limit\cite{Grisha}, the density matrix renormalization group (DMRG) identifies the QAH as the ground state. In the more realistic case, however, DMRG always produces a non-QAH state, even when the initial state for the algorithm is set to be the QAH state. This result is also confirmed by minimizing the energy of the trial wavefunction inspired by studying the correlations in the non-QAH state obtained in DMRG. Our variational analysis discovers three competing states: the QAH and two $C_2\mathcal{T}$ symmetric states with dramatically different fermion excitation spectra. Furthermore, applying the insights of recent work by Wu, Soluyanov and Bzdu{\v s}ek\cite{Tomas}, we can identify the mechanism of the transition between these two $C_2\mathcal{T}$ symmetric states via assignment of the non-Abelian topological charges to Dirac nodes once moire lattice translation symmetry is broken. This naturally explains why the gap can be opened while preserving $C_2\mathcal{T}$ and valley $U(1)$ symmetries. Interestingly, these $C_2\mathcal{T}$ symmetric states, with polarized spin and valley degrees of freedom, are variationally nearly degenerate with the QAH state even though they are not connected to the QAH by $U(4)$ symmetry~\cite{KangVafekPRL,Zalatel3} (or $U(4) \times U(4)$ symmetry in the chiral limit~\cite{Zalatel3}). As a consequence, the manifold of the low energy states in the realistic TBG appears to be larger than QAH-related states. We should also mention in passing that our earlier approach based on maximally localized Wannier states in all directions --relation to which we discuss in the section below-- did identify a period 2 stripe state as an insulating candidate for the odd integer filling\cite{KangVafekPRL}.

Although not gapped, the single particle excitation spectrum of the $C_2\mathcal{T}$ symmetric nematic state obtained variationally is also interesting in that it displays either a quadratic node or two close Dirac nodes\cite{Ashvin2} near $\Gamma$ point, i.e. the center of the moire mini-Brillouin zone, when the electron-electron interactions dominate the kinetic energy of the narrow band states.
This is in sharp contrast to the single particle spectrum obtained when the kinetic energy of the narrow bands dominates, in which case the two Dirac cones sit at the corners of the moire mini-Brillouin zone. In the latter case, the sequence of the Landau levels, restoring the spin and valley degeneracy, would be $\nu=\pm 4,\pm 12,\pm 20,\ldots$, inconsistent with the experimentally~\cite{Pablo1, Cory1} observed sequence near the magic angle ($\sim 1.1\degree$) $\nu = \pm 4, \pm 8, \pm 12, \pm 16 \ldots$.
In the former case, however, the quadratic node at the moire Brillouin zone center would indeed produce the experimentally observed sequence because the Landau levels are doubly degenerate at zero energy, and non-degenerate at all other energy levels\cite{McCannFalko2006} (not including the spin and valley degeneracy).
Two close Dirac nodes would also produce the experimentally observed sequence\cite{SenthilC3,Ashvin2}, except for a very small magnetic field below which the sequence would revert to the $\nu=\pm 4,\pm 12,\pm 20,\ldots$. In practice, no Landau quantization is seen at very small magnetic field, so two close nodes are also consistent with the data at the charge neutrality point (CNP).
Interestingly, because this explanation relies on the electron-electron Coulomb interaction dominating the kinetic energy of the narrow bands, it would suggest that a useful probe of their relative strength at different twist angles is the Landau level sequence. Indeed, at the higher twist angle ($\sim 1.8\degree$) the observed sequence reverts\cite{Pablo2016} to $\nu=\pm 4,\pm 12,\pm 20,\ldots$, suggesting that at this higher angle the kinetic energy dominates.

Note that our goal is not to identify strictly a {\it single} state that has the lowest energy for our Hamiltonian. Rather it is to identify a group of competing low energy states if they lie close in energy~\cite{LeonReview}. This is because small terms in the Hamiltonian beyond currently accepted theoretical models, and beyond control of the experimentalists, can tip the balance and select different ground state from this near degenerate group. There are experimental indications that this is indeed happening, in particular because nominally same fabrication protocols result in different phase diagrams, for example among the Columbia/UCSB and the Barcelona groups\cite{Cory1,Dmitry1}. Our strategy is therefore to identify the leading candidates for the ground state based on comparing the competing states' robust phenomenological properties with existing experiments.

We reach the above conclusions by starting with the (energy eigen-) Bloch states for the narrow band obtained from the Bistritzer-Macdonald (BM) model~\cite{BMModel}. This continuum model has two parameters, $w_0$ and $w_1$, related to interlayer $AA$ and $AB$ couplings respectively. Due to the lattice relaxation, $w_0$ is generally smaller than $w_1$, and  $w_0/w_1\sim 0.83$ as obtained by STM\cite{Yazdani}. Assuming both the spin and valley are polarized (i.e. spinless one valley model), we consider how the ground state at one-particle per unit cell could depend on this ratio. For each different value of the ratio, we solve the BM model to obtain the Bloch states, construct the hybrid WSs, and project the Coulomb interactions onto the hybrid WSs.
By neglecting the impact of the remote bands, the basis of the hybrid WSs allows us to run DMRG with projected interactions only.
In addition, we propose a trial wavefunction for the ground state based on the outcome of DMRG. Starting from this trial state, we minimize the energy to study the ground states and fermion excitations with both interactions and kinetic terms.

The rest of the paper is organized as follows: in the next section we describe the continuum model within which we compute the hybrid Wannier states, discuss their relation to the exponentially localized states in all directions\cite{KangVafekPRX,LiangPRX1}, and express the kinetic energy and the electron-electron Coulomb interaction in the hybrid Wannier basis. In Section III we describe the results of our DMRG calculation. In Section IV, we analyze the trial state inspired by the results from DMRG and compute its single fermion excitation spectrum. In section V we analyse improved trial states which further lower the energy. We also study their excitation spectrum and its evolution from gapless $C_2\mathcal{T}$ nematic to gapped $C_2\mathcal{T}$ stripe using the topological methods discussed above. Finally, Section VI is reserved for discussion. Various technical details of our calculations are presented in the Appendix.

\section{Continuum limit Hamiltonian and the narrow band hybrid Wannier states}
\begin{figure}[htbp]
	\centering
	\subfigure[\label{Fig:UnitCell:Lattice}]{\includegraphics[width=0.5\columnwidth]{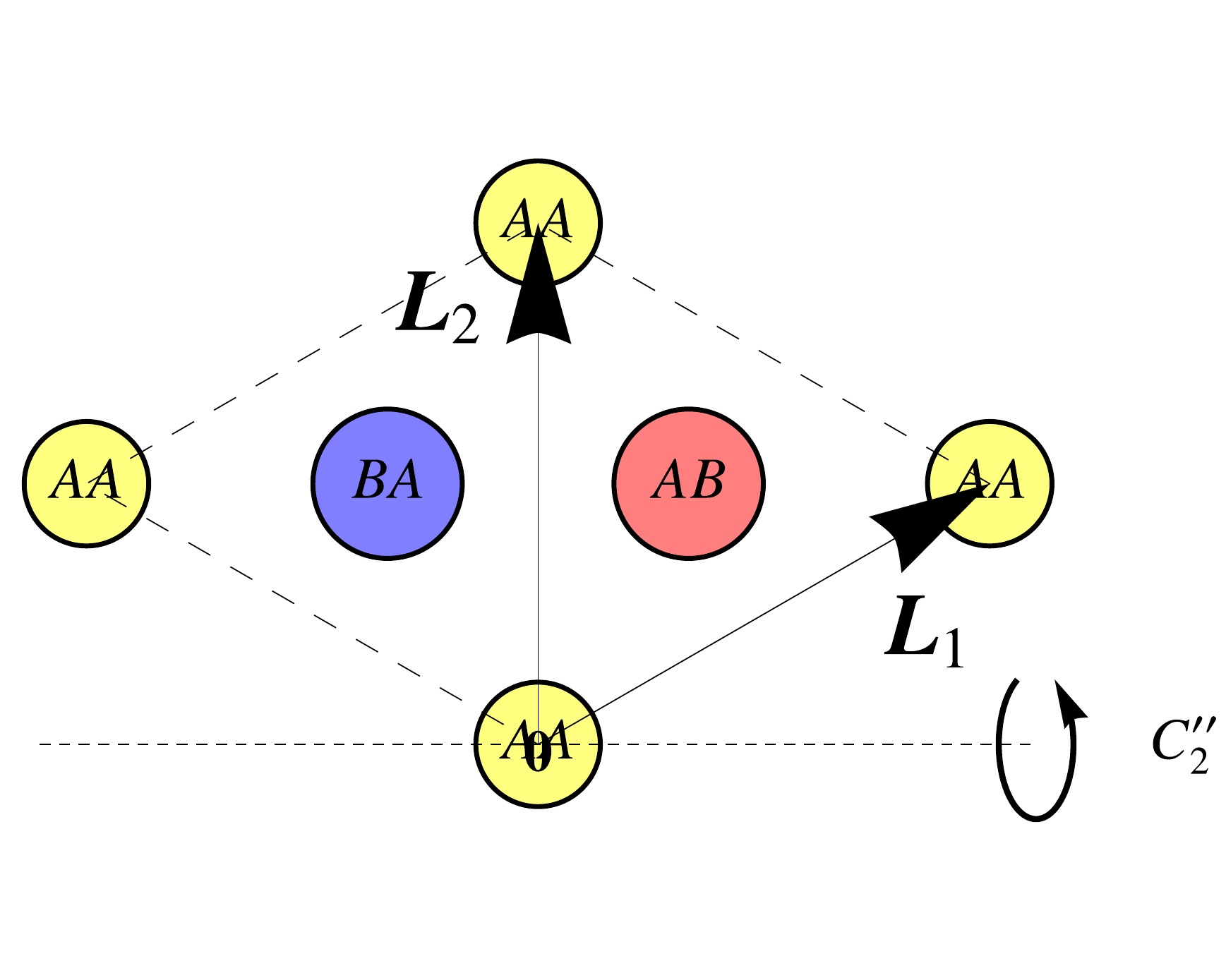}}
	\subfigure[\label{Fig:UnitCell:BZ}]{\includegraphics[width=0.45\columnwidth]{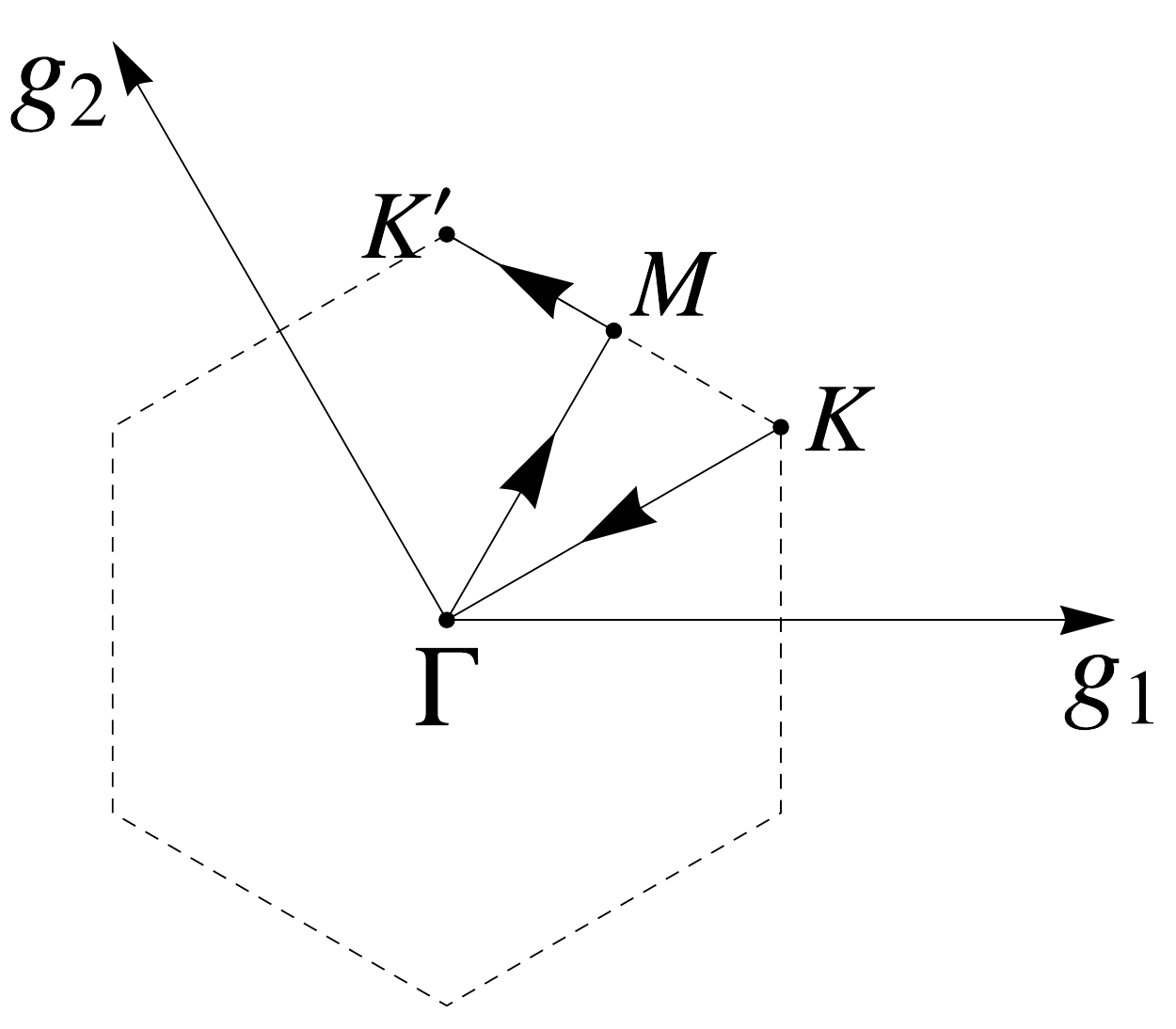}}
	\subfigure[\label{Fig:UnitCell:Band}]{\includegraphics[width=\columnwidth]{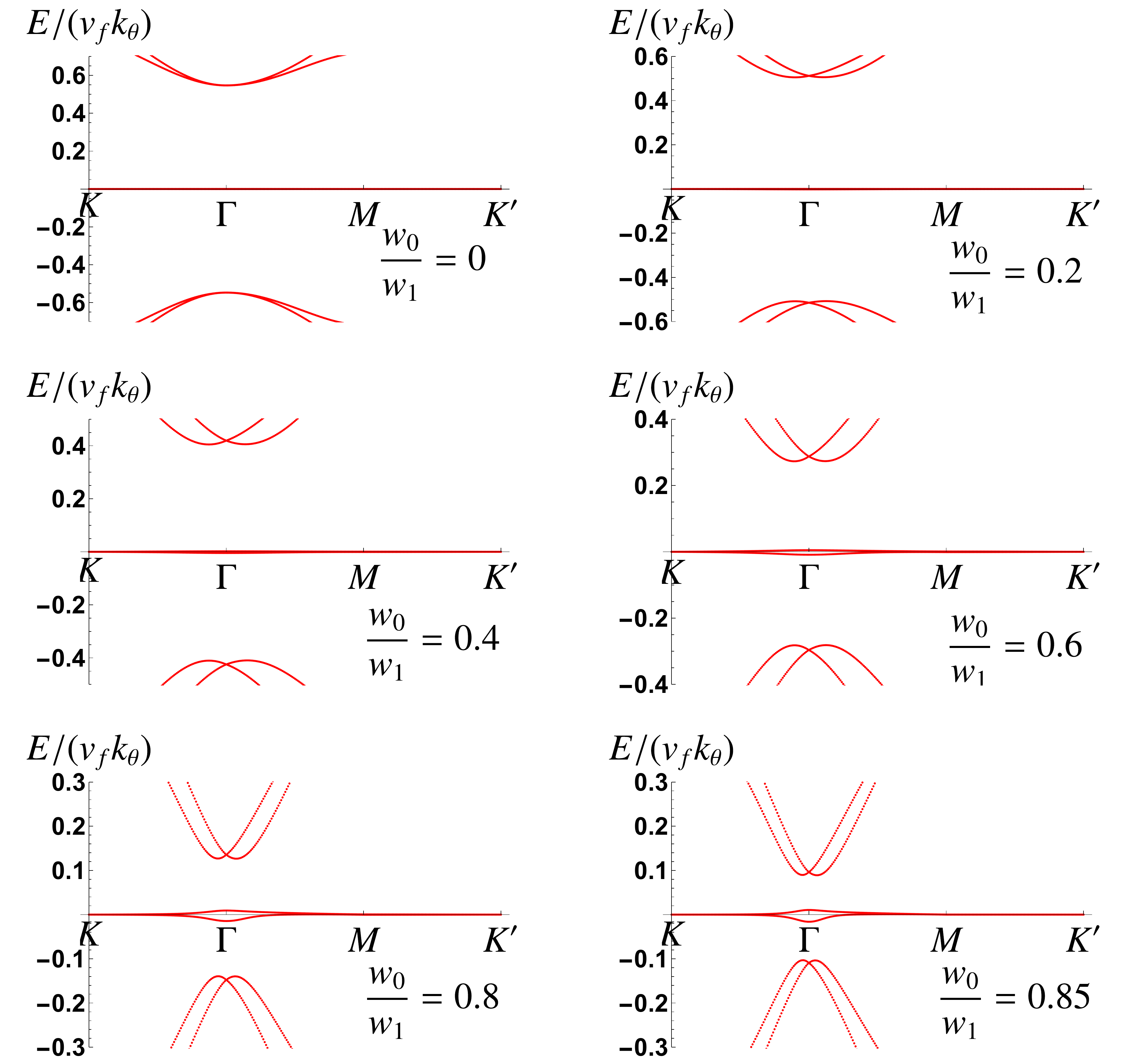}}
	\caption{(a) The schematic plot for the moire unit cell, with the lattice vectors of $\bL_1$ and $\bL_2$. At the `magic' angle, their length is about $13$nm. The colored circles refer to the AA, AB, and BA regions. Due to the lattice relaxation effects, the AB and BA regions become larger than AA region. (b) The schematic plot of the moire BZ. $\bg_1$ and $\bg_2$ are two reciprocal lattice vectors, and the region enclosed by dashed line is the first BZ. The band dispersions along the arrowed path with different $w_0/w_1$ are illustrated in (c). Experimentally relevant value\cite{Yazdani} is $w_0/w_1\sim 0.83$. The energies are normalized by $v_f k_{\theta}$, where $k_{\theta} = 2|\bK| \sin\theta/2$, $|\bK|=4\pi/(3a)$, $a=0.246nm$, and $\theta$ is the twist angle.}
	\label{Fig:UnitCell}
\end{figure}
The starting point of our analysis is the continuum Hamiltonian~\cite{BMModel,LiangPRX1,Senthil1}
\begin{eqnarray}\label{Eqn:BistritzerMacDonald}
H_{BM} & = & \hbar v_f \sum_{l = t, b} \sum_{\bp}  \psi_{l, \bp}^{\dagger} \fvec \sigma_{l, \theta} \cdot \bp \psi_{l, \bp} \nonumber \\
& & + \sum_{\bp} \sum_{j = 1}^3 \left( \psi^{\dagger}_{b, \bp + \bq_j} T_j \psi_{t, \bp} + h.c. \right),
\end{eqnarray}
where $\psi_{l, \bp}$ is the fermion operator that annihilates the state with the momentum of $\bp$ on layer $l$.  It contains two components corresponding the two sublattices on each layer.
\[ \fvec \sigma_{t/b,\theta} = e^{-i \frac{\theta}{4} \sigma_z} (\sigma_x, \sigma_y) e^{i \frac{\theta}{4} \sigma_z} ,  \]
with $\theta$ being the twist angle. Suppose $\fvec K_t$ and $\fvec K_b$ are the two Dirac points on two de-coupled layers. The second term in Eqn.~\ref{Eqn:BistritzerMacDonald} is the inter-layer coupling with $\bq_1 = \fvec K_b - \fvec K_t = k_{\theta}(0, -1)$ and $k_{\theta} = |\fvec K_t - \fvec K_b| = 2 |K_t| \sin\theta/2$, and $\bq_{2,3} = k_{\theta}( \pm \frac{\sqrt{3}}2, \half)$.  In addition,
\[ T_j = w_0 \sigma_0 + w_1 \left( \cos \frac{2\pi(j - 1)}3  \sigma_1 + \sin \frac{2\pi(j - 1)}3 \sigma_2 \right).    \]
The Bloch state is labeled by its crystal momentum, $\bk = \bp_l + \bK_l$, where $\bp_l$ is the momentum at the layer $l$. Although each Bloch state contains multiple $p_l$'s in the BM model, these momenta with the same layer index $l$ differ from each other only by reciprocal lattice vectors, and thus $\bk$ is uniquely defined if it is restricted in the Brillouin zone (BZ).

As discussed in Refs.~\cite{KoshinoLattice, LiangPRX1, Cantele}, the parameter $w_0$ is a measure of the tunneling within the AA regions while $w_1$ within AB/BA regions. In our calculation, $w_1$ is fixed to be $0.586 v_f k_{\theta}$ and $w_0$ is allowed to vary~\cite{Grisha}. The relative area of AA to AB/BA is $\sim 0.83$ as measured in STM~\cite{Yazdani}. The spectrum of this Hamiltonian is shown in the Figure.~\ref{Fig:UnitCell:Band} for a range of parameters $w_0/w_1$ starting from the chiral limit where $w_0=0$ and the narrow band is exactly flat~\cite{Grisha}.

We assume that the Coulomb interaction $\sim 25meV$ acts mainly in the subspace of the narrow bands where its effects may be non-perturbative~\cite{KangVafekPRL,Yazdani2}. We also assume that its mixing of the remote bands can be treated perturbatively due to the presence of the gap between the narrow band and the remote bands; its value is at least $\sim 30-40$meV as extracted from the transport activation gaps\cite{Pablo1,Cory1}. Therefore, in order to study the effects of the electron-electron interactions, we~\cite{KangVafekPRX} previously constructed a complete and orthogonormal basis for the narrow band which is exponentially localized in all directions~\cite{LiangPRX1}. To do this, we used a microscopic tight-binding model at a commensurate twist angle and $D_3$ symmetry~\cite{KangVafekPRX}. We found that the Coulomb interaction projected onto such basis leads to a homogeneous SU(4) ferromagnetic state at $\nu = \pm 2$. We also proposed a ferromagnetic period 2 stripe state at $\nu = 3$ as a good candidate for the insulating state observed at this filling without the hBN alignment.

The narrow bands obtained within the continuum Hamiltonian (\ref{Eqn:BistritzerMacDonald}) carry non-trivial (fragile) topology~\cite{SenthilTop,Bernevig1}. This is due to discarding the (expectedly small) mixing between the valleys, thus making the particle number within each valley separately conserved. $H_{BM}$ is indeed invariant under $U_v(1)$, and due to the invariance of $H_{BM}$ under the $C_2\mathcal{T}$ symmetry transformation, the non-trivial topology of the narrow bands is intimately linked to the combined symmetry protecting two Dirac cones with the same winding number~\cite{SenthilTop}. However, unlike in the case of a Chern insulator, the non-trivial topology here does {\em not} obstruct the construction of exponentially localized Wannier states (WS) in both directions\cite{Vanderbilt}, but it does obstruct such states from transforming in a simple way under {\em both} $U_v(1)$ and $C_2\mathcal{T}$. For example, if the WSs transform simply under $U_v(1)$ by acquiring an overall phase, then they cannot simply acquire a phase under $C_2\mathcal{T}$. Because the transformations which relate the Bloch states of $H_{BM}$ and such WSs are perfectly unitary, no information is lost, and the $C_2\mathcal{T}$ transformed WS can still be expressed exactly as a linear superposition of the exponentially localized WSs in both directions. The role of the non-trivial topology is to prevent this linear superposition to be confined to a single site. Instead, the $C_2\mathcal{T}$ transformed WS is reconstructed from a linear superposition of WSs whose centers lie within the region surrounding the transformed WS. The size of such region is determined by the exponential decay length of the WS, and the convergence towards full symmetry is achieved exponentially fast with increasing such region~\cite{Xiaoyu}.
In this respect it is perhaps helpful to reiterate that if the problem is solved on a microscopic tight-binding lattice with $\sim 10^4$ carbon sites within the unit cell\cite{KangVafekPRX} instead of in the continuum approximation, in, say, the $D_3$ configuration, then $U_v(1)$ are $C_2\mathcal{T}$ are emerging, but they are not exact; the exponentially localized WSs in both directions obtained in Ref.~\cite{KangVafekPRX} thus transform simply under all {\em exact} symmetries of the starting model. This approach based on exponentially localized WSs allowed us to obtain an explicit understanding of the form of the real space interaction and, importantly, to identify the generalized spin-valley {\it ferromagnetism} as the dominant ordering tendency in the strong coupling limit. We also linked this tendency to the nontrivial topological band properties~\cite{KangVafekPRL}.

In order to gain a clearer understanding of the effects of the Coulomb interaction on the $U_v(1)$ and $C_2\mathcal{T}$ symmetries of the low energy states, in this paper we chose to work in a Wannier basis which is localized only in one direction. In the other direction, our hybrid WSs behave as extended Bloch waves (see Fig.\ref{Fig:HybridWS}). Additional advantages of this basis are that the topology of the narrow bands of $H_{BM}$ is more transparent~\cite{Bernevig1}, and that states with broken translational symmetry in the localized direction can be readily described.
Moreover, in the basis of the hybrid Wannier orbitals the QAH state is completely unentangled. Because at $w_0/w_1=0$ the QAH state can be analytically shown to be the exact ground state of projected interactions\cite{Sau,Zalatel3}, and because it is gapped in this model, it is stable at small but finite $w_0/w_1$. We can therefore study within DMRG whether it `melts away' as $w_0/w_1$ increases beyond a critical value by initializing the DMRG with QAH. Since, as we will see, it does, we know with certainty that there is a quantum phase transition into a state different from QAH even for finite bond dimension which limits every numerical calculation, because low bond dimension would favor QAH.
The disadvantage is the complicated form the Coulomb interaction takes in the hybrid Wannier basis making its effect less transparent.

\subsection{Hybrid WSs for the narrow bands}
\begin{figure*}[htbp]
	\centering
	\subfigure[\label{Fig:HybridWS:k00}]{\includegraphics[width=\columnwidth]{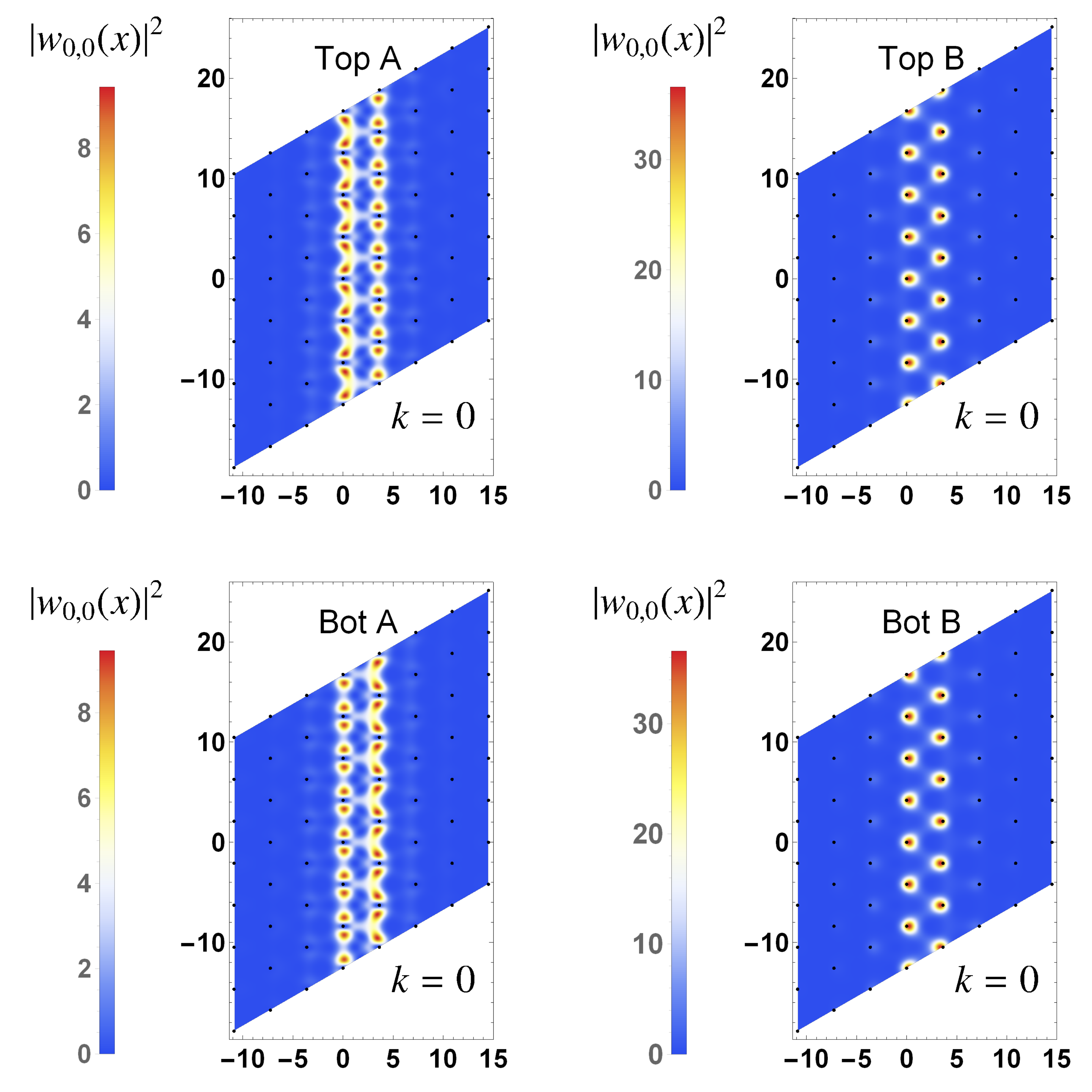}}
	\subfigure[\label{Fig:HybridWS:k05}]{\includegraphics[width=\columnwidth]{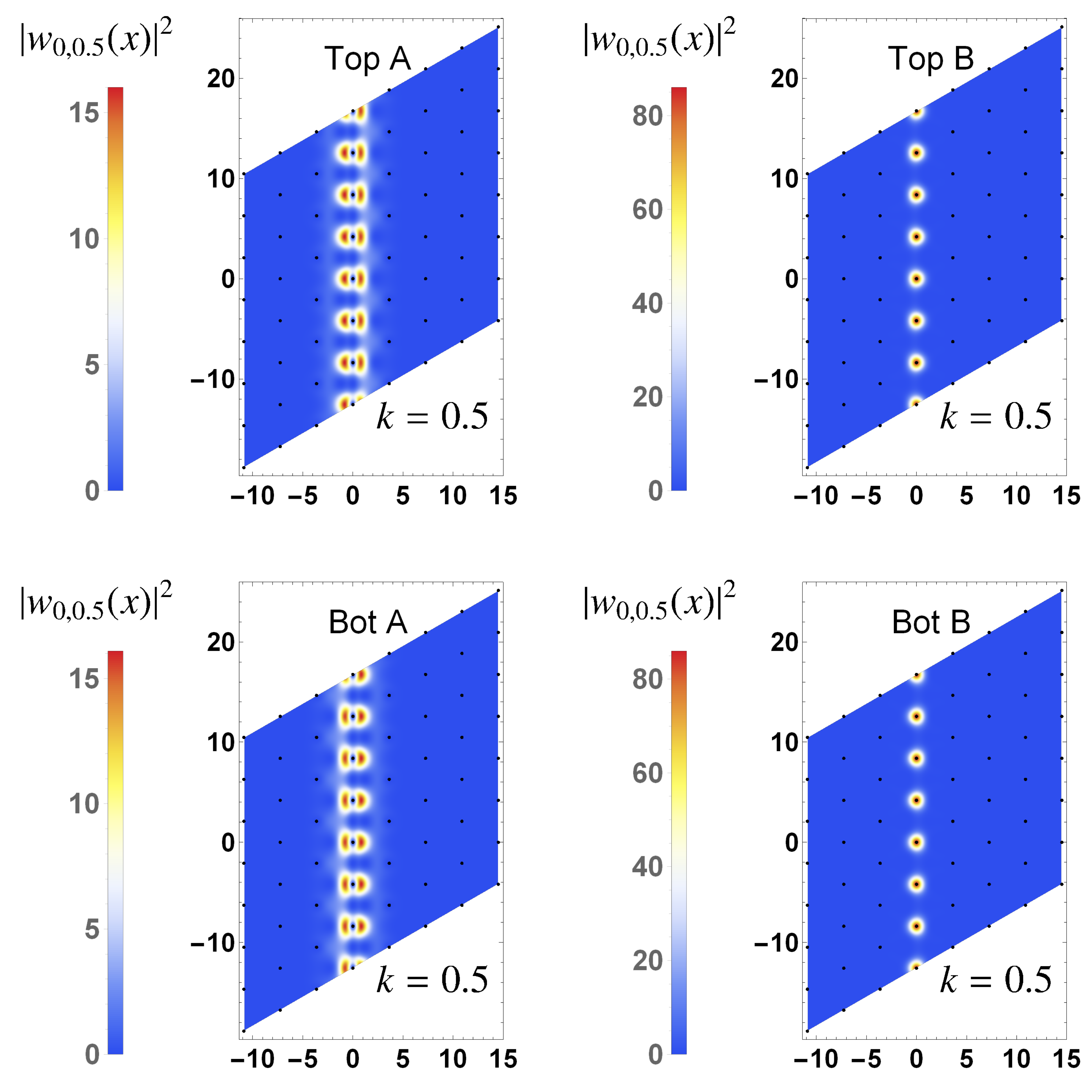}}
	\caption{The hybrid WSs in real space with $w_0/w_1 = 0.85$ and the Chern index of $-1$. ``Top/Bot'' refers to the top and bottom layer respectively. ``A/B'' refers to the sublattice A and B respectively. (a) With $k = 0$, the state contains two neighboring peaks in AA regions along the localized direction $\bL_1$ (See Fig.~\ref{Fig:UnitCell:Lattice}). (b) With $k = 0.5$, the state contains only one peak in the localized direction $\bL_1$. In both cases, the state mainly occupies sublattice B, showing that the Chern index is related to the sublattice polarization. }
	\label{Fig:HybridWS}
\end{figure*}
We follow the approach outlined in Ref.\cite{HybridWS} to construct the hybrid WS, which are maximally localized along the $\bL_1$-direction and extended Bloch waves along the $\bL_2$-direction (see Fig.~\ref{Fig:UnitCell}a).

Such hybrid WSs are the eigenstates of the projected position operator satisfying periodic boundary conditions~\cite{Resta}
\begin{eqnarray}
\hat O = \hat P e^{-i \frac1N_1 \bg_1 \cdot  {\bf r}} \hat P,
\end{eqnarray}
where $\hat P$ is the projection operator onto the narrow bands. $\bg_1$ is the primitive vector of the reciprocal lattice, and $N_1$ is the number of unit cells along the direction of $\bL_1$ in the entire lattice with periodic boundary conditions.
\begin{figure}[t]
	\centering
	\includegraphics[width=\columnwidth]{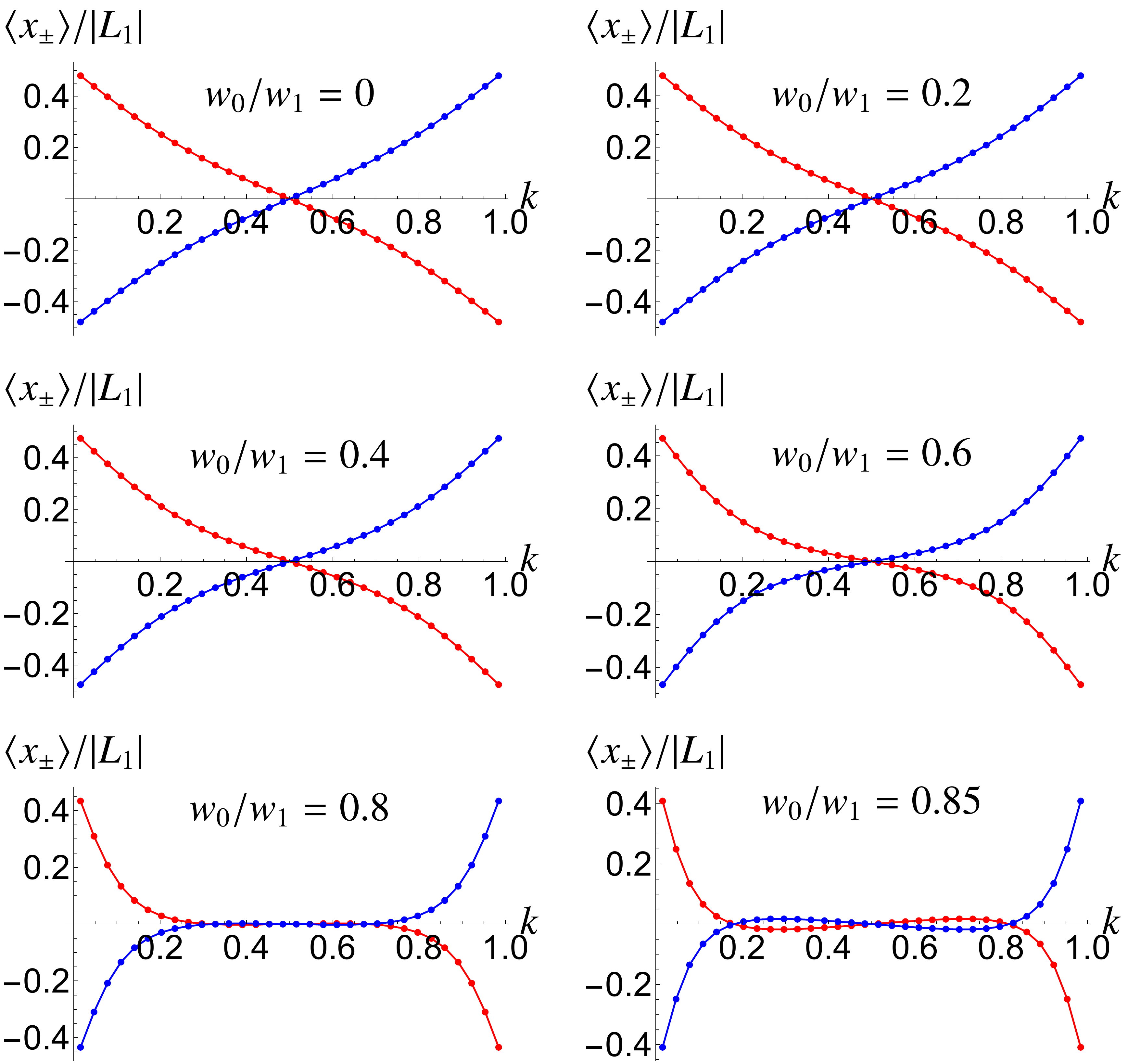}
	\caption{The phase of the eigenvalue of the Wilson loop operator of the two valley-polarized narrow bands with different values of $w_0/w_1$. Note that $\langle x \rangle$ has an odd winding number of $\pm 1$ as the momentum $k$ changes from $0$ to $1$, illustrating the nontrivial topological properties of the bands. Blue and red curves are $\langle x \rangle/|\bL_1|$ with Chern index of $+ 1$ and $-1$ respectively.}
	\label{Fig:Wilson}
\end{figure}
We thus have
\begin{eqnarray}
\hat O |w_{\pm}(n, k \bg_2) \rangle = e^{-2\pi i\frac1N_1\left(n+\langle x_\pm\rangle_k/|\bL_1|\right)}|w_{\pm}(n, k \bg_2) \rangle  .
\end{eqnarray}
The hybrid WSs $|w_{\alpha}(n, k \bg_2) \rangle$ are labeled by their momentum $k$ along $\bg_2$ which is conserved by $\hat O$ and the index $n$ of the unit cell along $\bL_1$ (see Appendix for details of the derivation~\cite{Appendix}); $\alpha=\pm 1$ labels their winding number. The amplitudes of the hybrid WSs in the real space are shown in the Fig.~\ref{Fig:HybridWS}. Unlike the familiar lowest Landau level wavefunctions in the Landau gauge, the shapes of our hybrid WSs for the narrow bands depend on the momentum index $k$. When $k$ is close to $k=0$ or $1$, the hybrid WSs contain two peaks centered around AA in the localized direction of $\bL_1$ so that $\langle x \rangle_{\pm} \approx 0.5 |\bL_1|$, whereas the hybrid WSs with $k$ close to $0.5$ contain only one peak in AA along the direction of $\bL_1$.

The $\langle x_\pm \rangle_k$ physically represents the average of the position operator within each 1D unit cell whose dependence on the conserved momentum $k$ is shown in the Fig.~\ref{Fig:Wilson}. Such shapes were previously obtained in Ref.\cite{Bernevig1}. The two curves display the winding numbers of $\pm 1$ as the momentum $\bk$ increases from 0 to $\bg_2$
i.e. the average position of one set of states slides to the right and the other set of states to the left under the increase of the wavenumber $k$, similar to Landau gauge Landau level states in opposite magnetic field~\cite{Dai1, Zalatel1}. This makes the nontrivial topology of the system explicit: within each valley, the two narrow bands of $H_{BM}$ can be decomposed into one Chern $+1$ band and one Chern $-1$ band.

Although the two narrow bands with different values of $w_0/w_1$ are topologically the same, the shapes of $\langle x_\pm \rangle_k$ clearly differ for different $w_0/w_1$. As seen in the Fig.~\ref{Fig:Wilson}, the slope of $\langle x\rangle$ near $k=0.5$ decreases with increasing $w_0/w_1$. In the chiral limit where $w_0/w_1=0$, the slope is almost the same as for Landau states, while near the more realistic value $w_0/w_1=0.8$, the slope at $k = 0.25$ nearly vanishes and the curve is very flat and thus insulating-like throughout most of the BZ. It is only close to the BZ boundary that the winding numbers are established. As shown below, the nature of the many-body ground state in the strong coupling limit is sensitive to the shape of the $\langle x_\pm \rangle_k$ curves, not just their topology.

We carefully choose the phases of the hybrid WSs so that the states $| w_{\pm}(n, k\bg_2) \rangle$ are continuous functions of the momentum $k$ and also satisfy the following properties:
\begin{align}
& | w_{\pm}(n, (k + 1)\bg_2 \rangle = | w_{\pm}(n \pm 1, k \bg_2) \rangle \label{Eqn:hybridkshift} \\
& C_2 \mathcal{T} | w_{\pm}(n, k \bg_2 \rangle = | w_{\mp}(-n, k \bg_2) \rangle  \label{Eqn:C2TSym}\\
& C_2'' | w_{\pm}(n, k \bg_2 \rangle = e^{-2\pi i n k} | w_{\mp}(n, ( 1 - k ) \bg_2) \rangle  \label{Eqn:C2ppSym}\\
& \hat{T}_{\bL_1} |w_\pm(n,k \bg_2) \rangle = |w_\pm(n + 1, k \bg_2) \rangle \\
& \hat{T}_{\bL_2} |w_\pm(n, k \bg_2) \rangle = e^{-2\pi i k}|w_\pm(n,k \bg_2) \rangle,
\end{align}
where $C_2''$ is the two fold rotation around the in-plane $x$ axis as shown in Fig.~\ref{Fig:UnitCell:Lattice} and $\hat{T}_{\bL_{1,2}}$ are translation operators by $\bL_{1,2}$. Note that the phase on the right side of Eqn.~\ref{Eqn:C2ppSym} cannot be removed, because $C_2''$ does not commute with $\hat{T}_{\bL_1}$ and, as a consequence, the extra phase becomes necessary as long as the unit cell index $n$ is non-zero.


\subsection{Kinetic energy and the Chern Bloch states}
The kinetic energy can be written in the hybrid Wannier basis as
\bea
H_{kin} & = & \sum_{n n' k} \sum_{\alpha \alpha' = \pm}  t_{\alpha \alpha'}(n - n', k)  d^\dagger_{\alpha, n, k} d_{\alpha', n', k} \ ,
\eea
where the 1D hopping matrix elements are
\bea
t_{\alpha \alpha'}(n - n', k) =\langle w_{\alpha}(n, k \bg_2) | H_{BM} | w_{\alpha'}(n', k \bg_2) \rangle,
\eea
and where $d^{\dagger}_{\alpha, n, k}$ creates the hybrid WS with the Chern index $\alpha$, the unit cell $n$, and the momentum $k\bg_2$. The kinetic energy operator is diagonal in $k$, but not in $n$. Due to $C_2 \mathcal{T}$ symmetry whose action on our basis follows (\ref{Eqn:C2TSym}), and the fact that $H_{BM}$ is Hermitian, it is straightforward to show that
\beq
\label{Eqn:nosigma3}
t_{++}(\delta n, k) = t_{--}^*(-\delta n, k) = t_{--}(\delta n, k) \   .
\eeq
There are no additional constraints on the hopping constants $t_{+ -}$ imposed by $C_2 \mathcal{T}$. Also, because of $C_2''$ symmetry, we have
\beq
\label{Eqn:C2pponhopping}
t_{+-}(\delta n, k) =  e^{2\pi i k \delta n} t_{-+}(\delta n, 1-k) \ .
\eeq

The expression (\ref{Eqn:nosigma3}) guarantees that the 2$\times$2 matrix $t_{\alpha \alpha'}(n - n', k)$ does not contain the Pauli matrix $\sigma_3=\left(\begin{array}{cc}1 & 0\\ 0 & -1\end{array}\right)$. This in turn allows us to study the winding number of the two Dirac points in $H_{kin}$ by defining Bloch states via the Fourier transform of the hybrid WSs:
\begin{eqnarray}
| \phi_{\pm}(q, k) \rangle
=\frac{1}{\sqrt{N_1}}\sum_n e^{2\pi i q n}| w_{\pm}(n, k\bg_2) \rangle,
\label{Eqn:chernBloch}
\end{eqnarray}
and expressing the kinetic energy operator in this Bloch basis.
It is important to emphasize that the states (\ref{Eqn:chernBloch}) are not kinetic energy eigenstates, but they do satisfy Bloch condition as can be seen by acting with the translation operators:
\begin{eqnarray}
\hat{T}_{\bL_1}| \phi_{\pm}(q, k) \rangle
&=&\frac{1}{\sqrt{N_1}}\sum_n e^{2\pi i q n} | w_{\pm}(n + 1, k\bg_2) \rangle \nonumber\\
&=& e^{- 2\pi i q}| \phi_{\pm}(q, k) \rangle \label{Eqn:chernBlochL1}
\end{eqnarray}
and
\begin{eqnarray}
\hat{T}_{\bL_2}| \phi_{\pm}(q, k) \rangle
&=& e^{-2\pi i k}| \phi_{\pm}(q, k) \rangle. \label{Eqn:chernBlochL2}
\end{eqnarray}
As is seen from Eq.(\ref{Eqn:chernBloch}), the states $\phi_{\pm}(q, k)$ are smooth and {\it periodic} functions of $q$ with the period $1$.
Moreover, because the hybrid WSs were constructed to be continuous functions of $k$ and satisfy (\ref{Eqn:hybridkshift}), we also have
\begin{eqnarray}
| \phi_{\pm}(q, k+1) \rangle
&=& e^{\mp 2\pi i q}| \phi_{\pm}(q, k) \rangle. \label{Eqn:chernBlochk}
\end{eqnarray}
This means that $| \phi_{\pm}(q, k) \rangle$ are Bloch states and carry Chern numbers $\pm 1$.

Defining the annihilation operators for the Chern Bloch states as
\begin{eqnarray}
b_{\alpha, q, k} = \frac{1}{\sqrt{N_1}} \sum_n e^{-i 2\pi q n} d_{\alpha, n, k}, \label{Eqn:bOprMom}
\end{eqnarray}
we can now express the kinetic energy as
\begin{eqnarray}
\label{Eqn:HkinGauge}
&&H_{kin}
=  \sum_{\alpha \alpha' = \pm} \sum_{q, k} t_{\alpha \alpha'}(q, k) b^{\dagger}_{\alpha, q, k}  b_{\alpha', q, k} \  \\
& = & \sum_{k, q} \begin{pmatrix} b_{+,q,k} \\ b_{-,q,k} \end{pmatrix}^{\dagger}
\left( \sum_{\mu = 0}^3 n_{\mu}(q, k) \sigma_{\mu}   \right)
\begin{pmatrix}  b_{+,q,k} \\ b_{-,q,k} \end{pmatrix}, \label{Eqn:KinMatrix}
\end{eqnarray}
where $t_{\alpha \alpha'}(q, k)  =  \sum_{\delta n} t_{\alpha \alpha'}(\delta n, k) e^{2\pi i q \delta n }$.  As pointed out above $n_3  = 0$. From (\ref{Eqn:C2pponhopping}) we also find
\bea
n_1(q, k) & = & n_1(q + k, 1 - k), \nonumber \\
n_2(q, k) & = & -n_2 (q + k, 1 - k).
\eea
\begin{figure}[t]
	\centering
	\includegraphics[width=0.7\columnwidth]{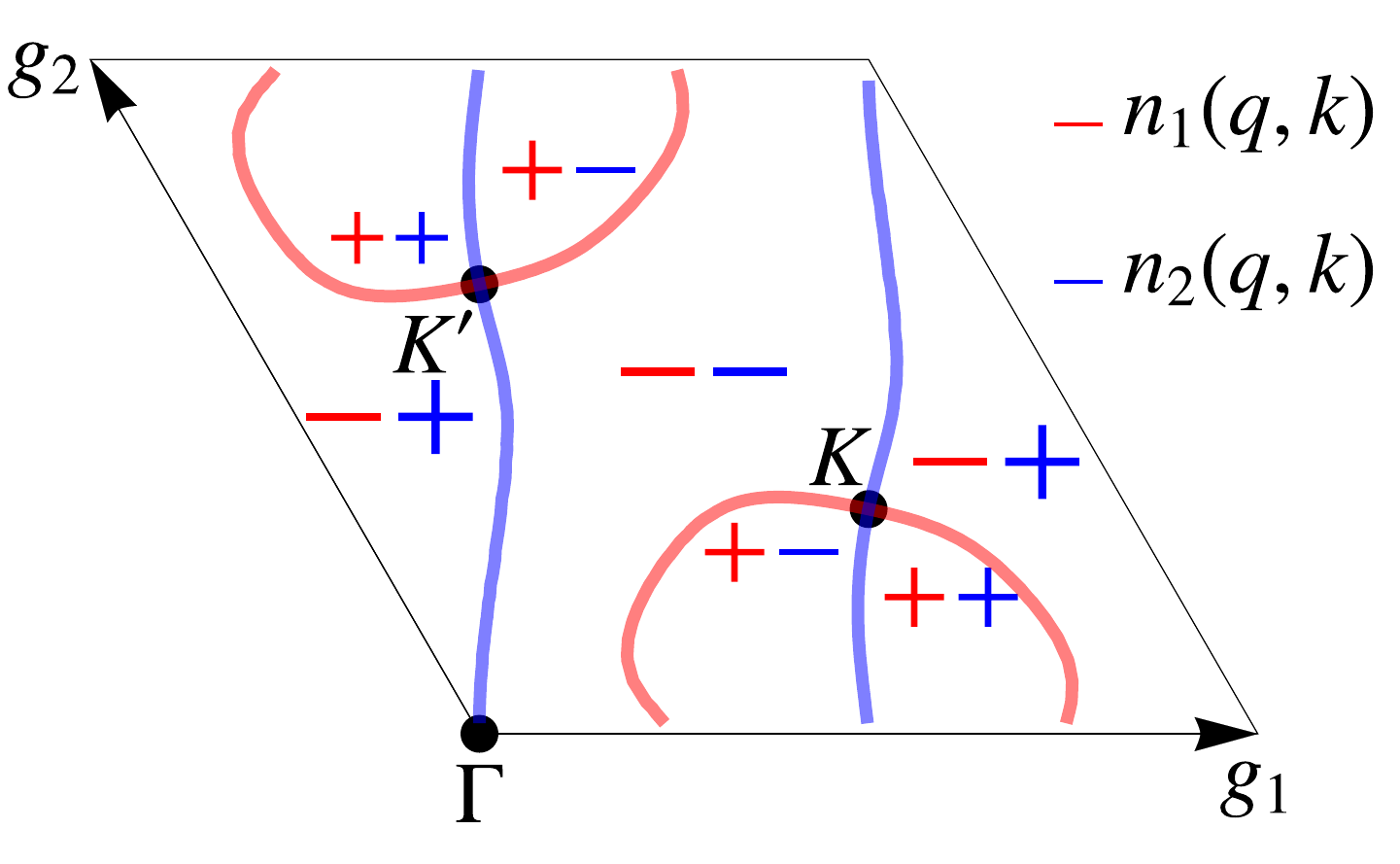}
	\caption{Chiralities of two Dirac points. The bold red and blue curves show where $n_1$ and $n_2$, defined in Eqn.~\ref{Eqn:KinMatrix}, vanish respectively, and the colored ``$\pm$'' shows the sign of corresponding $n_1$ or $n_2$ in the region separated by the bold curves. The two Dirac points $\fvec K$ and $\fvec K'$ are at the intersections of the two colored curves and have the same chiralities, as can be determined by how $n_1$ and $n_2$ change their signs going around the points.}
	\label{Fig:Dirac}
\end{figure}
Fig.~\ref{Fig:Dirac} shows the sign of $n_1$ (red) and $n_2$ (blue) as a function of the momentum $q$ and $k$ in the BZ. We see that the two Dirac points have the {\em same} chirality\cite{Senthil1,SenthilTop} in that going from, say, $++$ to $+-$, we encircle either one of the Dirac nodes clockwise.
Naively, this seems to violate the fermion doubling theorem based on which we expect opposite chirality of the Dirac nodes~\cite{VafekVishwanath2014,BJYangPRX}. However, this theorem assumes not only that the Hamiltonian $H_{kin}(\bk)$ in Eqn.~\ref{Eqn:KinMatrix} is smooth, but also that it is periodic in the momentum space. Periodicity in $q$ is guaranteed by (\ref{Eqn:chernBloch}), but not in $k$ as shown by Eqn.~\ref{Eqn:chernBlochk}. Indeed, $n_2$ would suffer a sign change at a step discontinuity if we were to identify $k=0$ with $k=1$, as can be seen in Fig.~\ref{Fig:Dirac}. The same chirality of two Dirac points characterizes the nontrivial topology of the narrow bands. It prevents construction of exponentially localized WSs in both directions {\it if} we also insist that each originates within a single valley (i.e.~no valley mixing) and with a simple transformation under $C_2\mathcal{T}$, because in such case the kinetic energy would be smooth, periodic in the BZ, and the $\sigma_3$ matrix would be absent~\cite{Senthil1,SenthilTop}. However, as discussed above, such states can be constructed if we relax the mentioned requirements.

\subsubsection{Symmetry and Fermion Spectrum}
Before proceeding to the detailed calculations, we first summarize the impact of various symmetry breaking on the fermion spectrum and thus provide a qualitative understanding of our results. The kinetic Hamiltonian $H_{kin}$ in Eqn.~\ref{Eqn:KinMatrix}  produces two $C_2\mathcal{T}$ symmetry protected Dirac nodes with the same chirality at the corner of the BZ\cite{Senthil1,SenthilTop}. Without breaking the $C_2\mathcal{T}$ or the translation symmetry, the system is metallic even in the strong coupling regime. Because the $x$-direction electrical current density $j_x$ and the perpendicular electric field $E_y$ have opposite parities under $C_2\mathcal{T}$ transformation, the Hall conductivity, defined by the formula $j_x = \sigma_{xy} E_y$, always vanishes in a $C_2\mathcal{T}$ symmetric system. Breaking $C_2\mathcal{T}$ symmetry can open a gap in our single flavor model at half filling, corresponding to $\nu = 3$ if the fermions with different spins or different valleys are assumed to be filled. This gapped phase could be either QAH if the masses ($n_3$ in Eqn.~\ref{Eqn:HkinGauge}) at two nodes are the same, or a topologically trivial phase if these two masses are opposite, consistent with the flipped Haldane model picture of Refs.~\cite{Senthil1,SenthilTop}. As shown later in the text, our numerical calculation can only find the QAH phase when $C_2\mathcal{T}$ symmetry is spontaneously broken, suggesting that the phase with opposite masses is not energetically favored by the interactions. Furthermore, a $C_2 \mathcal{T}$ symmetric stripe phase is also found to be energetically favored by the interactions and gapped; as mentioned it must have vanishing Hall conductivity.

\subsection{Coulomb interaction energy in the hybrid Wannier basis}
We start from the gate-screened Coloumb interaction, with two metallic gates placed distance $\xi$ above and below the TBG,
\bea
\hat{V} & = &\half  \sum_{\br_1 \br_2} \sum_{\mu \nu} \left( \sum_l V_{intra}(\br_1 - \br_2) : \hat{\rho}_{l \mu}(\br_1) \hat{\rho}_{l \nu}(\br_2): + \right. \nonumber \\
& & \left. \sum_{l \neq l'} V_{inter}(\br_1 - \br_2) : \hat{\rho}_{l \mu}(\br_1) \hat{\rho}_{l' \nu}(\br_2): \right) \ ,
\eea
where $V_{intra}(\br)$ ($V_{inter}(\br)$) is the gate-screened Coulomb potential for two point charges separated by in-plane distance $r$, and located in the same (different) graphene layers. The graphene layers in the TBG are assumed to be separated by a small distance $d_\perp$ of the order of a couple of carbon lattice spacings\cite{BMModel}. $l$ is the layer index, and $\mu$ is the index combining spin and sublattice degrees freedom (as mentioned, we ultimately study a spinless model, so this is just for generality). $\hat \rho(\br)$ is the charge density at $\br$ and $: \hat \rho \hat \rho :$ is the normal ordered operator $\hat \rho \hat \rho$. The Fourier transform of such gate-screened Coulomb interactions is~\cite{Appendix}
\beq
V_{intra}(q) \approx V_{inter}(q) \approx \frac{e^2}{4 \pi \epsilon} \frac{2\pi}q \tanh\frac{q \xi}2  \   \label{Eqn:Coulomb}
\eeq
for $q d_\perp \ll 1$. At large momentum $q d_\perp \gtrsim 1$, the charge density of our single valley model, $\rho(q)$, becomes negligibly small, and thus, the Coulomb interaction with large momentum transfer can be safely neglected. (In the two valley case, it also peaks at the momentum difference between the valleys and there the decrease of $V_{intra/inter}(q)$ with increasing $q$ makes such terms smaller, see e.g. Ref\cite{Zalatel3}). Eqn.~\ref{Eqn:Coulomb} is used in all the following analysis and numerical calculations with $\xi$ set to $10$nm.

To obtain the projected Coulomb interaction, we first project the bare fermion creation and annhilation operator to the constructed hybrid WSs:
\bea
& &c^{\dagger}_{\mu}(\fvec r) \quad \longrightarrow \quad \sum_{\beta = \pm} \sum_{k, n} w^*_{\beta, n, k}(\mu, \fvec r) d^{\dagger}_{\beta, n, k} \nonumber \\
& = & \sum_{\beta = \pm} \sum_{k, n}  w^*_{\beta, 0, k}(\mu, \fvec r - n \bL_1) d^{\dagger}_{\beta, n, k} \ ,
\eea
where $d^{\dagger}_{\beta, n, k}$ creates the hybrid WS $| w_{\beta}(n, k \bg_2) \rangle$ with the wavefunction of $\langle {\bf r}| w_{\beta}(n, k \bg_2) \rangle = w_{\beta, n, k}(\fvec r)$. Note that we now absorb the layer index $\mu$ and the sublattice index apparent in $H_{BM}$, Eq. (\ref{Eqn:BistritzerMacDonald}), into the four component `spinor' $w_{\beta, n, k}(\fvec r)$. The projected interaction becomes
\begin{eqnarray}
\hat{V}_{int} & = & \sum_{\substack{\beta \beta'\\ \gamma \gamma'}} \sum_{\substack{n_1 n_2\\ n_3 n_4}} \sum_{\substack{k_2 k_2'\\ p_2 p_2'}}  J_{\beta \beta', n_1 n_2, k_2 k_2'}^{\gamma \gamma', n_3 n_4, p_2 p_2'} d_{\beta n_1 k_2}^{\dagger} d_{\gamma n_2 p_2}^{\dagger} d_{\gamma' n_3 p_2'} d_{\beta' n_4 k_2'} \nonumber \\
& & \sum_{i \in Z} \delta_{k_2 + p_2, p_2' + k_2' + i}.  \label{Eqn:ProjectedInt}
\end{eqnarray}
The term in the last line ensures the momentum conservation along $\bg_2$. It is worth emphasizing that the obtained interaction in Eqn.~\ref{Eqn:ProjectedInt} has been numerically found to be sizable even if the difference of unit cell indices $|n_i - n_j| \leq 2$ for all pairs of $(i, j)$. Different from the wavefunction of the lowest Landau level (LLL),   the constructed hybrid Wannier state, shown in Fig.~\ref{Fig:HybridWS}, contains two peaks along $\fvec L_1$, leading to significant overlap between two hybrid Wannier states with consecutive unit cell indices. Correspondingly, the projected Coulomb interactions decays exponentially only when $|n_i - n_j| \geq 2$, leading to a rather complicated interaction form.

\section{DMRG}
We consider a system with the size of $N_1\bL_1 \times N_2\bL_2$ and choose the open boundary condition along $\bL_1$ and anti-periodic boundary condition along $\bL_2$.  Therefore, the momentum indices of the hybrid WSs $| w_{\pm}(n, k \bg_2) \rangle$ take the values:
\[ k = \frac{i + \half}{N_2} \quad \mbox{with} \quad i = 0, 1, \cdots, N_2 -1 \]
and $n = -\frac{N_1}2, -\frac{N_1}2, \cdots, \frac{N_1}2$. Since we study the quasi-1D system with DMRG, the hybrid WSs are arranged in a one-dimensional chain with each site indexed as $i + n N_2$. Also, each site contains two hybrid WSs, labeled by $\beta = \pm 1$. In the DMRG calculation, $N_2 = 6$, and $N_1 = 30$, the bond dimension set to be $2000$ and the truncation error no more than $10^{-4}$.
Calculations were performed using the ITensor Library\cite{ITensor} to study the ground states at the half filling of the spinless one-valley model, i.e. at the average occupation of one particle per unit cell. The Hamiltonian studied in DMRG includes only the electron-electron interactions with the kinetic terms $H_{kin}$ neglected. Because of the complicated form of the projected interaction, ITensor produces a rather large matrix product operator (MPO), $3-4$Gb at bond dimension $2000$. During each sweep, ITensor saves the MPO and the matrix product state, and therefore places an upper limit on the bond dimension we can reach with our resources.

The obtained ground state is expected to depend only on the two parameters $w_0$ and $w_1$ in the BM model (\ref{Eqn:BistritzerMacDonald}). When $w_0 = 0$, i.e.~the interlayer intra-sublattice hopping vanishes, the system is in the chiral limit~\cite{Grisha} with two Chern bands located on different sublattices. Consequently, the ground state has been shown to be the QAH state~\cite{Sau,Zalatel3}. As $w_0 /w_1$ increases, the two Chern bands start to spatially overlap, leading to the scattering among them and frustrating the QAH. Nevertheless, because QAH is a gapped phase, it is stable with respect to a small increase of $w_0 /w_1$ from the chiral limit. Whether or not it collapses and a different state is favored as $w_0 /w_1$ reaches $\sim0.83$ is the purpose of our DMRG calculation. We should note, however, that the increased propensity towards a many body insulating state with increasing $w_0 /w_1$ could also be intuited from the shapes of the phases of Wilson loop eigenvalues. In Fig.(\ref{Fig:Wilson}) we see that they progressively flatten, suggesting that a good correlation hole can be built when the two hybrid WSs are coherently (and equally) distributed among the Chern $+1$ and Chern $-1$ branches. Such a state then need not break $C_2\mathcal{T}$ and possibly insulate.

This intuitive picture turns out to be consistent with our DMRG calculation. With $N_2$ up to $6$, DMRG finds QAH as the ground state when $w_0/w_1 \leq 0.7$, i.e.~the many-body ground state turns out to be a product state of the hybrid WSs $| w_{\beta, n, k} \rangle$ with the same Chern index $\beta$:
\beq
| \Psi \rangle_{GS} = \prod_{n, k} d^{\dagger}_{+, n, k} | \emptyset \rangle  \quad \mbox{or} \quad  \prod_{n, k} d^{\dagger}_{-, n, k} | \emptyset \rangle \ .
\label{Eqn:DMRGQAH}
\eeq
With  $w_0/w_1=0.8$ and $w_0/w_1=0.85$, DMRG always produces a non-QAH state with translation symmetry breaking even if the initial state is set to be the QAH state. In this parameter regime, however, the DMRG calculation does not result in a fully converged ground state, in that the details of the final state are sensitive to the choice of the bond dimension; this is despite the entanglement entropy through the middle bond never going above $0.82$. However, several interesting properties are found to be {\it common} among all the obtained states which we now discuss.

\begin{figure}[t]
	\centering
	\includegraphics[scale=0.4]{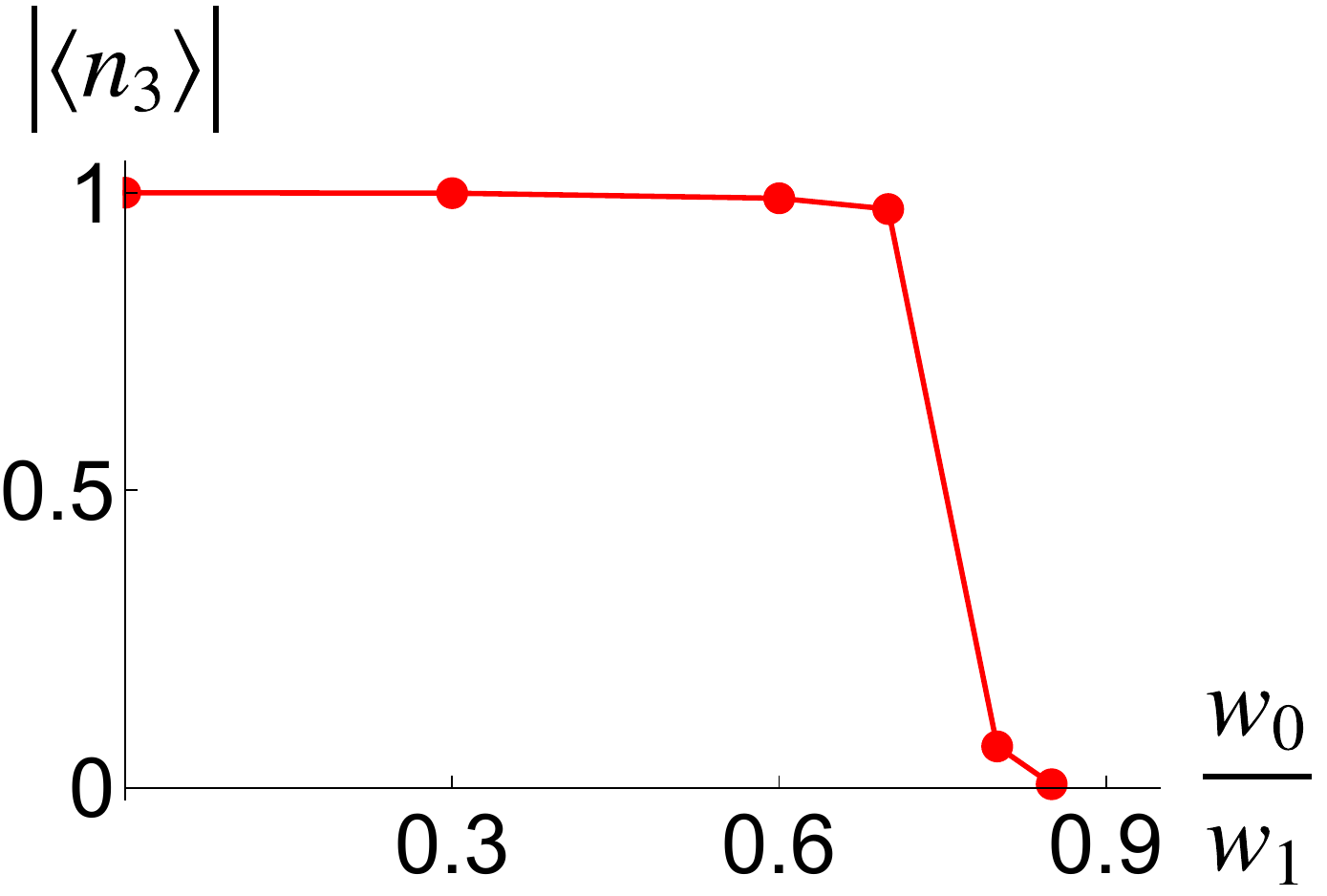}
	\caption{The order parameter $\langle n_3\rangle $ given by Eqn.~\ref{Eqn:n3DMRG} of the ground state as obtained by DMRG. $|\langle n_3\rangle | = 1$ when the ground state is QAH for $w_0/w_1 \lesssim 0.7$. However, $\langle n_3\rangle \approx 0$ when $w_0/w_1 \gtrsim 0.8$, suggesting the vanishing Hall conductivity in the system.}
	\label{Fig:DMRGn3}
\end{figure}

\begin{figure*}[htbp]
	\centering
	\subfigure[\label{Fig:DMRGCorr:ChernPlus}]{\includegraphics[width=0.48\textwidth]{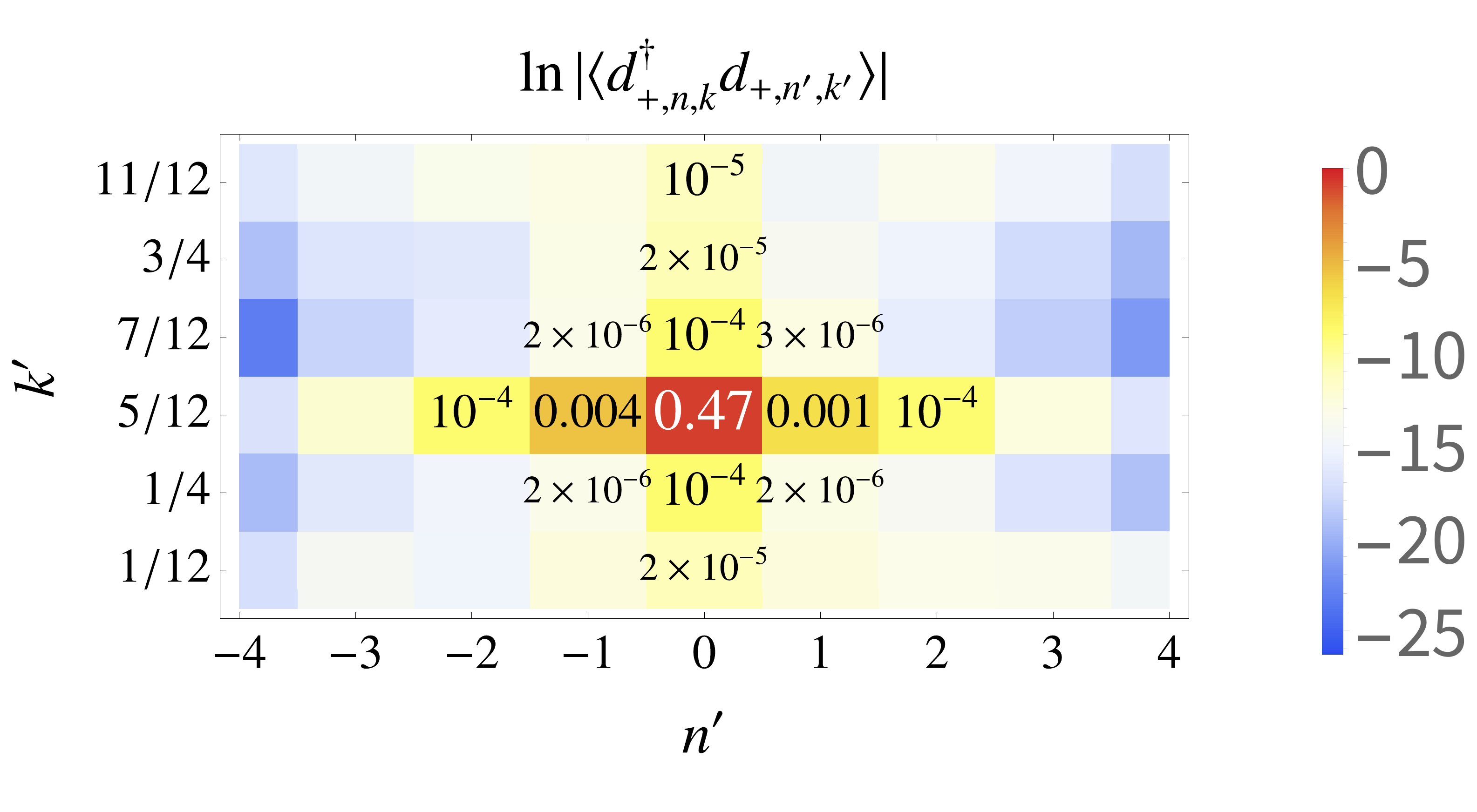}}
	\subfigure[\label{Fig:DMRGCorr:ChernTailPlus}]{\includegraphics[width=0.48\textwidth]{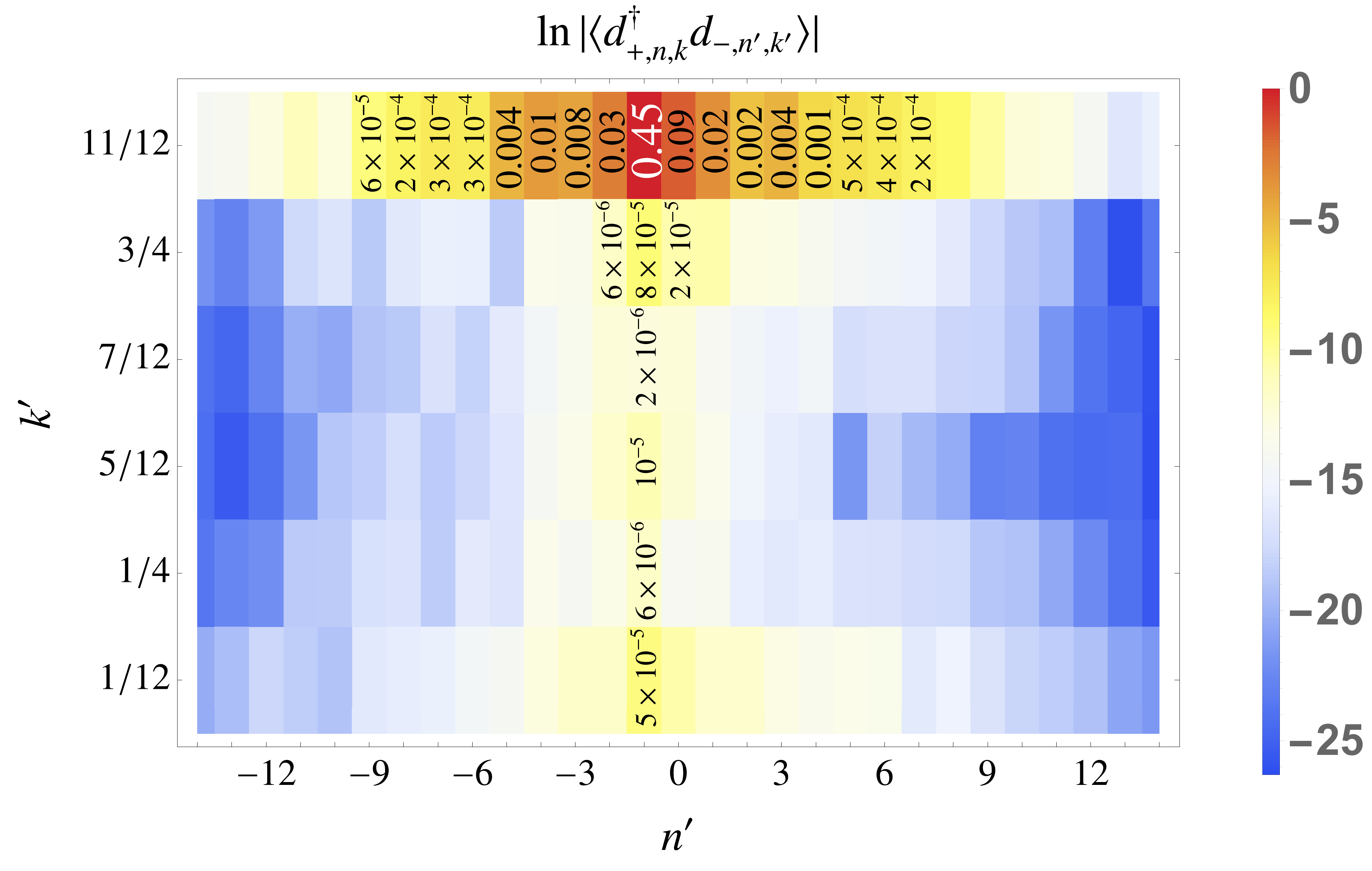}}\\
	\subfigure[\label{Fig:DMRGCor:ChernMinus}]{\includegraphics[width=0.48\textwidth]{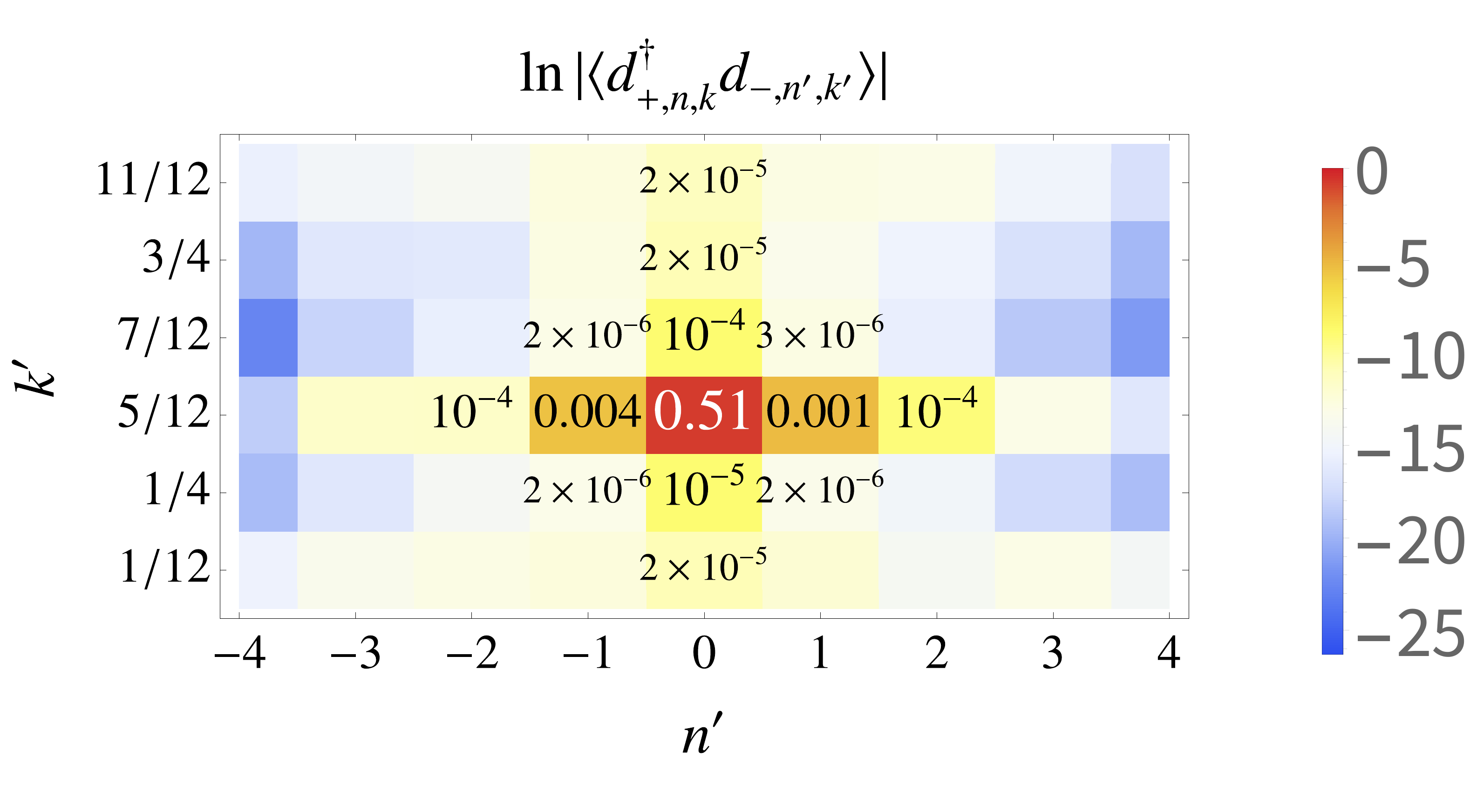}}
	\subfigure[\label{Fig:DMRGCorr:ChernTailMinus}]{\includegraphics[width=0.48\textwidth]{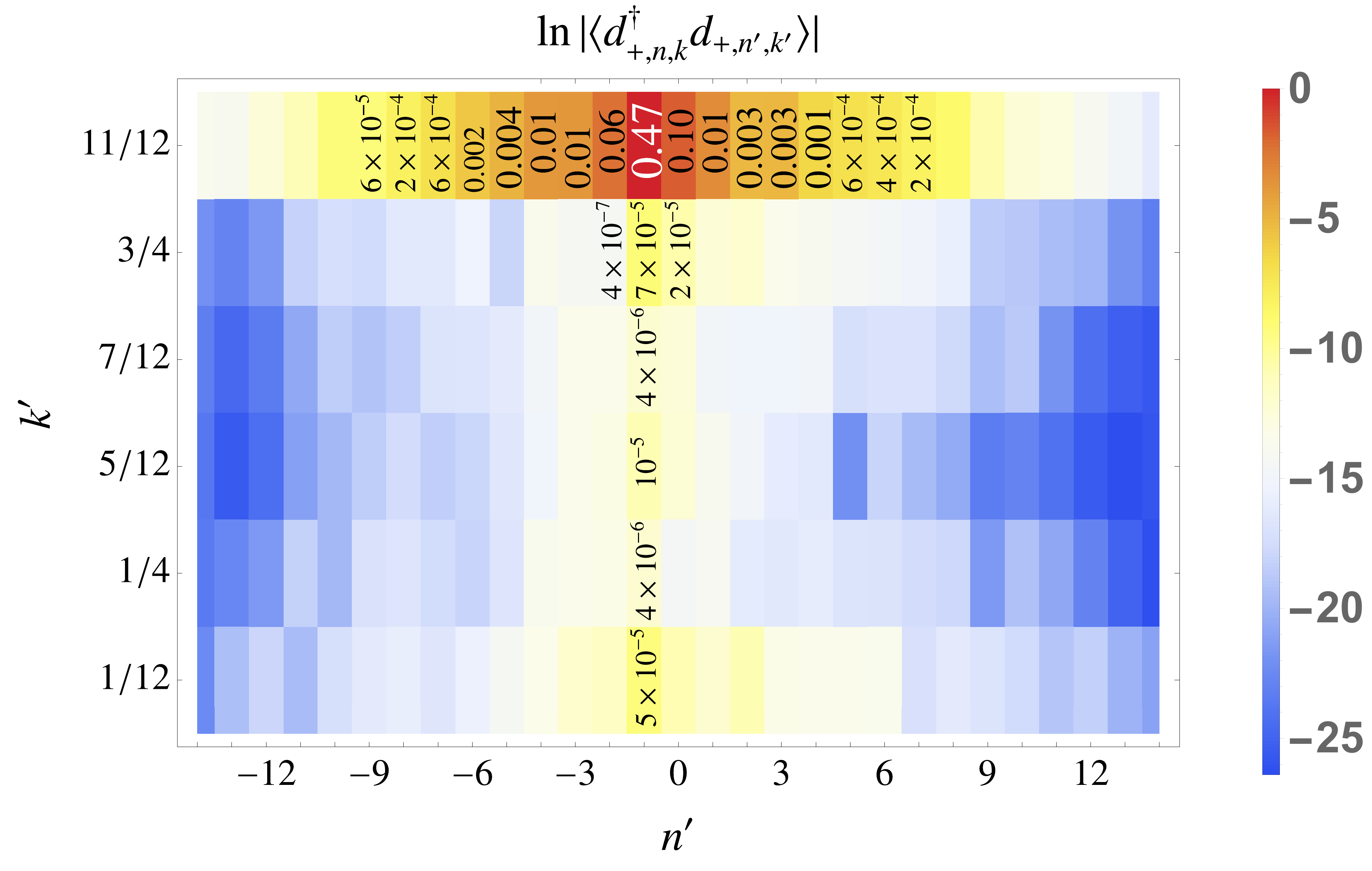}}
	\caption{The fermion correlation between different sites with $w_0/w_1 = 0.85$ in the state obtained by DMRG. We used 6 k-points and 31 n-points with the bond dimension of 2000 and the truncation error of $10^{-4}$.}
	\label{Fig:DMRGCorr}
\end{figure*}

\emph{Vanishing $\langle n_3\rangle$}: To illustrate the difference between the QAH states and the state obtained when $w_0/w_1 \geq 0.8$, we define the order parameter $\langle n_3 \rangle$:
\beq
\langle n_3 \rangle = \frac1N \sum_{n, k} \langle d^{\dagger}_{+, n, k} d_{+, n, k} - d^{\dagger}_{-, n, k} d_{-, n, k}  \rangle_{GS} \ ,
\label{Eqn:n3DMRG}
\eeq
where $N = N_1 \times N_2$ is the total number of particles in the system.  We found that $\langle n_3\rangle $ changes dramatically when $w_0/w_1$ is between $0.7$ and $0.8$. When $w_0/w_1 \leq 0.7$, $\langle n_3\rangle  \approx \pm 1$, consistent with the QAH states described in Eqn.~\ref{Eqn:DMRGQAH}. With $w_0/ w_1 \geq 0.8$, $\langle n_3\rangle $ quickly drops to $0$, suggesting that DMRG gives a topologically trivial state with vanishing Hall conductivity.

\emph{Product State}: On each site labeled by the two indices $n$ and $k$, we also find that the fermion occupation number in the ground state is almost $1$ when $0 \leq w_0/w_1 \leq 0.85$. This is obviously true for the QAH state described in Eqn.~\ref{Eqn:DMRGQAH}. Table.~\ref{Tab:DMRGFNum} lists the probability of having zero, one, and two particles on several typical sites $(n, k)$ when   $w_0/w_1 = 0.85$, where the particle number operator $\hat{N}_{n, k}$ on site $(n, k)$ is $d^{\dagger}_{+, n, k} d_{+, n, k} + d^{\dagger}_{-, n, k} d_{-, n, k}$. The probability of having $0$, $1$, and $2$ particles on the site $(n, k)$ is calculated with the following formula:
\begin{align}
P(\hat{N}_{n, k} = 0) = & \half \langle (1 - \hat{N}_{n, k}) (2 - \hat{N}_{n, k}) \rangle_{GS} \\
P(\hat{N}_{n, k} = 1) = & \langle \hat{N}_{n, k} (2 - \hat{N}_{n, k}) \rangle_{GS} \\
P(\hat{N}_{n, k} = 2) = & \half \langle \hat{N}_{n, k} (\hat{N}_{n, k} - 1) \rangle_{GS}.
\end{align}

\begin{table}[htbp]
	\begin{ruledtabular}
		\begin{tabular}{|c|c|c|c|c|c|c|}
			$(n, k)$ & $(0, \dfrac1{12})$ &  $(0, \dfrac14)$ & $(0, \dfrac5{12})$ & $(0, \dfrac7{12})$ & $(0, \dfrac34)$ & $(0, \dfrac{11}{12})$ \\  
			$P(\hat{N}_{n, k} = 0)$	& $0.038$ & $0.017$ & $0.018$ & $0.018$ & $0.018$ & $0.036$ \\  
			$P(\hat{N}_{n, k} = 1)$	& $0.922$ & $0.966$ & $0.965$ & $0.964$ & $0.965$ & $0.924$ \\  
			$P(\hat{N}_{n, k} = 2)$	& $0.040$ & $0.017$ & $0.017$ & $0.018$ & $0.017$ & $0.040$ \\  
		\end{tabular}
	\end{ruledtabular}
	\caption{The probability $P$ of having zero, one, or two particles on several typical sites labeled by $(n, k)$ when $w_0/w_1 = 0.85$. The small probability of having zero or two particles shows the negligible charge fluctuations away from the one-particle occupancy.}
	\label{Tab:DMRGFNum}
\end{table}
As shown in the Table.~\ref{Tab:DMRGFNum}, the probability of having $0$ or $2$ particles on each site are negligible, suggesting that the DMRG produced state can be well approximated by the product of one-particle state on each site. To further justify this statement, we calculated the equal-time fermion correlation between different sites, i.e.~$\langle d^{\dagger}_{\alpha, n, k} d_{\beta, n', k'} \rangle_{GS}$. This correlation is shown in Fig.~\ref{Fig:DMRGCorr} when $w_0/w_1 = 0.85$. In Fig.~\ref{Fig:DMRGCorr:ChernPlus} and \ref{Fig:DMRGCor:ChernMinus}, we fix $(n, k) = (0, \frac5{12})$, and list the absolute value of the fermion correlation with various $(n', k')$. When $(n, k)$ is fixed to be $(0, \frac{11}{12})$, the correlation is also listed in Fig.~\ref{Fig:DMRGCorr:ChernTailPlus} and \ref{Fig:DMRGCorr:ChernTailMinus}. Although this correlation increases with $k$ close to $0$ or $1$, the off-site correlation is generally found to be tiny. In addition, we found that that the correlation exponentially decays as a function of $|n - n'|$, suggesting that this is a gapped phase.

Overall, the dominant one-particle occupancy and tiny off-site correlation of the DMRG produced state suggest that the DMRG produced wavefunction can be well approximated by the following formula:
\beq
| \Psi_{GS} \rangle \approx \prod_{n, k} \left(u_{n, k} d^{\dagger}_{+, n, k} + v_{n, k} d^{\dagger}_{-, n, k}  \right) | \emptyset \rangle \label{Eqn:TrialGS}
\eeq
with $|u_{n, k}|^2 + |v_{n, k}|^2 = 1$ for all the sites labeled by $(n, k)$.

\emph{Phase of $\langle d^{\dagger}_{+, n, k} d_{-, n, k}\rangle $}: We also found that the phase of  $\langle d^{\dagger}_{+, n, k} d_{-, n, k}\rangle$ in the non-QAH phase can be described by a function
\[ \arg\left( \langle d^{\dagger}_{+, n, k} d_{-, n, k}\rangle  \right) \approx \arg\left( \frac{v(n, k)}{u(n, k)} \right) \approx f(n) \frac{\pi}2 \ , \]
where $f(n) = \pm 1$. Obviously, the phase of $\langle d^{\dagger}_{+, n, k} d_{-, n, k}\rangle$ depends on the choice of the phase of the constructed hybrid WSs, and thus is not $U(1)$ gauge invariant. Once the $U(1)$ phase of the hybrid WSs is fixed~\cite{Appendix}, $f(n)$ is found to depend only on $n$ but does not show any regular pattern in DMRG produced final state. In the next section, we will see that the magnitude of this phase, $\frac{\pi}2$, is reproduced by minimizing the $\langle H \rangle$ with the trial state in Eqn.~\ref{Eqn:TrialGS}.

As stated previously, the DMRG produced state does not converge and is very sensitive to the bond dimension. Additionally, we found that the $C_2 \mathcal{T}$ local order parameter $\langle d^{\dagger}_{+, n, k}d_{+, n, k}- d^{\dagger}_{-, -n, k}d_{-, -n, k}\rangle $ strongly depends on $n$ and does not show any spatially periodic pattern. This may come from the strong competition between various low energy states, such as QAH and other $C_2 \mathcal{T}$ symmetric states. As suggested by the drop of $\langle n_3\rangle$ in Fig.~\ref{Fig:DMRGn3} and analysis in the following sections, the system undergoes a first order phase transition from QAH to a non-QAH state when $w_0/w_1$ reaches approximately $0.8$. Since the global order parameter $\langle n_3\rangle $ vanishes in the non-QAH state, we suspect that $C_2 \mathcal{T}$ symmetry may be still locally conserved in this state, leading to zero Hall conductivity. When $w_0/w_1$ becomes slightly larger than $0.8$, the system is in the non-QAH regime but close to the phase transition point. As a consequence, these states are still almost degenerate, leading to strong competition among them. Furthermore, the QAH state, as discovered in DMRG, is favored at the open ends of the system. Therefore, the small advantange of the non-QAH states in the bulk and the superiority of the QAH state at the boundary may drive the system into the intermediate phase without regular spatial patterns in the DMRG produced state.

\section{Analysis of the trial state in Eqn.(\ref{Eqn:TrialGS})}

\subsection{Ground State}
\begin{figure*}[htbp]
	\centering
	\includegraphics[width=0.99\textwidth]{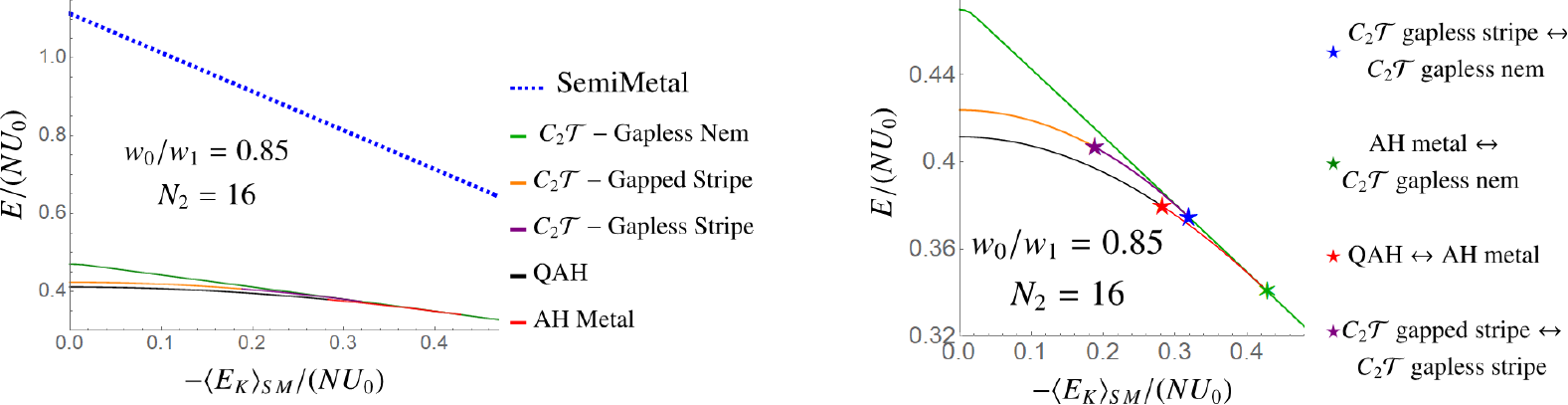}
	\caption{The energy of various correlated states with the trial function in Eqn.~\ref{Eqn:TrialGS}. The energies are normalized by $U_0 = e^2/(4\pi \epsilon L_m)$, where $L_m=|{\bf L}_1|$. Left: the energies of four different states: $C_2 \mathcal{T}$ broken state, $C_2\mathcal{T}$ symmetric nematic state, $C_2 \mathcal{T}$ symmetric period-2 stripe state, and the semi-metal; the semi-metal is defined as the (non-interacting) state obtained by diagonalizing the kinetic energy only. The results are plotted vs $\langle E_{K}\rangle_{SM}\equiv\lambda \langle \hat H_{kin}\rangle_{SM}$ as $\lambda$ increases from $0$ and $\langle \ldots \rangle_{SM}$ is the expectation value in the semi-metal state.
		Right: the energies of three nearly degenerate states, and the transition between them are marked with colored stars. As seen here, at $w_0/w_1 = 0.85$, these three states are nearly degenerate. In contrast, at $w_0/w_1 = 0.3$, the near degeneracy is lifted in favor of QAH whose energy is $0.17 U_0 \approx 3$meV below the two $C_2\mathcal{T}$ symmetric states (see the corresponding Fig.~\ref{FigS:DMRGEne030} in the Appendix~\cite{Appendix}).}
	\label{Fig:DMRGEne}
\end{figure*}
Because the trial ground state (Eqn.~(\ref{Eqn:TrialGS})) suggested by DMRG is a product state at each site $(n, k)$, it is straightforward to analyze its energy variationally by implementing the Wick's theorem. This allows us to increase $N_2$, the number of $k$-points, which is limited to $N_2\leq 6$ in DMRG with our computing resources. The ground state at half filling of the spinless one-valley problem is thus obtained by minimizing
\begin{eqnarray}
E_N=\langle \Psi_{GS}| \lambda\hat H_{kin}+\hat{V}_{int} | \Psi_{GS} \rangle,   \label{Eqn:ScaledHam}
\end{eqnarray}
with the constraint $|u(n, k)|^2 + |v(n, k)|^2 = 1$ for every $n$ and $k$, and allowing $\lambda$ to increase continuously form $0$.
Furthermore, we seek a solution periodic in the unit cell index $n$, i.e.
\[ u(n, k) = u(n + n_p, k),  \quad v(n, k) = v(n + n_p, k)  \ ,  \]
where $n_p$ is the period. The $C_2 \mathcal{T}$ symmetric state satisfies the constraint that
\begin{align}
u(n, k) & = v^*(-n, k) e^{i \theta(n, k)} \nonumber \\
v(n, k) & = u^*(-n, k) e^{i \theta(n, k)}
\end{align}
and the $C_2''$ symmetric state should satisfy
\begin{align}
u(n, k) & = v(n, 1 - k) e^{i \theta'(n, k)} \nonumber \\
v(n, k) & = u(n, 1 - k) e^{i \theta'(n, k)}.
\end{align}
We then optimize the energy allowing both $C_2\mathcal{T}$ and $C_2''$ symmetries to be broken and allowing the period $n_p$ to be as large as $8$. Numerically, we found three types of the solutions with the symmetry listed in Table.~\ref{Tab:DMRGTrialSym}.

\begin{table}[htbp]
	\begin{ruledtabular}
		\begin{tabular}{|c|c|c|c|}
			Solution & Translation &  $C_2\mathcal{T}$ & $C_2''$ \\  \hline
			$C_2\mathcal{T}$ broken	& Conserved & Broken & Broken  \\  \hline
			$C_2\mathcal{T}$	& \multirow{2}{*}{Conserved}  & \multirow{2}{*}{Conserved} & Broken if $\lambda \lesssim 0.8$ \\
			nematic &   & & Conserved if $\lambda \gtrsim 0.8$ \\ \hline
			$C_2\mathcal{T}$	& Broken  & \multirow{2}{*}{Conserved} & Broken  \\
			Stripe &  ($n_p = 2$) & & ($T_{\bL_1} C_2''$ Conserved) \\ 
		\end{tabular}
	\end{ruledtabular}
	\caption{The symmetry of three obtained solutions with the trial states given by Eqn.~\ref{Eqn:TrialGS}, where $\lambda$, defined in Eqn.~\ref{Eqn:ScaledHam} is the scaling factor of the kinetic terms. Whether the one fermion spectrum of the states is gapless or gapped is specified later in the text and figures; we only focus on the symmetry breaking here. }
	\label{Tab:DMRGTrialSym}
\end{table}

For $N_2 = 6$, the $C_2\mathcal{T}$ broken state is found to be the ground state for $w_0/w_1\lesssim 0.8$ and is identified as the QAH state, while the $C_2\mathcal{T}$ symmetric state with broken translation symmetry is found to be the ground state for $w_0/w_1 \approx 0.85$, corresponding to the $C_2\mathcal{T}$-symmetric period-2 stripe.  Although this stripe state breaks $C_2''$ symmetry, the combination of translation along $\bL_1$ and $C_2''$, i.e.~$T_{\bL_1} C_2''$ is still conserved. This result is consistent with the one obtained using DMRG for the same value of $N_2$, in that the $\langle n_3 \rangle$ vanishes at roughly the same values of $w_0/w_1$. Moreover, up to a $\pi$, the $k$- dependence of the $\arg(v(n, k)/u(n, k))$ obtained by the variational method is the same as the one by DMRG. The $k$ and $n$ dependence of this phase will be more thoroughly discussed in the next subsection.

We can also obtain another variational state by imposing the translational symmetry ($n_p=1$) {\em and} $C_2\mathcal{T}$ invariance. We refer to this state as $C_2\mathcal{T}$-nematic. The best variational energies of these states are compared in Fig.~\ref{Fig:DMRGEne} as we change $\lambda$, plotting the result per particle as a function of the kinetic energy in the half-filled non-interacting semi-metallic state,
$\langle E_{K}\rangle_{SM}\equiv\lambda \langle \hat H_{kin}\rangle_{SM}$, in units of $U_0=e^2N/(4\pi \epsilon L_m)$.
The interaction constant $U_0  \approx 17$meV, where $L_m$ is the moire superlattice constant and $\epsilon$ is the dielectric constant of hBN. Because the simple non-interacting semi-metal state diagonalizes the kinetic energy --and therefore optimizes it-- we include in the plot the expectation value of $\lambda\hat H_{kin}+\hat{V}_{int}$ in this state. Although, the non-interacting semi-metal is clearly not competitive in the range of the parameters of interest to us, its expected energy does provide us with a measure of near degeneracy among the competing states.

For $N_2 = 16$ and $w_0/w_1 = 0.85$ the Fig.~\ref{Fig:DMRGEne} thus compares the energies of three different variational states: $C_2 \mathcal{T}$ broken state, $C_2 \mathcal{T}$ symmetric period-2 stripe phase, and $C_2\mathcal{T}$ symmetric nematic state. As seen, all these three states are nearly degenerate.
As a measure of how close the energies of the three states are, we divide the energy difference between the competitive states by the energy difference between the ground state and the non-interacting semi-metal. Without the kinetic terms in the Hamiltonian, we find that the normalized energy difference between QAH and $C_2\mathcal{T}$ nematic state is only $0.015U_0/(0.8 U_0) \approx 0.02$, and the normalized energy difference between QAH and $C_2\mathcal{T}$ stripe state is $0.05/0.80 \approx 0.06$, in favor of QAH. As seen in Fig.\ref{Fig:DMRGEne}, the energies of the competitive states are even closer when the kinetic energy terms are included.

We find that the ground state is always translationally invariant. Additionally, the $C_2\mathcal{T}$ symmetry is broken when the kinetic energy $\langle E_K/N \rangle_{SM} < 0.4 U_0$, and fully gapped when $\langle E_K/N \rangle_{SM}  < 0.31 U_0$, suggesting that the state we found is QAH for small kinetic energy and turns into an anomalous Hall metal when $ 0.31 < |\langle E_K \rangle_{SM}| /N U_0 < 0.4$. It eventually evolves into a normal metal with vanishing Hall conductivity when $|\langle E_K \rangle_{SM}| /N U_0 > 0.4$. Nevertheless, the energies of the two $C_2\mathcal{T}$ symmetric states in Table.~\ref{Tab:DMRGTrialSym} are very close to the energy of the QAH state in all the parameter regimes we have calculated. As we will discuss later, the energy of the $C_2 \mathcal{T}$ symmetric states can be further lowered by improving the form of the variational states. This near degeneracy necessitates the inclusion of all three different states as the candidates for the ground state at odd integer filling.

Because the anomalous Hall state seems to have been ruled out in experiments on magic angle TBG at $\nu=3$ without the alignment with hBN, and because as we will see below $C_2\mathcal{T}$ period-2 stripe state can be fully gapped, we consider it as a candidate for the Chern-0 insulating state experimentally observed at $\nu=3$. QAH state, on the other hand, can be favored by breaking $C_2$ symmetry, and thus is the state discovered at the same filling but aligning the system with the hBN substrate. In addition, the $C_2\mathcal{T}$ nematic state, being gapless, simultaneously breaks $C_3$ rotation symmetry and possesses the interesting pattern of the Landau level degeneracy. Therefore, after including the spin and valley degree of freedom, we propose the $C_2\mathcal{T}$ nematic state as a candidate for the gapless state at the charge neutrality point (CNP).

\subsection{$C_2''$ Symmetry}
The QAH states can be approximated as
\begin{align}
| \Psi_{QAH} \rangle \approx \prod_{n, k} d_{+, n, k}^{\dagger} | \emptyset \rangle \quad \mbox{or} \quad \prod_{n, k} d_{-, n, k}^{\dagger} | \emptyset \rangle.
\end{align}
Since the hybrid states transform as Eqn.~\ref{Eqn:C2ppSym} under $C_2''$, this state obviously breaks $C_2''$ symmetry (in addition to, of course, $C_2\mathcal{T}$).

If the $C_2 \mathcal{T}$ symmetric state is translationally invariant or has the period of $2$ unit cells, the $u(n, k)$ and $v(n, k)$ in the trial wavefunction Eq.~\ref{Eqn:TrialGS} can be written as
\begin{align}
u(n, k) = \frac1{\sqrt{2}} e^{i \phi(n, k)} \quad v(n, k) = \frac1{\sqrt{2}} e^{-i \phi(n, k)}  \ . \label{Eqn:PhiC2TState}
\end{align}
If the state is further $C_2''$ symmetric,
\begin{align}
& e^{i\phi(n, k)} = \pm e^{-i \phi(n, 1-k)} \nonumber  \\
\Longrightarrow \ & \phi(n, k) + \phi(n, 1-k) = 0 \ \text{or} \ \pi \ .
\end{align}
\begin{figure}[htbp]
	\centering
		\subfigure[\label{Fig:C2ppSym:Nem}]{\includegraphics[width=0.45\textwidth]{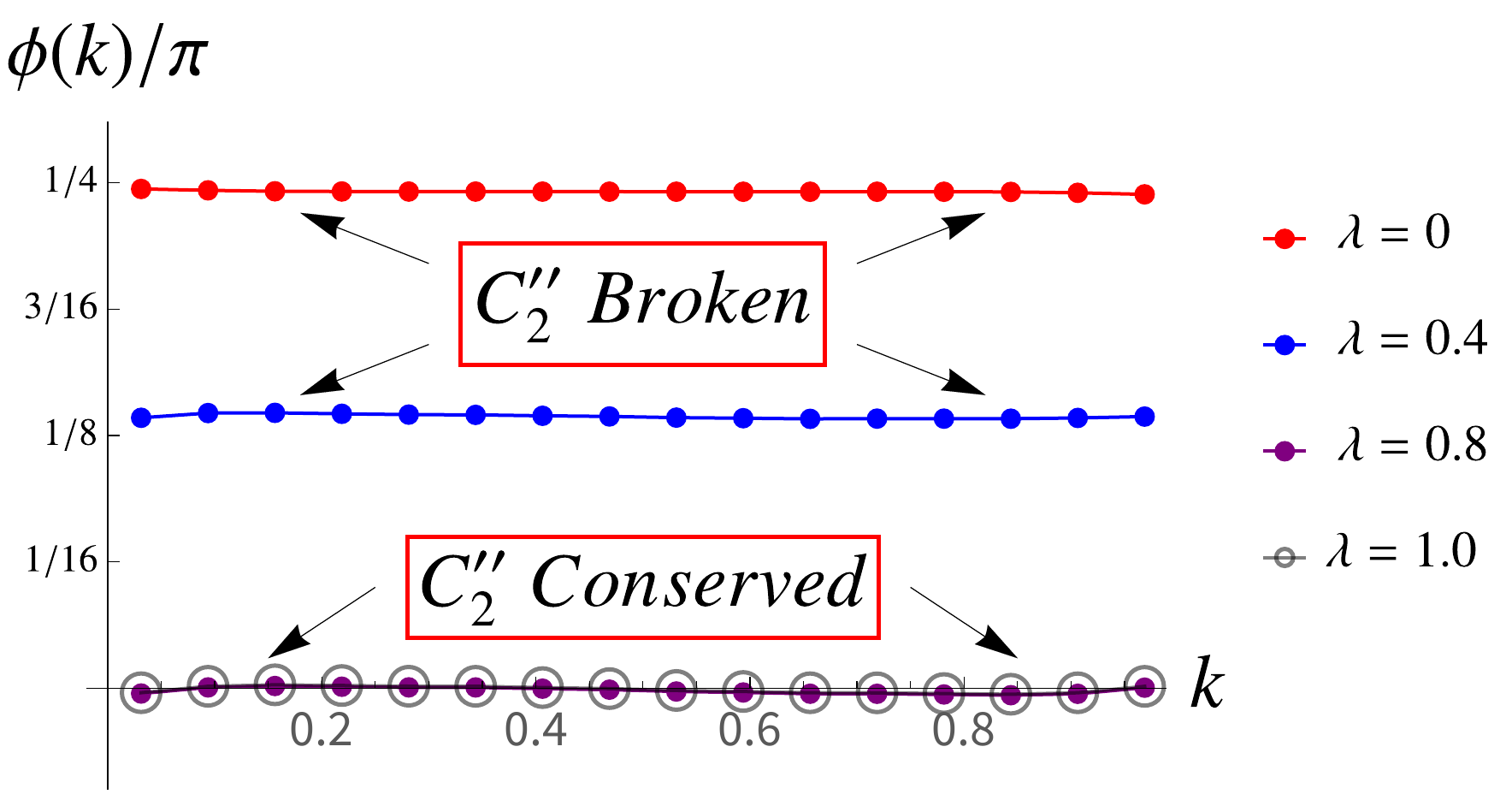}}
		\subfigure[\label{Fig:C2ppSym:Stripe}]{\includegraphics[width=0.48\textwidth]{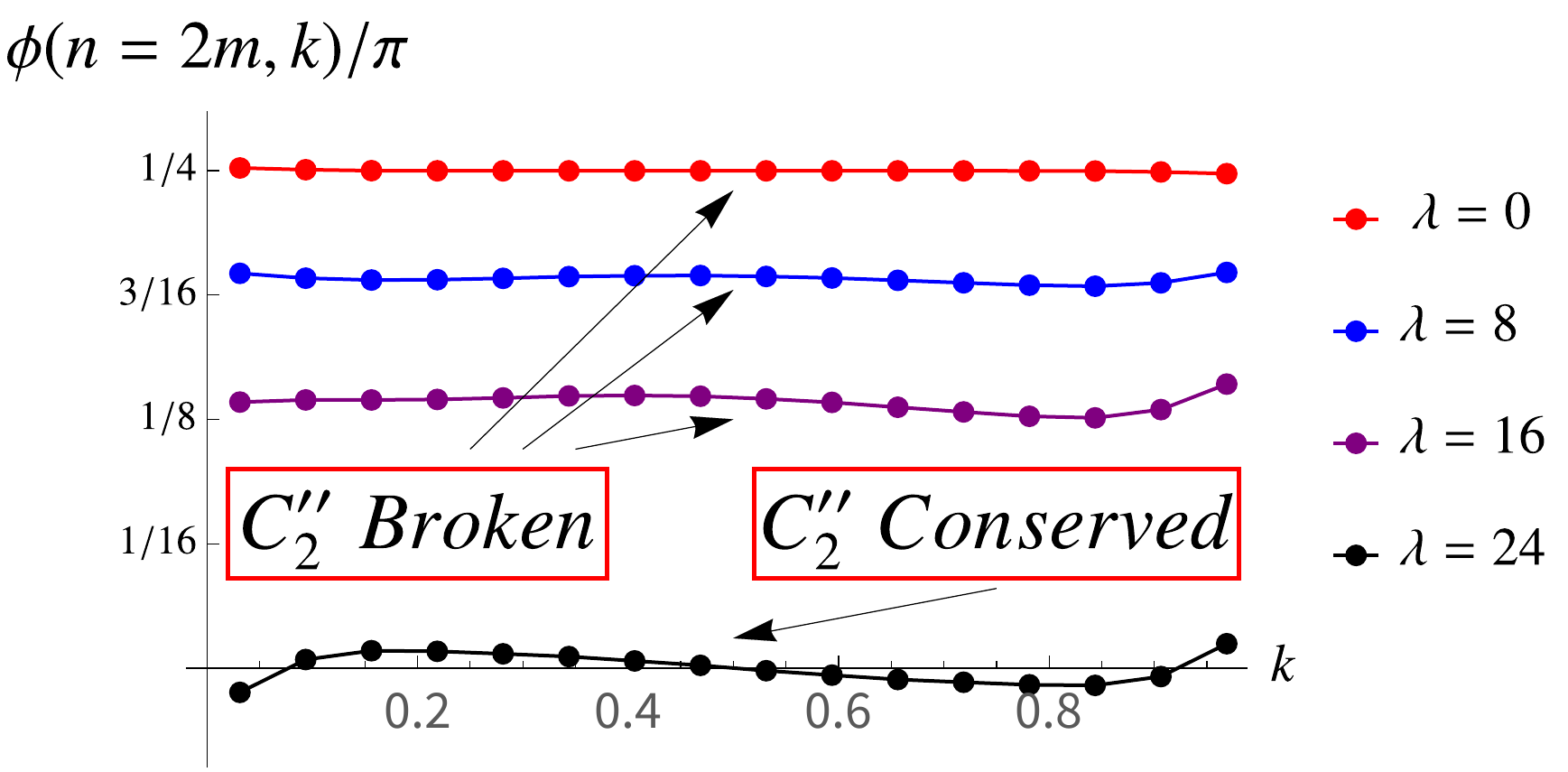}}
	\caption{(a) $\phi(n, k)$, defined in Eqn.~\ref{Eqn:PhiC2TState},  in  the $C_2\mathcal{T}$ nematic state with the trial function in Eqn.~\ref{Eqn:TrialGS} and $N_2 = 16$. $\lambda$ is the scaling factor of the kinetic energy, defined in Eqn.~\ref{Eqn:ScaledHam}.  (b) $\phi(n, k)$ in the $C_2 \mathcal{T}$ stripe state when $n$ is even. If $n$ is odd, $\phi(n, k)$ can be obtained from the relation in Eqn.~\ref{Eqn:C2ppTL1}. }
	\label{Fig:C2ppSym}
\end{figure}

Fig.~\ref{Fig:C2ppSym} illustrates the $k$ dependence of the phase $\phi(n, k)$ in the two $C_2\mathcal{T}$ symmetric states. In the $C_2\mathcal{T}$ nematic state, due to the translation symmetry, $\phi(n, k)$ is independent of $n$. With the interactions only,  $\lambda$, defined in Eqn.~\ref{Eqn:ScaledHam}, vanishes, and as the Fig.~\ref{Fig:C2ppSym:Nem} shows, the state can be approximated as
\beq
| \Psi_N \rangle \approx \prod_{n, k} \frac1{\sqrt{2}} \left( e^{i \frac{\pi}4} d^{\dagger}_{+, n, k} +   e^{-i \frac{\pi}4} d^{\dagger}_{-, n, k} \right)  | \emptyset \rangle.
\eeq
Obviously, this state breaks the $C_2''$ symmetry. With increasing $\lambda$, the phase $\phi(k)$ becomes smaller, and eventually vanishes when $\lambda \gtrsim 0.8$, and thus the $C_2''$ symmetry is recovered. In particular, for the BM model including both the interaction and kinetic terms without any scaling, $\lambda = 1$, and thus the state can be approximated as
\beq
| \Psi_N \rangle \approx \prod_{n, k} \frac1{\sqrt{2}} \left( d^{\dagger}_{+, n, k} + d^{\dagger}_{-, n, k} \right)  | \emptyset \rangle.
\eeq

Our calculation shows that the $C_2\mathcal{T}$ stripe state is always invariant under $T_{\fvec L_1} C_2''$ transformation. This leads to the relation
\beq  \phi(n, k) = -\phi(n + 1, 1 -k)  \label{Eqn:C2ppTL1} \eeq
relating the phase $\phi(n, k)$ with even $n$ and the phase with odd $n$. When $\lambda$ vanishes, Fig.~\ref{Fig:C2ppSym:Stripe} shows that the wavefunction of the stripe state can be approximated as
\begin{align}
| \Phi_N^s \rangle \approx & \prod_{m, k} \frac1{\sqrt{2}} \left( e^{i \frac{\pi}4} d^{\dagger}_{+, 2m, k} +   e^{-i \frac{\pi}4} d^{\dagger}_{-, 2m, k} \right) \times \nonumber \\
& \frac1{\sqrt{2}} \left( e^{-i \frac{\pi}4} d^{\dagger}_{+, 2m + 1, k} +   e^{i \frac{\pi}4} d^{\dagger}_{-, 2m + 1, k} \right) | \emptyset \rangle.
\end{align}
Obviously, $\phi(n, k) \neq -\phi(n, 1 - k)$ and thus the $C_2''$ symmetry is broken in this state.

Similar to the nematic phase, the magnitude of $\phi(n, k)$ decreases with increasing $\lambda$. When $\lambda \gtrsim 24$, the  stripe state satisfies the relation  $\phi(n, k) = -\phi(n, 1 - k)$ and thus becomes $C_2''$ symmetric.

\subsection{Excitation spectrum}
If the trial state (Eqn.~\ref{Eqn:TrialGS}) does not break the translation symmetry, it can be written as
\begin{align}
| \Psi_N \rangle = &
\prod_{q, k} \left( u(k)  b^{\dagger}_{+, q, k} +  v(k) b^{\dagger}_{-, q, k} \right) | \emptyset \rangle \ ,
\label{Eqn:TRTrialBloch}
\end{align}
where $b_{\pm, q, k}$ is defined in Eqn.~\ref{Eqn:bOprMom} as the Fourier transform of the fermion operator $d_{\pm, n, k}$.

To construct the one particle and hole excited states, we delocalize a linear combination of $d^\dagger_{+,n,k}$ and $d^\dagger_{-,n,k}$:
\begin{eqnarray}
&& | \Psi_{N + 1}(q, k) \rangle = \left( v^*(k)  b^{\dagger}_{+, q, k} - u^*(k) b^{\dagger}_{-, q, k} \right) | \Psi_N \rangle \label{Eqn:NP1State} \\
&& | \Psi_{N - 1}(q, k) \rangle = \left( u^*(k)  b_{+, q, k} +  v^*(k) b_{-, q, k} \right) | \Psi_N \rangle. \label{Eqn:NM1State}
\end{eqnarray}
The variational energies of these excited states are
\begin{eqnarray}
E_{N \pm 1}(q,k) = \langle \Psi_{N \pm 1}(q, k) | H | \Psi_{N \pm 1}(q,k) \rangle, \label{Eqn:TRStateSpec}
\end{eqnarray}
and the gap is given by
\begin{eqnarray}
\Delta & = & \min(E_{N+1}(q,k) - E_N) - \nonumber \\
& & \quad \max(E_N - E_{N - 1}(q', k'))  \ .  \label{Eqn:DGapDef}
\end{eqnarray}

\begin{figure}[t]
	\centering
	\subfigure[\label{Fig:DMRGHFDGap:QAH}]{\includegraphics[width=\columnwidth]{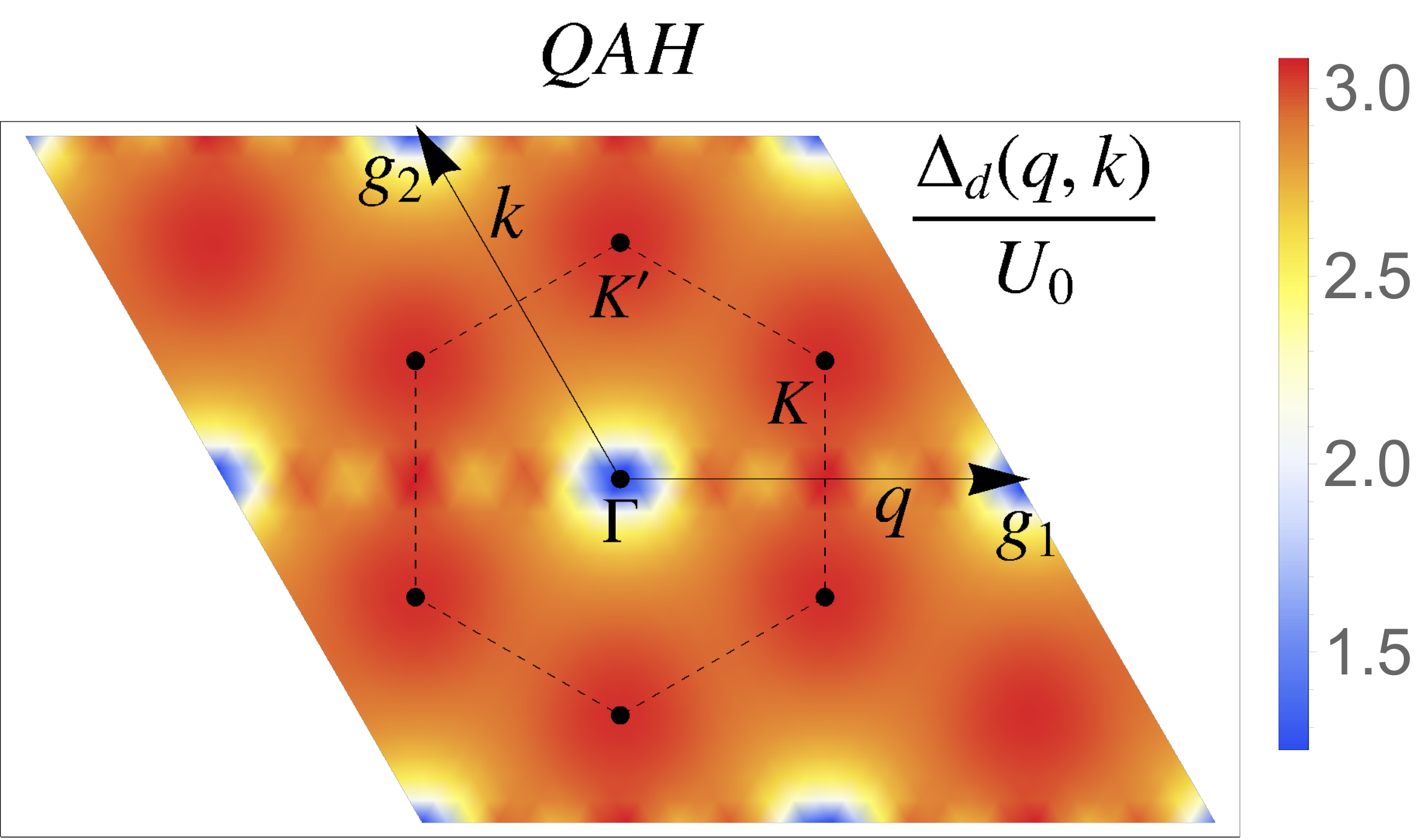}}
	\subfigure[\label{Fig:DMRGHFDGap:C2TNematic}]{\includegraphics[width=\columnwidth]{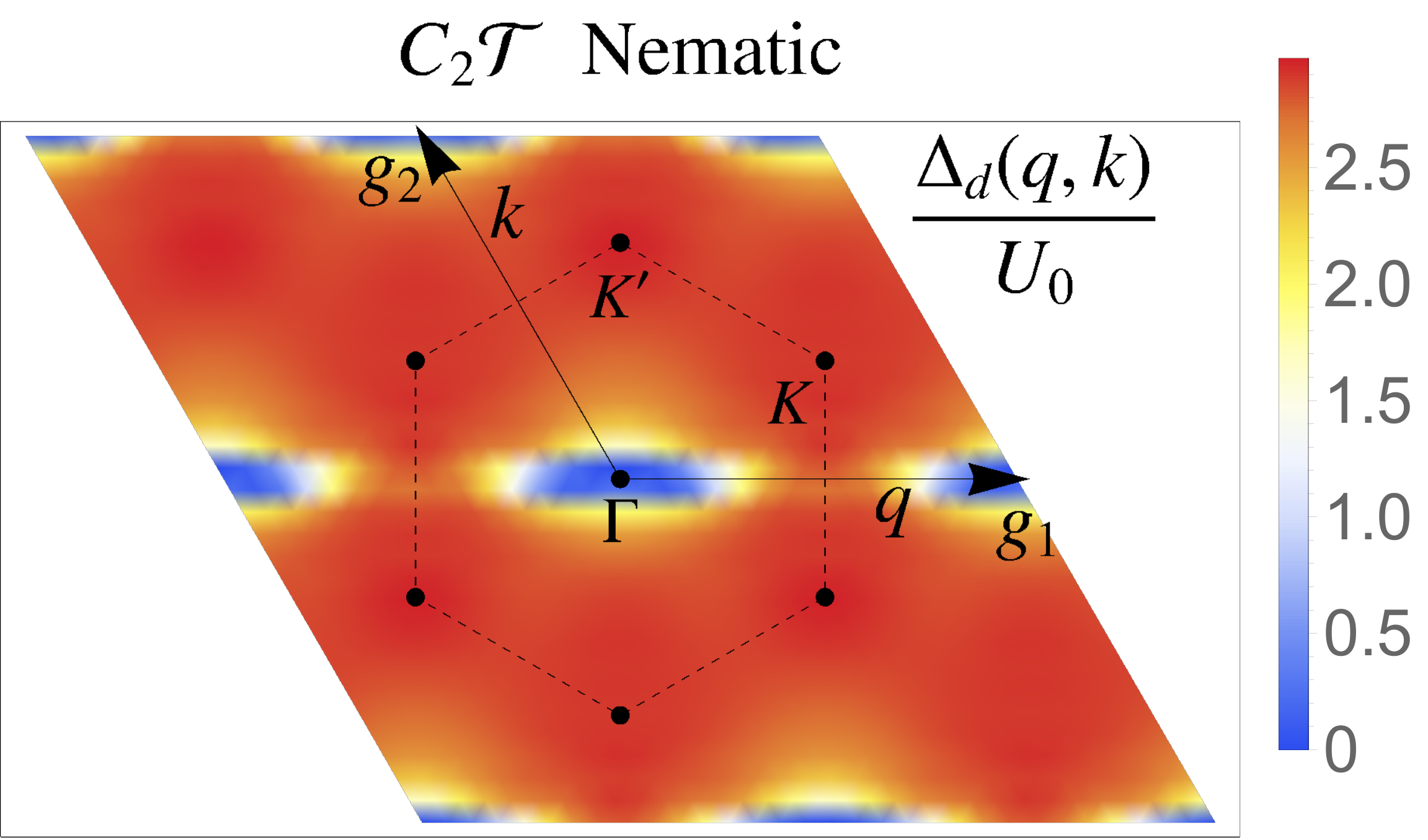}}
	\subfigure[\label{Fig:DMRGHFDGap:C2TStripe}]{\includegraphics[width=\columnwidth]{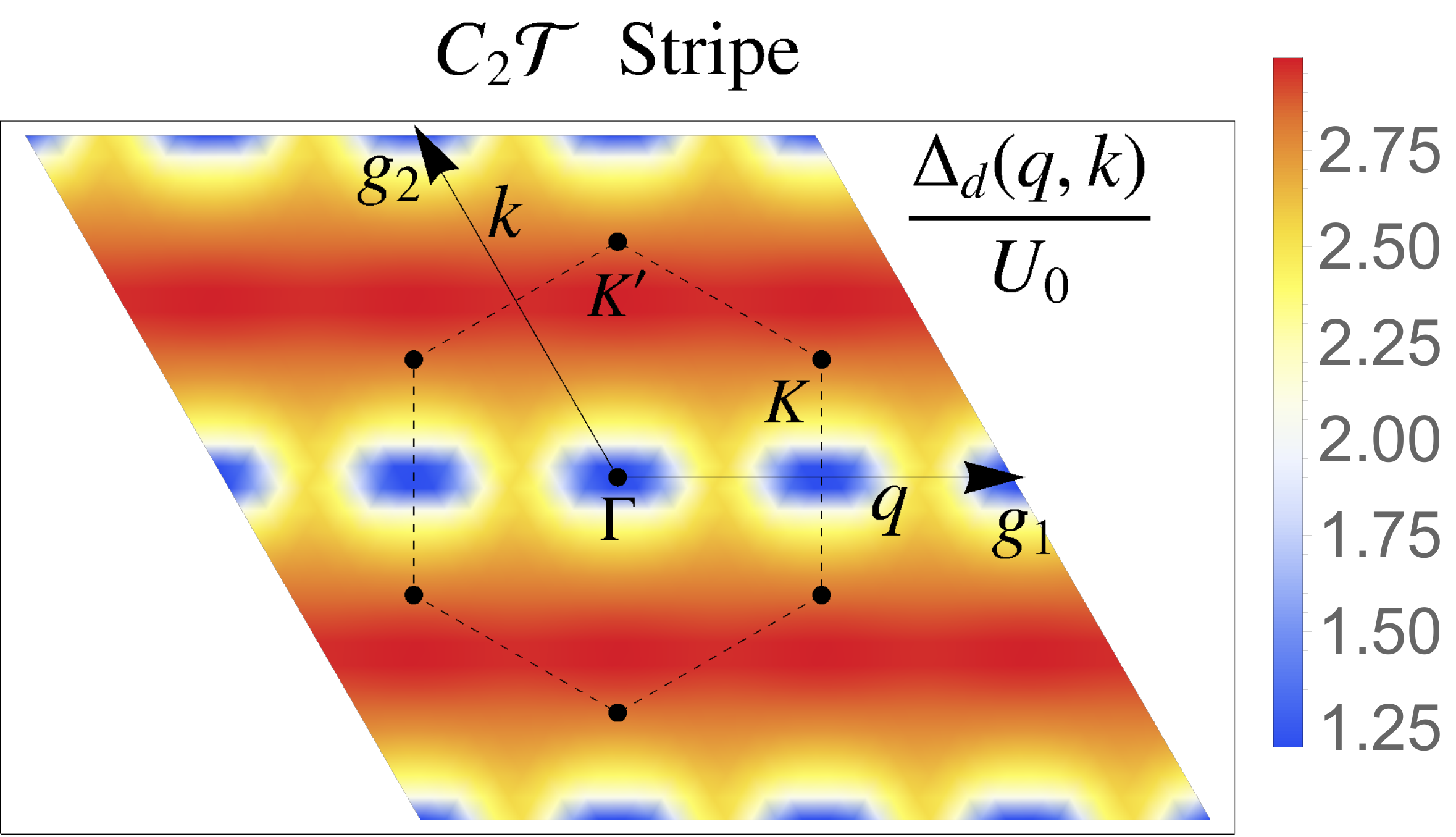}}
	\caption{The direct gap $\Delta_d(q, k)$ of three states with the trial state in Eqn.~\ref{Eqn:TrialGS} with $w_0/w_1 = 0.85$. The kinetic energy is set to be $\langle E_K \rangle_{SM} = 0$. The energies are normalized by $U_0 = e^2/(4\pi \epsilon L_m)$.}
	\label{Fig:DMRGHFDGap}
\end{figure}

Fig.~\ref{Fig:DMRGHFDGap} illustrates the direct gap $\Delta_d$, defined as
\[\Delta_d(q, k) = (E_{N+1}(q,k) - E_N) - (E_N - E_{N - 1}(q, k)) \ , \]
in the BZ when $w_0/w_1 = 0.85$ and the kinetic terms are set to be $0$. Interestingly, we found the $C_2 \mathcal{T}$ symmetric nematic phase is gapless with nodes around $\fvec \Gamma$. The robustness of the nodes has been discussed in Ref.~\cite{Zalatel1} and additional properties will be presented in the next section. The gap can be opened by breaking $C_2\mathcal{T}$ symmetry. A typical example of this case is the QAH state having a gap of order $U_0$ as shown in Fig.~\ref{Fig:DMRGHFDGap:QAH}. In the next subsection, we will show that a gapped $C_2 \mathcal{T}$ symmetric state can be obtained by breaking the translation symmetry.

\begin{figure}[htbp]
	\centering
	\includegraphics[width=0.9\columnwidth]{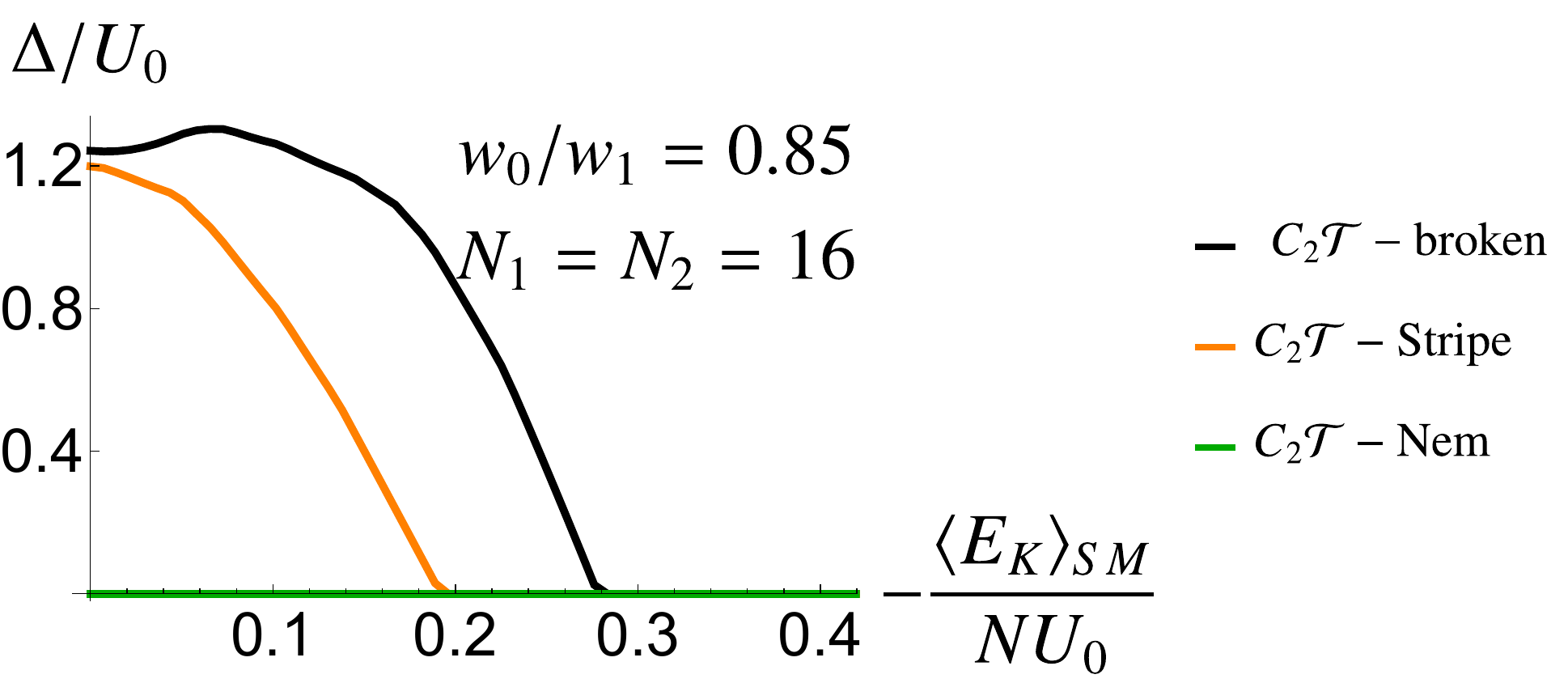}
	\caption{The single fermion excitation gap of three nearly degenerate states. The energies are normalized by $U_0 = e^2/(4\pi \epsilon L_m)$.}
	\label{Fig:HFGap}
\end{figure}

As mentioned before, we obtained three different types of solutions. While the former two are translationally invariant, the last one is $C_2\mathcal{T}$ symmetric but breaks the translation symmetry with the period of $2$. Consequently, the coefficients $u$'s and $v$'s in the trial state satisfy
\begin{align}
& u(2n, k) = u(0, k) & & v(2n, k) = v(0, k) \\
& u(2n + 1, k) = u(1, k) & & v(2n + 1, k) = v(1, k).
\end{align}
The corresponding stripe state can be described by the following wavefunction:
\begin{align}
& | \Psi_N^{s} \rangle =  \prod_{n, k} \prod_{m = 0}^1  \left( u(m, k) d_{+, 2 n + m, k}^{\dagger} + v(m, k) d_{-, 2 n + m, k}^{\dagger}    \right)  | \emptyset \rangle. \label{Eqn:StripeHWS}
\end{align}
Similar to the translationally invariant state, we can also describe this stripe state in the basis of Chern Bloch states. For notational convenience, we introduce another set of fermion operators:
\begin{align}
f_{\pm, q, k, 0} & = \frac1{\sqrt{2}} \left( b_{\pm, q, k} + b_{\pm, q + \half, k}  \right) \\
f_{\pm, q, k, 1} & = \frac1{\sqrt{2}} \left( b_{\pm, q, k} - b_{\pm, q + \half, k}  \right),
\end{align}
with $0 \leq q < \half$. Up to an overall phase, the stripe state in Eqn.~\ref{Eqn:StripeHWS} can be written as
\begin{align}
| \Psi_N^s \rangle = \prod_{\substack{k \in [0, 1)\\ q \in [0, 1/2)}} \prod_{m = 0}^1  \left( u(m, k) f^{\dagger}_{+, q, k, m} + v(m, k) f^{\dagger}_{-, q, k, m} \right) | \emptyset \rangle.
\label{Eqn:StripeTrialBloch}
\end{align}
Similar to the case of the translationally invariant ground state,  the one particle and hole excited states are built with delocalized linear combination of fermion operators $d^{\dagger}_{\pm, n, k}$:
\begin{eqnarray}
& & | \Psi^s_{N + 1, m}(q, k) \rangle \nonumber \\
& = & \left( v^*(m, k) f^{\dagger}_{+, q, k, m} - u^*(m, k) f^{\dagger}_{-, q, k, m} \right) | \Psi^s_N \rangle \\
& & | \Psi^s_{N - 1, m}(q, k) \rangle \nonumber \\
& = & \left( u^*(m, k) f_{+, q, k, m} + v^*(m, k) f_{-, q, k, m} \right) | \Psi^s_N \rangle.
\end{eqnarray}
To obtain the fermion spectrum, we construct the two $2 \times 2$ matrices for the electron and hole excited states respectively:
\begin{equation}
\left( H^s_{N \pm 1}(q, k) \right)_{m_1 m_2}
= \langle \Psi^s_{N \pm 1, m_1}(q, k) | H | \Psi^s_{N \pm 1, m_2}(q, k) \rangle.
\end{equation}
The energy and the wavefunction of the electron and hole excites states are obtained by diagonalizing these matrices $H^s_{N\pm1}(q, k)$. At each momentum, we obtain two eigenvalues $E^s_{N\pm 1, 1}(q, k)$ and $E^s_{N\pm 1, 2}(q, k)$. The gap of the period-2 stripe ground state is given by
\begin{eqnarray}
\Delta & = & \left(\min(E^s_{N + 1, 1}(q, k), E^s_{N + 1, 2}(q, k)) - E^s_N \right) - \nonumber \\
& & \left( E^s_N -  \min(E^s_{N - 1, 1}(q', k'), E^s_{N - 1, 2}(q', k')) \right).
\end{eqnarray}
Fig.~\ref{Fig:HFGap} shows the magnitude of the gap in the $C_2 \mathcal{T}$ symmetric stripe state. This gap is found to be $ \gtrsim U_0$ at vanishing kinetic energy, but decreases with increasing kinetic energy. It vanishes before evolving into the $C_2 \mathcal{T}$ nematic phase. The opening and closing of this gap will be discussed in much more detail in the next section, where we expose the non-Abelian topological\cite{Tomas} aspects of this process.

Fig.~\ref{Fig:DMRGHFDGap:C2TStripe} plots the direct gap $\Delta_d$, defined as
\begin{eqnarray}
\Delta_d(q, k) & = &\big( \min(E_{N+1, 1}(q,k), E_{N+1, 2}(q, k) )  - E_N \big) -  \nonumber \\
& & \big( E_N - \min(E_{N - 1, 1}(q, k), E_{N - 1,2 }(q, k) )     \big)   \  ,  \label{Eqn:DGapStripe}
\end{eqnarray}
making it is obvious that $\Delta_d(q, k) = \Delta_d(q + \half, k)$ due to the breaking of the translation symmetry with the period of $2$.

\section{Generalized Trial States}

\subsection{Translationally Invariant State}
The trial function in Eqn.~\ref{Eqn:TRTrialBloch} is not the most general form for the translationally invariant state, as the coefficients $u$'s and $v$'s are independent of the momentum component $q$. This comes from the complete absence of the correlations between hybrid WSs on different sites in our trial state ( Eqn.~\ref{Eqn:TrialGS} ). To improve the trial state, we consider the following wavefunction
\begin{align}
| \Psi_N \rangle = \prod_{q, k} \left( u(q, k) b^{\dagger}_{+, q, k} + v(q, k) b^{\dagger}_{-, q, k}  \right) | \emptyset \rangle, \label{Eqn:GeneralTrial}
\end{align}
where $u$ and $v$ depend on both $q$ and $k$ and satisfy $|u(q, k)|^2 + |v(q, k)|^2 = 1$. If this state is $C_2 \mathcal{T}$ symmetric, $u$'s and $v$'s should have the same magnitude, i.e.~$|u(q, k)| = |v(q, k)| = 1/\sqrt{2}$.

Similar to the approach in the previous section, the ground state is obtained by minimizing
\[ E_N(u, v) = \langle \Psi_N | H | \Psi_N \rangle \]
with respect to $u$'s and $v$'s at various momenta. The momentum mesh in the BZ is chosen to be
\[ q = \frac{i + 1/2}{N_1} \quad \mbox{and} \quad k = \frac{j + 1/2}{N_2}  \]
with $i = 0, 1, \cdots, N_1 -1$ and $j = 0, 1, \cdots, N_2 -1$. $N_1$ and $N_2$ are taken to be $16$. Our calculations still find two different solutions, $C_2\mathcal{T}$ broken and $C_2\mathcal{T}$ nematic solution. Compared with the trial state in Eqn.~\ref{Eqn:TrialGS}, both solutions have lower energies. Interestingly, the $C_2 \mathcal{T}$ nematic state now has a lower energy than the $C_2 \mathcal{T}$ broken state, although the difference between these two solutions are again tiny. The $C_2 \mathcal{T}$ broken solution is obtained by searching a local minimum near the product state with $u(q, k) = 1$ and $v(q, k) = 0$.

\begin{figure}[t]
	\centering
	\includegraphics[width=\columnwidth]{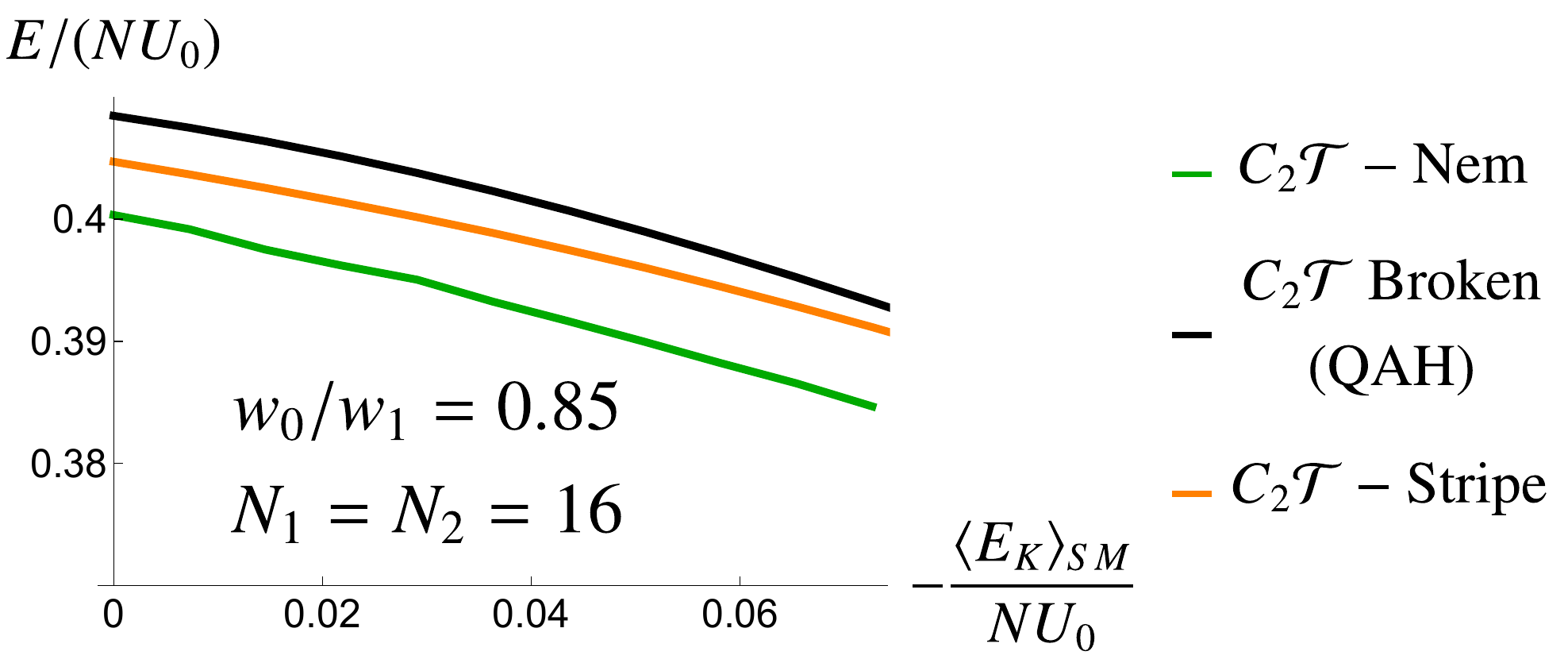}
	\caption{The energies of three nearly degenerate states at $w_0/w_1 = 0.85$ with the more general trial wavefunction given by Eqn.~\ref{Eqn:GeneralTrial}, except for the stripe state which is given by the wavefunction (Eqn.~\ref{EqnS:C2TStripeState}) in the Appendix~\cite{Appendix}.}
	\label{Fig:GenEne}
\end{figure}

\begin{figure}[t]
	\centering
	\includegraphics[width=\columnwidth]{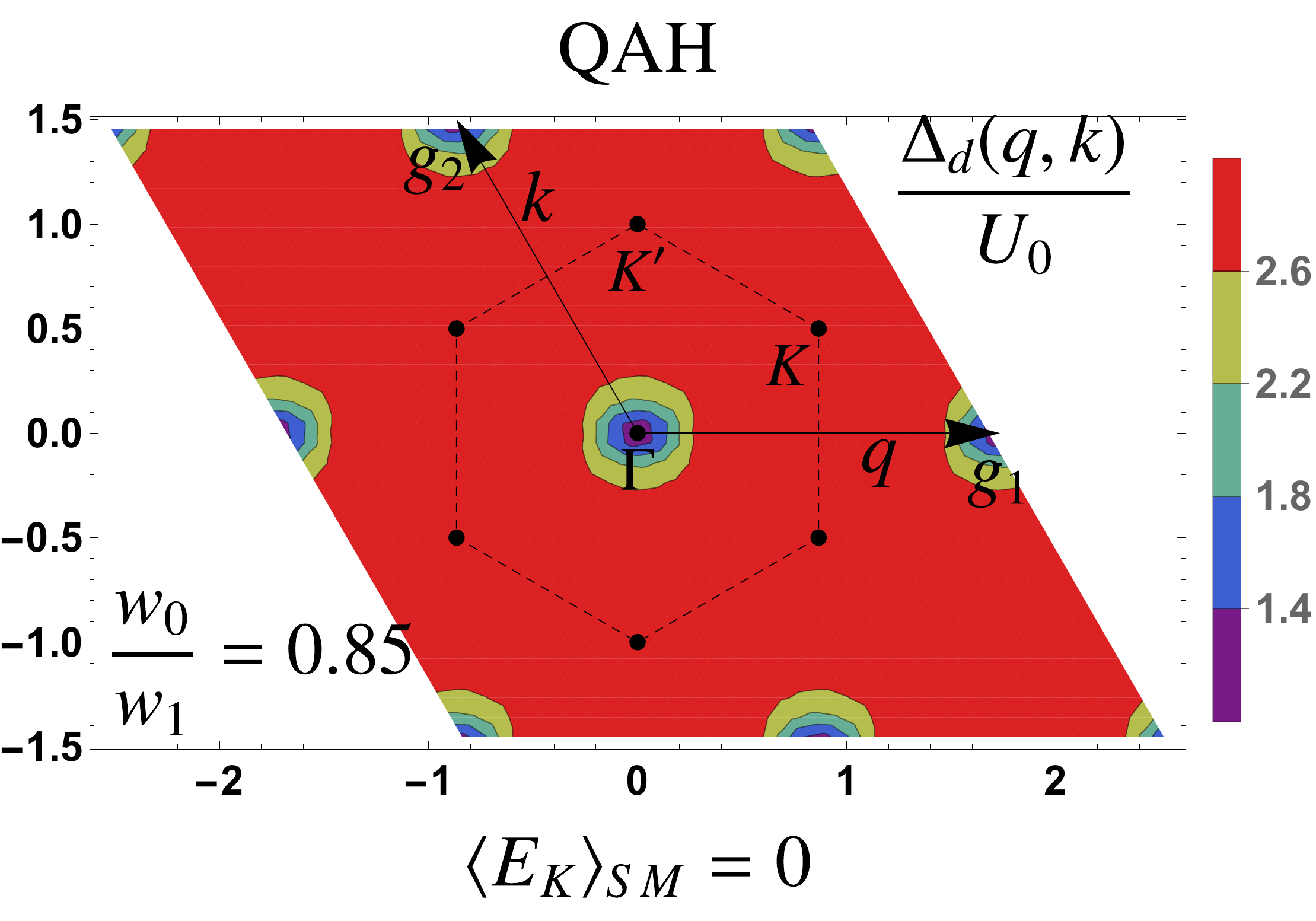}
	\caption{The direct gap $\Delta_d(q, k)$ of the QAH state. This state is given by Eqn.~\ref{Eqn:GeneralTrial}.}
	\label{Fig:GenDGapQAH}
\end{figure}

The fermion spectrum is also calculated in the same way as shown by Eqn.~\ref{Eqn:NP1State} -- \ref{Eqn:DGapDef}.
Fig.~\ref{Fig:GenDGapQAH} shows the direct gap of the $C_2\mathcal{T}$ broken state inside the BZ. With small kinetic energy, this state is fully gapped over the whole BZ. The global $C_2 \mathcal{T}$ breaking order parameter, defined as
\begin{eqnarray}
\langle n_3 \rangle  = \frac1N \sum_{q, k} \langle \Psi_N | b^{\dagger}_{+, q, k}  b_{+, q, k} - b^{\dagger}_{-, q, k}  b_{-, q, k} | \Psi_N \rangle  \ ,
\end{eqnarray}
is found to be almost $1$ in this state. Therefore, it is identified as the QAH state. Interestingly, the gap minimum is always located at the $\Gamma$ point. This is dramatically different from the non-interacting state, in which the gap is largest at $\Gamma$ and closes at $\fvec K$ and $\fvec K'$.

\begin{figure}[htbp]
	\centering
	\subfigure[\label{Fig:QBTC2TNem085:EK0}]{\includegraphics[width=\columnwidth]{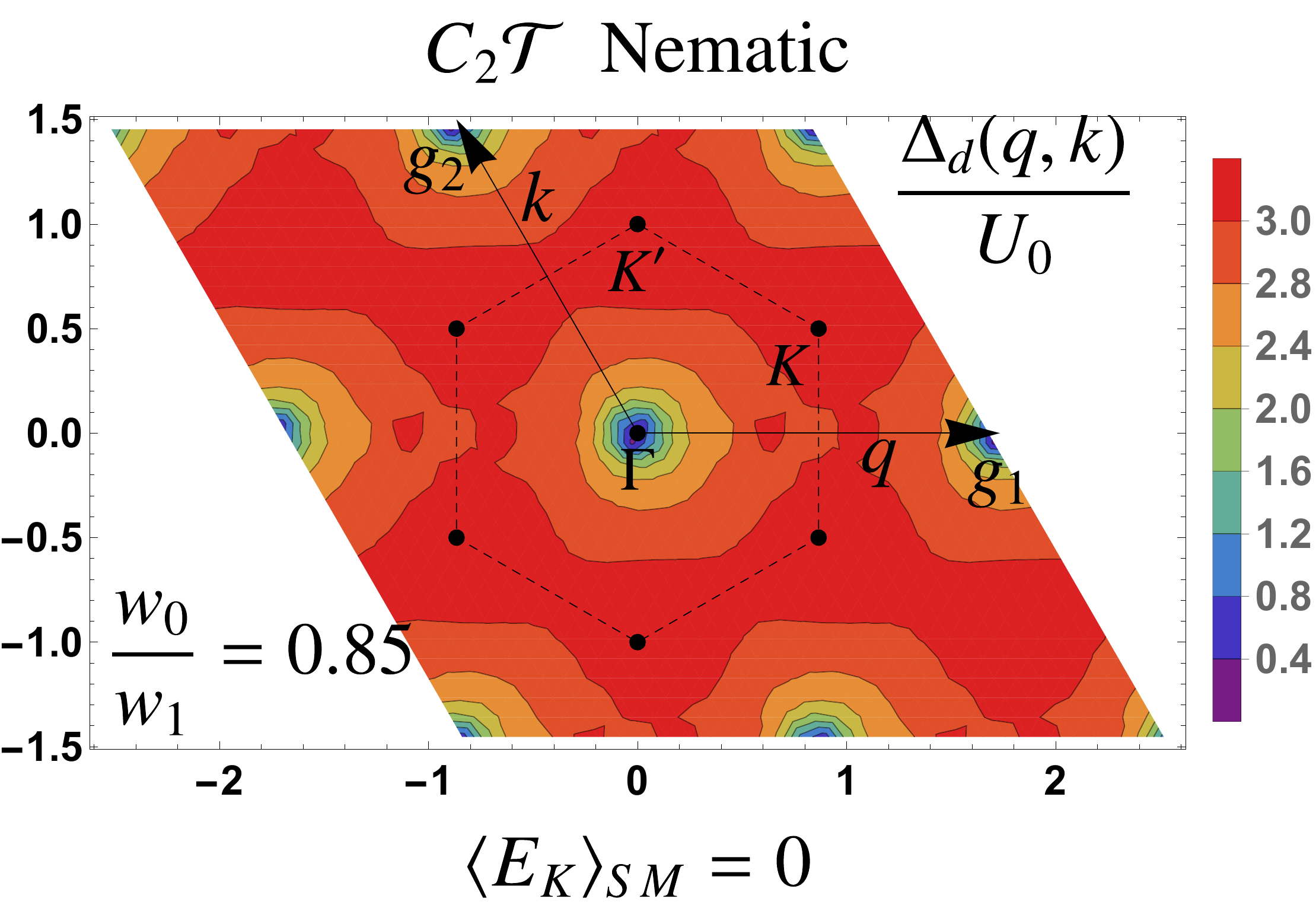}}
	\subfigure[\label{Fig:QBTC2TNem085:EK25}]{\includegraphics[width=\columnwidth]{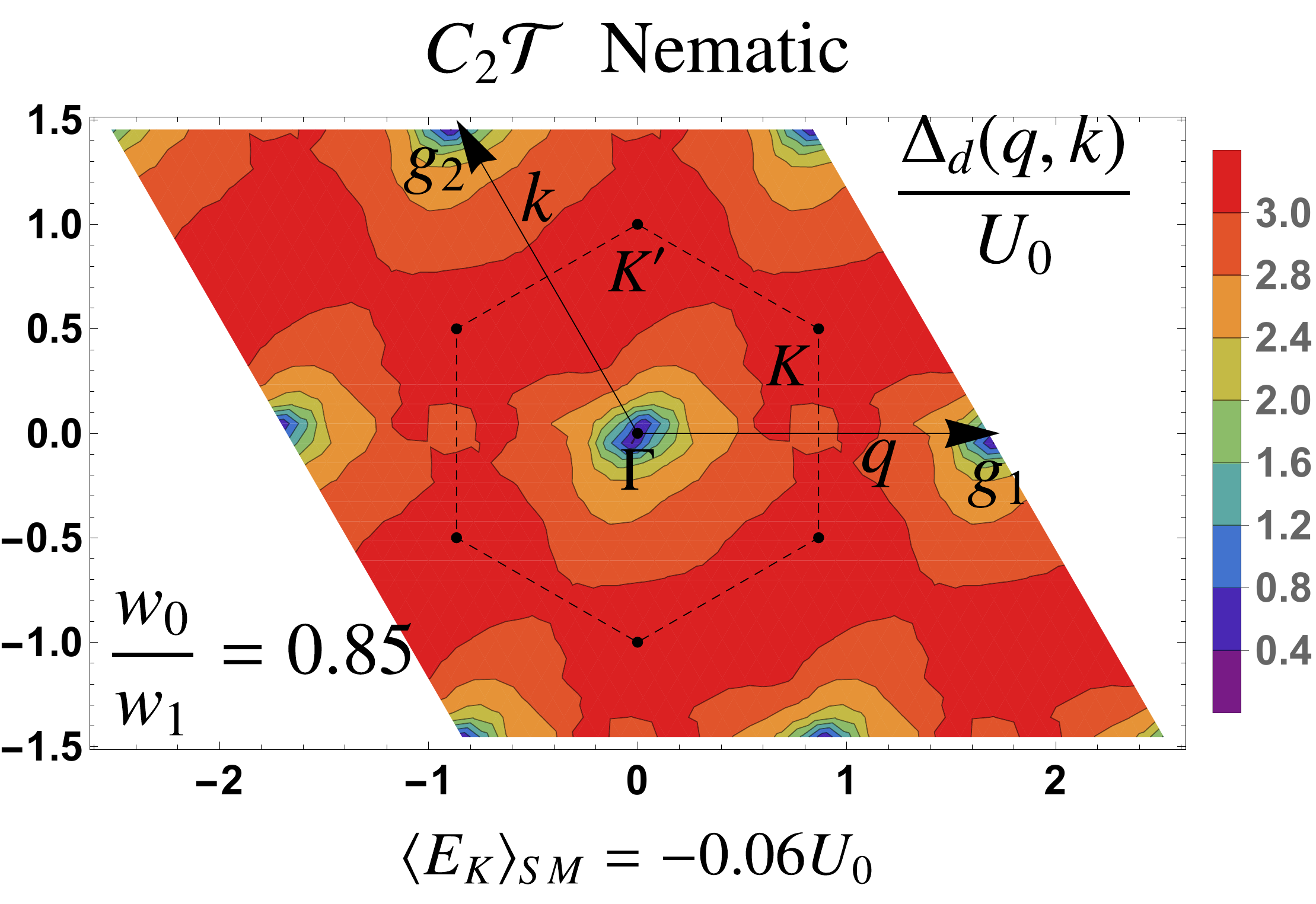}}
	\caption{The direct gap $\Delta_d(q, k)$ for $C_2\mathcal{T}$ nematic state with $w_0/w_1 = 0.85$ at (a) $\langle E_K \rangle_{SM} = 0$ and (b) $\langle E_K \rangle_{SM} = -0.06 U_0$.}
	\label{Fig:QBTC2TNem085}
\end{figure}

In contrast, the $C_2 \mathcal{T}$ nematic state is always gapless, and thus, can never be the experimentally observed insulating phase at $\nu=3$. The plot of the direct gap in the BZ in Fig.~\ref{Fig:QBTC2TNem085} has shown a node (or nodes) either at, or very close to, $\Gamma$ point if the kinetic terms are set to be $0$. Additionally, the direct gap quickly increases to $\sim U_0$ once the momentum is away from $\Gamma$. To understand the properties of nodes in the $C_2\mathcal{T}$ nematic phase and the gap opening in the $C_2 \mathcal{T}$ broken phase, we consider the self-consistent equations obtained as
\begin{eqnarray}
\frac{\delta \langle \Psi_N|  H | \Psi_N \rangle}{\delta u^{*}(q, k)} & = & \mathcal{E}(q, k) u(q, k) \\
\frac{\delta \langle \Psi_N|  H | \Psi_N \rangle}{\delta v^*(q, k)} & = & \mathcal{E}(q, k) v(q, k) \ .
\end{eqnarray}
This is equivalent to the minimization of $\langle \Psi_N|  H | \Psi_N \rangle$. The $\mathcal{E}(q, k)$ is the Lagrange multiplier, needed because of the constraints $|u(q, k)|^2 + |v(q, k)|^2 = 1$ for each $(q, k)$. This equation can be written in the matrix form,
\begin{equation}
H_{eff}(q, k) \begin{pmatrix} u(q, k) \\ v(q, k) \end{pmatrix} = \mathcal{E}(q, k) \begin{pmatrix} u(q, k) \\ v(q, k) \end{pmatrix} \ ,  \label{Eqn:SelfConsistent}
\end{equation}
where $H_{eff}(q, k)$ is a Hermitian $2 \times 2$ matrix and is a functional of $u$'s and $v$'s. $\mathcal{E}(q, k)$ is an eigenvalue of this matrix, and, by Koopman's theorem, the direct gap $\Delta_d(q, k)$ is calculated as the difference between the two eigenvalues of the matrix $H_{eff}(q, k)$.

The Hermitian matrix $H_{eff}$ is obtained as
\beq
\left( H_{eff}(q, k) \right)_{\beta \beta'} = \langle \phi_{\beta}(q, k) | \mathcal{F} | \phi_{\beta'}(q, k) \rangle \ ,  \label{Eqn:SelfConsistentO}
\eeq
where $| \phi_{\beta}(q, k) \rangle$ is the Bloch state defined in Eqn.~\ref{Eqn:chernBloch}, and  $\mathcal{F}$ is a Hermitian operator independent of the momentum~\cite{Appendix}. Note that the Bloch state $| \phi_{\pm}(q, k) \rangle$ has the winding number of $\pm1$ going around the BZ. Since the operator $\mathcal{F}$ has no winding number, the matrix element $\big( H_{eff}(q, k) \big)_{+-}$ has the winding number of $-2$ around the BZ\cite{Zalatel1}. As a consequence, it contains, at least, either a quadratic node or two Dirac nodes inside the BZ\cite{Zalatel1}. For a $C_2\mathcal{T}$ symmetric state, $\left( H_{eff}(q, k) \right)_{++} = \left( H_{eff}(q, k) \right)_{--}$, and thus a node appears as long as the off-diagonal matrix element $\left( H_{eff}(q, k) \right)_{+-} = \left( H_{eff}(q, k) \right)^*_{-+}$ vanishes. This state, therefore, must be a gapless state. On the other hand, the QAH state breaks the $C_2 \mathcal{T}$ symmetry, and the two diagonal elements $\left( H_{eff}(q, k) \right)_{++}$ and $\left( H_{eff}(q, k) \right)_{--}$ become unequal, leading to the opening of a gap.

It is interesting to investigate phenomenological consequences of the nodes in this state. The arguments above suggest the fermion spectrum contains either a quadratic band touching point or two close Dirac nodes with linear dispersion. As shown in Fig.~\ref{Fig:QBTC2TNem085:EK0}, our numerical calculation suggests the existence of a quadratic node when the weight of kinetic terms $\langle E_K \rangle_{SM}$ vanishes. In this case, the Landau levels are doubly degenerate at zero energy and non-degenerate at all other energy levels\cite{McCannFalko2006}. It is worth noting that such Landau level degeneracy, plus the degeneracy brought by valley and spin degrees of freedom, produces the filling patterns of Landau fan observed in the experiments $\nu = \pm 4, \pm 8, \pm 12,\ldots$~\cite{Pablo1, Cory1}. We should also point out that the possibility of two very close Dirac nodes in this system cannot be ruled out due to the insufficient resolution of the momentum mesh. But this does not affect such Landau level filling pattern as long as the two Dirac nodes are close enough and the magnetic field is not too small. Moreover, Fig.~\ref{Fig:QBTC2TNem085:EK0} illustrates that the density of states monotonically increases as a function of energy as the filling changes away from the neutrality point, a feature also qualitatively consistent with experiments\cite{Pablo2}.

Besides the Landau fan pattern, the Fig.~\ref{Fig:QBTC2TNem085:EK0} also shows that the fermion spectrum in this state breaks $C_3$ symmetry and therefore we refer to it as the $C_2 \mathcal{T}$ nematic phase. To understand why $C_3$ symmetry is broken, we expand the effective Hamiltonian around $\Gamma$. It is more convenient to express the momenta as the complex numbers and introduce $z = k_x - i k_y$. Up to the quadratic terms, the effective Hamiltonian can be approximated as
	\begin{equation}
	H_{eff}(\fvec k) = \begin{pmatrix}
	0 & \lambda (z - z_0)(z - z_1)   \\ \lambda^* (z - z_0)^*(z - z_1)^* & 0
	\end{pmatrix}   \ , \label{Eqn:EffHamGamma}
	\end{equation}
where $\lambda$ is a complex constant. It is obvious that the the Hamiltonian contains two nodes at $z_0$ and $z_1$ with the same chirality. Under $C_3$ rotation, the two Bloch states at $\Gamma$ transform trivially~\cite{KangVafekPRX,Senthil1,LiangPRX1}, but the momentum $z = k_x - i k_y$ obtains a phase of $e^{-i 2\pi/3}$. Therefore, $H_{eff}(\fvec k)$ must break the $C_3$ rotation, and the resulting gapless phase is nematic.

\begin{figure}[t]
	\centering
	\subfigure[\label{Fig:QBTC2TNem030:EK0}]{\includegraphics[width=\columnwidth]{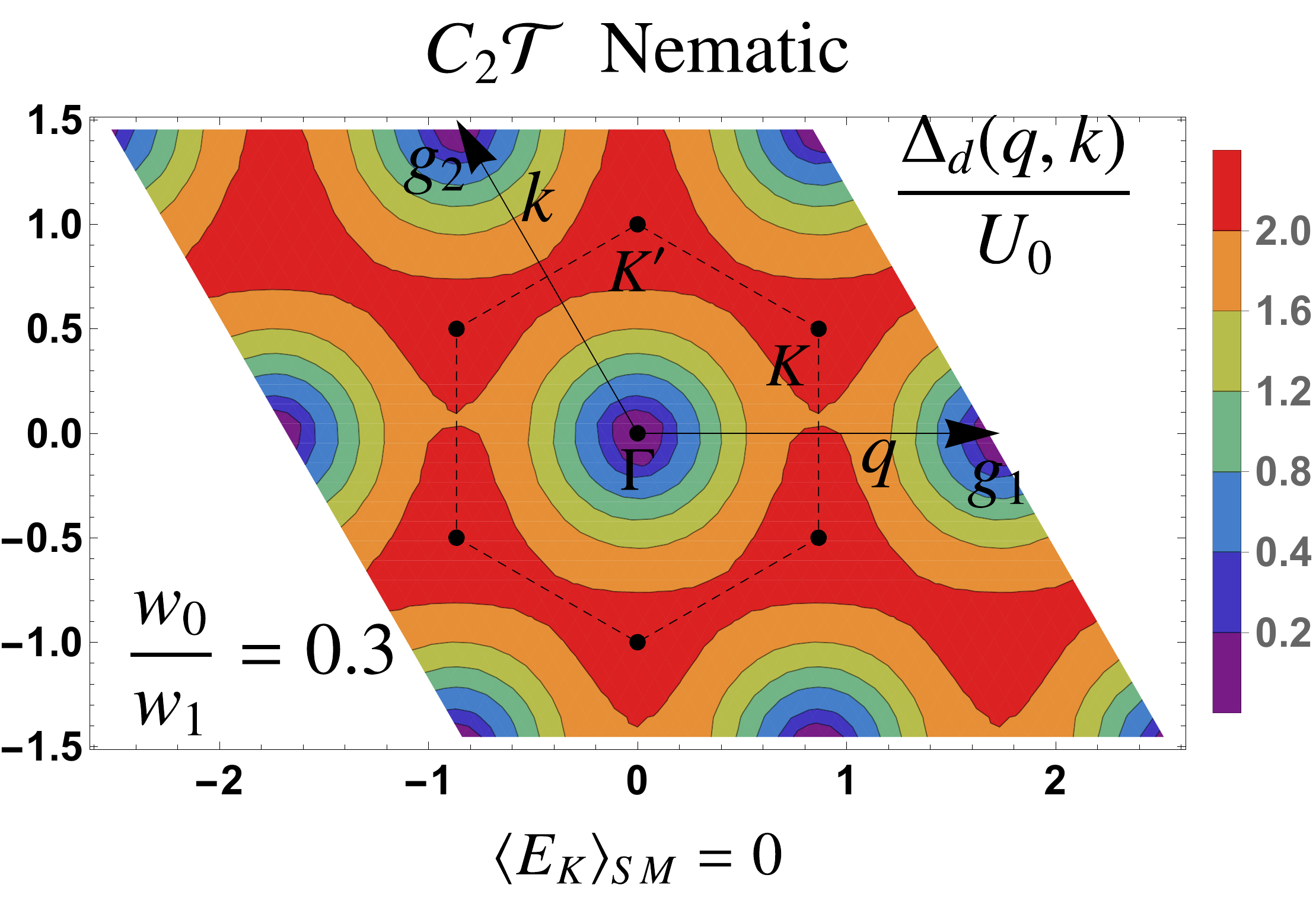}}
	\subfigure[\label{Fig:QBTC2TNem030:EK25}]{\includegraphics[width=\columnwidth]{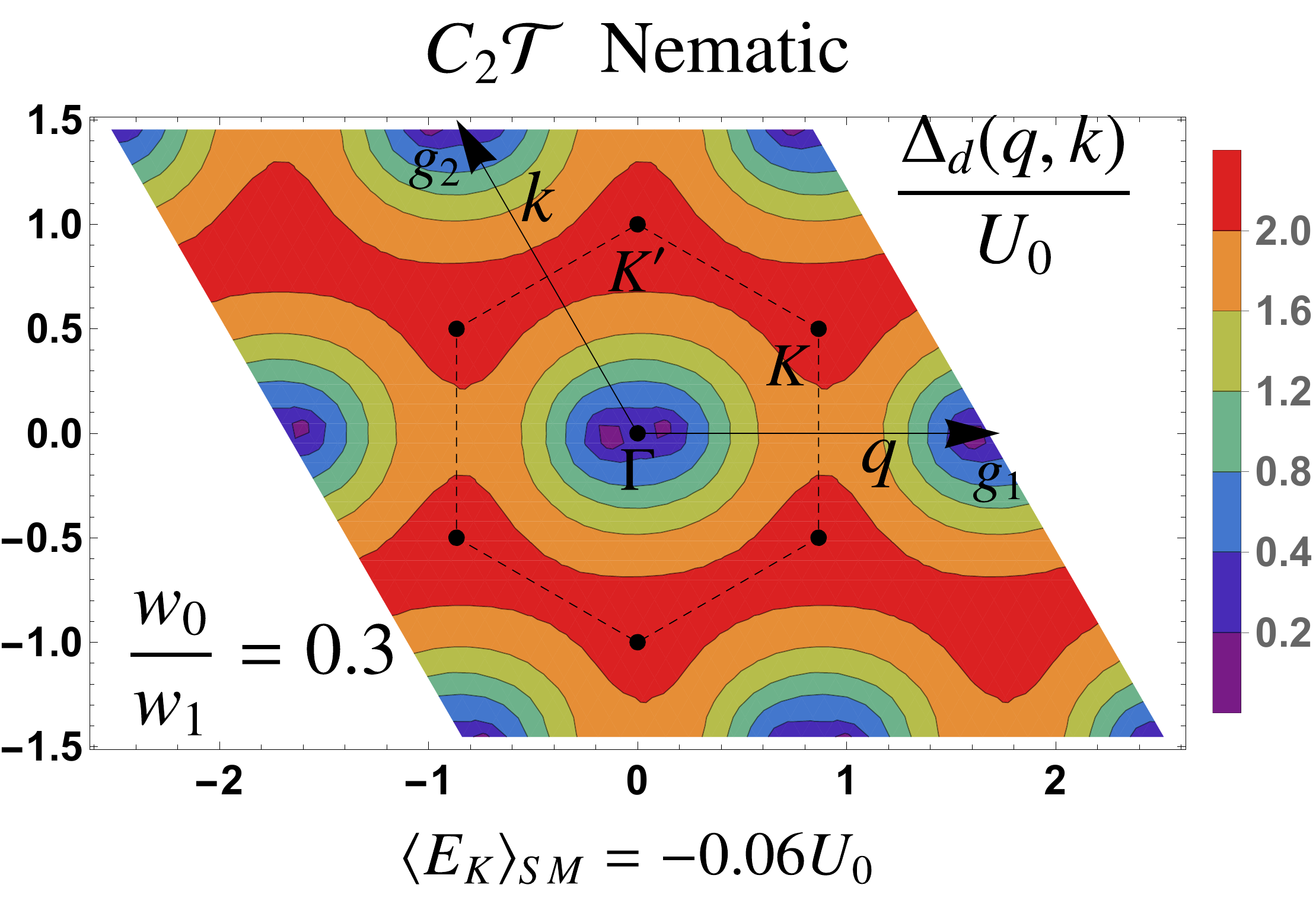}}
	\caption{The direct gap $\Delta_d(q, k)$ of the $C_2\mathcal{T}$ nematic state with $w_0 / w_1 = 0.3$ at (a) $\langle E_K \rangle_{SM} = 0$ and (b) $\langle E_K \rangle_{SM}= -0.06 U_0$. Note the splitting of the node near $\Gamma$.}
	\label{Fig:QBTC2TNem030}
\end{figure}

Another interesting feature is the location and robustness of the node close to $\Gamma$. Fig.~\ref{Fig:QBTC2TNem030} illustrates the direct gap of the $C_2 \mathcal{T}$ nematic state when $w_0/w_1 = 0.3$, a system close to the chiral limit. Although $\sim3$meV per particle above the QAH state (see Fig.~\ref{FigS:DMRGEne030} in the Appendix~\cite{Appendix}), this $C_2 \mathcal{T}$ nematic state contains a quadratic node at or very close to $\Gamma$ when the weight of the kinetic term vanishes, but evolves into two well-separated Dirac nodes with a small weight of kinetic term. At the larger ratio, $w_0/w_1=0.85$, and the same weight of kinetic terms $\langle E_K \rangle_{SM} = 0.06 U_0$, no splitting of nodes or the movement of nodes can be identified in Fig.~\ref{Fig:QBTC2TNem085:EK25}. It seems that the nodes are trapped in a deep potential well at $\Gamma$ with large $w_0/w_1$. This may be related with the steep slope of the Wilson loop eigenvalue at $k$ close to $0$ with $w_0/w_1 \gtrsim 0.8$, or more explicitly, the high peak of the Berry curvature of $\pm 1$ Chern Bloch states at $\Gamma$.

\subsection{$C_2\mathcal{T}$ symmetric period-$2$ stripe state}
In this subsection, we investigate the properties of the $C_2\mathcal{T}$ symmetric period-$2$ stripe phase. Here, we only consider a subset of the such states, $| \Psi^s \rangle$, that can be written in the form of a product states, so that the Wick's theorem applies. The most general form of such states can be written in a relatively simple expression with four free parameters specifying two points on an abstract unit sphere at each momentum~\cite{Appendix}. By minimizing $E^s = \langle \Psi^s | \hat H | \Psi^s \rangle$, we found a local minimum of $E^s$ with the state $| \Psi^s \rangle$ breaking the translation symmetry. As shown in Fig.~\ref{Fig:GenEne}, this state has the energy $E^s$ slightly higher than the $C_2\mathcal{T}$ nematic phase, and still lower than the QAH state. But the energy differences between this stripe state and other states are found to be very small, no more than $0.005 U_0 \approx 0.1$meV. Given the uncertainty in the starting Hamiltonian which almost certainly exceeds this value and the fact that we neglect the valley and spin degrees of freedom, and given the phenomenology of the magic angle twisted bilayer graphene, this state is therefore still a strong candidate for the insulating state experimentally observed at $\nu = 3$. With some degree of valley mixing it may also be possible to further lower the energy of such a state.

\begin{figure}[t]
	\centering
	\includegraphics[width=\columnwidth]{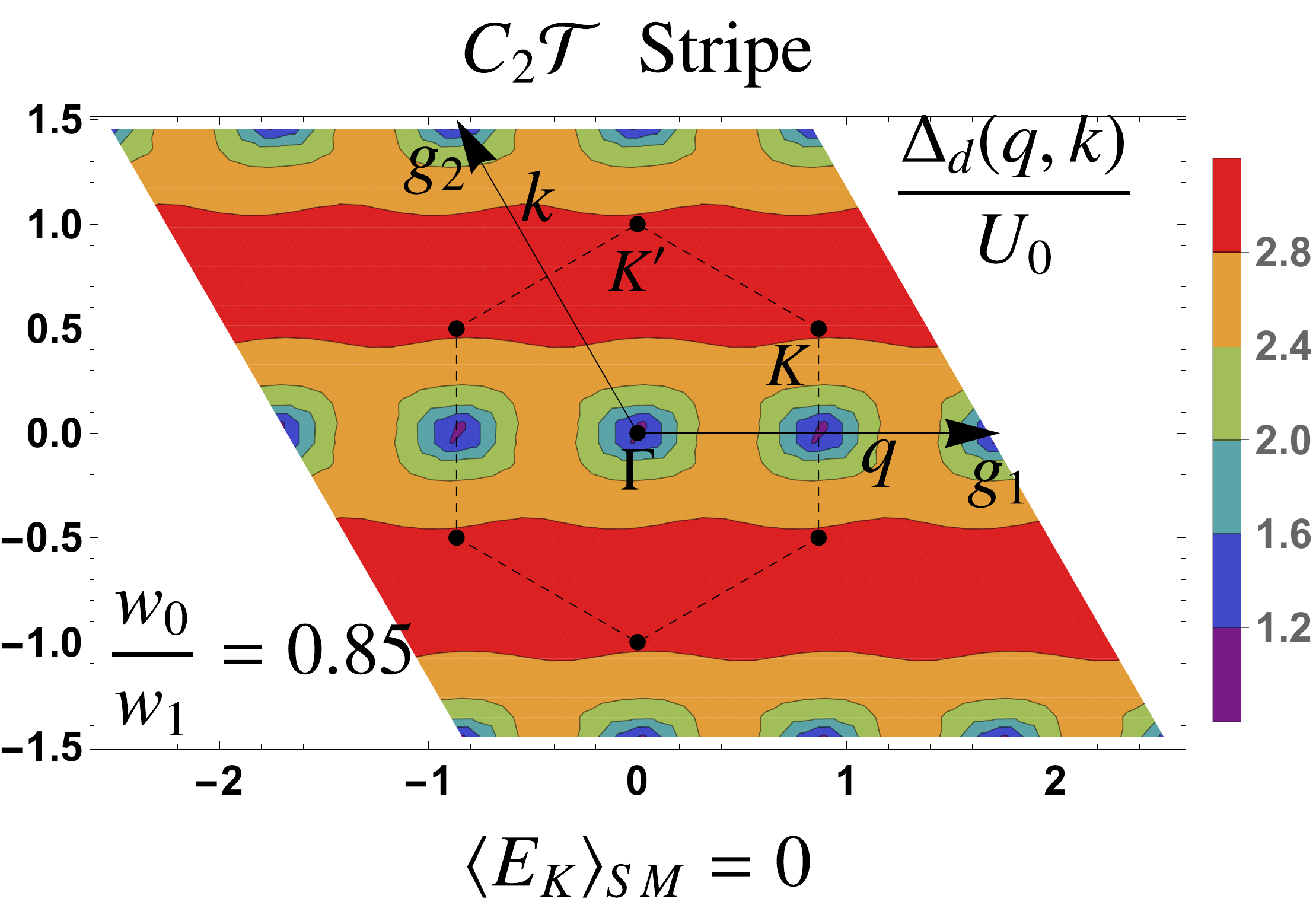}
	\caption{The direct gap $\Delta_d(q, k)$ of the $C_2\mathcal{T}$ stripe phase at $E_K = 0$. It is clear that this state does not contain any nodes.}
	\label{Fig:QBTC2TStripe085}
\end{figure}

Fig.~\ref{Fig:QBTC2TStripe085} shows the momentum dependent direct gap, defined in Eqn.~\ref{Eqn:DGapStripe} but calculated with general trial wavefunction Eqn.~\ref{EqnS:C2TStripeState}. It is obvious that the gap is periodic $\Delta_d(q, k) = \Delta_d(q + \half, k)$. To have a deeper understanding of the fermion spectrum in this phase, we consider the self-consistent equations, which can be written as
\begin{eqnarray}\label{Eqn:StripeEffHam}
& & H_{eff}^s(q, k) \begin{pmatrix} u(q, k) \\ u(q+\half, k) \\ v(q, k) \\ v(q + \half, k) \end{pmatrix} = E(q, k) \begin{pmatrix} u(q, k) \\ u(q+\half, k) \\ v(q, k) \\ v(q + \half, k) \end{pmatrix} \ ,
\end{eqnarray}
with $H_{eff}^s$ being a $4 \times 4$ Hermitian matrix. For notational convenience, we can define a four-component state
\bea
| \eta(q, k) \rangle & = & \left( | \phi_+( q, k) \rangle \ , \ |\phi_+(q + \half, k) \rangle \ , \right. \nonumber \\
& & \left.  | \phi_-(q, k) \rangle \ , \ |\phi_-(q + \half, k) \rangle \right) \ .
\eea
The matrix element of the effective Hamiltonian of the stripe state can then be written as
\begin{eqnarray}
\left( H^s_{eff}(q, k) \right)_{i j} = \langle \eta_i(q, k) | \mathcal{F}^s | \eta_j(q, k) \rangle \ ,  \label{Eqn:SelfC2TStripeO}
\end{eqnarray}
where $\mathcal{F}^s$ is an operator independent of the momentum~\cite{Appendix}. In addition, $C_2 \mathcal{T}$ symmetry leads to the form
\begin{align}
& H_{eff}^s(q, k)  =  \nonumber \\
& \begin{pmatrix}  \epsilon(q, k) & \delta(q, k) & \Delta_1(q, k) & \Delta_2(q, k) \\ \delta^*(q, k) & \epsilon(q + \half, k) & \Delta_2(q, k) & \Delta_1'(q, k) \\ \Delta_1^*(q, k) & \Delta_2^*(q, k) & \epsilon(q, k) & \delta^*(q, k) \\  \Delta_2^*(q, k) &  (\Delta'_1(q, k))^* & \delta(q, k) & \epsilon(q+ \half, k)
\end{pmatrix} \label{Eqn:SelfC2TStripe}
\end{align}
where $\Delta_1'(q, k) = \Delta_1(q + \half, k)$. The ground state is obtained by solving the eigenvalue problem of the Hermitian matrix $H_{eff}^s$. Being a $4 \times 4$ matrix, it contains $4$ eigenvalues, with $E_1(q, k) \leq E_2(q, k) \leq E_3(q, k) \leq E_4(q, k)$. The direct gap $\Delta_d(q, k)$ can be calculated as $E_3(q, k) - E_2(q, k)$.

With the gauge chosen in Eqn.~\ref{Eqn:chernBlochk}, we find the matrix elements satisfy the following conditions~\cite{Appendix}:
\begin{eqnarray}
\Delta_2(q + \half, k) & = & \Delta_2(q, k) \nonumber \\
\Delta_1(q + \half, k) & = & \Delta'_1(q, k) \nonumber \\
\Delta'_1(q + \half, k) & = & \Delta_1(q, k)  \nonumber \\
\Delta_1(q, k + 1) & = & e^{i 4\pi  q} \Delta_1(q, k) \nonumber \\
\Delta'_1(q, k + 1) & = & e^{i 4 \pi q} \Delta_1'(q, k) \nonumber \\
\Delta_2(q, k + 1) & = & - e^{i 4 \pi q} \Delta_2(q, k) \nonumber \\
\epsilon(q + 1, k) & = & \epsilon(q, k)   \ . \label{Eqn:StripeMatBC}
\end{eqnarray}
Based on the boundary conditions above, it is easy to see that the matrix element $\Delta_2(q, k)$ has the winding number of $1$ around the stripe BZ ($0 \leq q < \half$ and $0 \leq k < 1$).

\subsubsection{Gapped Spectrum}
To understand the gapped fermion spectrum of this stripe phase, we first consider a special case in which $\Delta_1 = \delta = 0$. Then,
\begin{align}
&   H_{eff}^s(q, k) =  \nonumber \\
& \begin{pmatrix}  \epsilon(q, k) & 0 & 0 & \Delta_2(q, k) \\ 0 & \epsilon(q + \half, k) & \Delta_2(q, k) & 0 \\ 0 & \Delta_2^*(q, k) & \epsilon(q, k) &  0 \\  \Delta_2^*(q, k) &  0 & 0 & \epsilon(q + \half, k) \end{pmatrix}. \label{Eqn:EffStripeGap}
\end{align}
It is obvious that this effective Hamiltonian matrix can be decomposed into two $2\times 2$ matrices with the same set of eigenvalues, and therefore, contains two doubly degenerate bands. For notational convenience, we define $\epsilon'(q, k) = \frac12(\epsilon(q, k) - \epsilon(q + \half, k))$. The four energies $E_1(q,k),\ldots E_4(q,k)$ are
\beq
\mathcal{E}^{\pm}_{1,2} = \frac{\epsilon(q, k) + \epsilon(q + \half, k)}2 \pm  \sqrt{\left| \Delta_2(q, k) \right|^2 + \left( \epsilon'(q, k) \right)^2}, \label{Eqn:StripeGap}
\eeq
where the subscript $1$($2$) is the index of the degenerate bands. Therefore, the direct gap can be calculated as $\Delta_d(q, k) = 2\sqrt{\left| \Delta_2(q, k) \right|^2 + \left( \epsilon'(q, k) \right)^2}$. Since $\Delta_2(q, k)$ has the winding number of $1$ around the stripe BZ, it must contain a zero point at a momentum $(q_0, k_0)$. Because, in general, $\epsilon(q_0, k_0) \neq \epsilon(q_0 + \half, k_0)$, this state is fully gapped. The addition of small $\delta(q, k)$ and $\Delta_1(q, k)$ will not close this gap.

It is also interesting to study how the double degeneracy between the two low (high) energy bands is lifted by $\delta(q, k)$ and $\Delta_1(q, k)$. For this purpose, we first write down the eigenstates of the Hamiltonian in Eqn.~\ref{Eqn:EffStripeGap}:
\begin{align}
| \psi^+_1 \rangle & = \left( \cos\frac{\theta}2 \ , 0 \, \ , \ 0\ , \ \sin\frac{\theta}2 e^{-i \phi_2} \right)^T \nonumber \\
| \psi^-_1 \rangle & = \left( -\sin\frac{\theta}2 e^{i \phi_2} \ , 0 \, \ , \ 0\ , \ \cos\frac{\theta}2 \right)^T \nonumber \\
| \psi^+_2 \rangle & = \left( 0 \ , \ \sin\frac{\theta}2 e^{i \phi_2} \ , \ \cos\frac{\theta}2 \ , \ 0 \right)^T \nonumber \\
| \psi^-_2 \rangle & = \left( 0 \ ,\ \cos\frac{\theta}2 \ , \ -\sin\frac{\theta}2 e^{-i \phi_2} \ , 0 \right)^T  \label{Eqn:StripeGapStates}
\end{align}
with $e^{i \phi_2} = \Delta_2/|\Delta_2|$ and $\cos\theta = \epsilon'/\sqrt{|\Delta_2|^2 + \epsilon'^2}$. Applying the first order perturbation theory with degenerate states, we obtain
\begin{align}
H^+_{ij} & =  \langle \psi^+_i | H_1 | \psi^+_j \rangle  \label{Eqn:PerturbPlus}  \\
H^-_{ij} & =  \langle \psi^-_i | H_1 | \psi^-_j \rangle  \label{Eqn:PerturbMinus} \\
H_1 & = \begin{pmatrix} 0 & \delta & \Delta_1 & 0 \\ \delta^* & 0 & 0 & \Delta_1' \\ \Delta_1^* & 0 & 0 & \delta^* \\ 0 & \Delta_1'^* & \delta & 0
\end{pmatrix}   .     \label{Eqn:PerturbH1}
\end{align}
For simplicity, consider the effects of $\Delta_1$ and $\Delta_1'$ only. We obtain
\begin{eqnarray}
H^+_{12} & = & \Delta_1\cos^2\frac{\theta}2 + \Delta_1'^* \left( \frac{\Delta_2}{|\Delta_2|} \right)^2 \sin^2\frac{\theta}2 \\
H^-_{12} & = & \Delta_1'^*\cos^2\frac{\theta}2 + \Delta_1 \left( \frac{\Delta_2^*}{|\Delta_2|} \right)^2 \sin^2\frac{\theta}2
\end{eqnarray}
and $H^+_{11} = H^+_{22}$ and $H^-_{11} = H^-_{22}$ because of the $C_2 \mathcal{T}$ symmetry. Applying the boundary conditions listed in Eqn.~\ref{Eqn:StripeMatBC}, we obtain
\begin{eqnarray}
\theta(q + \half, k) & = & \pi - \theta(q, k) \\
H^+_{12}(q + \half, k) & = & \big( H^+_{12}(q, k) \big)^* e^{2i \phi_2(q, k)} \\
H^+_{12}(q, k + 1) & = & e^{4\pi i q} H^+_{12}(q, k) \ .
\end{eqnarray}
Although the boundary conditions cannot determine the exact winding number of $H^+_{12}$ around the stripe BZ, they restrict the parity of winding number to be even~\cite{Appendix}. As a consequence, the two bands above the CNP can have winding numbers of $0, \pm2, \pm4, \cdots$. This conclusion is still valid with the inclusion of $\delta$ terms~\cite{Appendix}.

Finally, we should  point out that the set of the eigenstates in Eqn.~\ref{Eqn:StripeGapStates} is ill-defined if, at a particular momentum $(q', k')$ in the stripe BZ, $\Delta_2(q', k') = 0$ and $\epsilon'(q', k') < 0$ (because we would sit at the south pole which, in this `gauge', contains the famous Dirac string singularity). As mentioned above, $\Delta_2$ has the winding number of $1$ in the stripe BZ, and thus must vanish at a momentum $(q_0, k_0)$ in the stripe BZ. If this is the only momentum at which it vanishes, and $\epsilon'(q_0, k_0) < 0$, we can choose another stripe BZ ($\half \leq q < 1$ and $0 < k < 1$), and notice that $\Delta_2(q_0 + \half, k_0) = 0$ and $\epsilon'(q_0 + \half, k_0) = - \epsilon'(q_0, k_0) > 0$ (which is where the north pole is located without any singularity). As a consequence, the states in Eqn.~\ref{Eqn:StripeGapStates} are well defined in this stripe BZ. And therefore, we can follow the above analysis and obtain the same conclusion. If $\Delta_2(q, k)$ accidentally vanishes at multiple momenta, the conclusions are still valid~\cite{Appendix}.

Our $C_2\mathcal{T}$ stripe state obtained variationally is found to be close to this limiting case, in that $\Delta_2(q, k)$ dominates over other matrix elements in most of the BZ. Furthermore, when $\Delta_2$ vanishes at the momentum $(q_0, k_0)$, the direct gap $\Delta_d$ comes from $\epsilon'(q_0, k_0)$, and $\delta$ and $\Delta_1$ are negligible close to $(q_0,k_0)$.

\subsubsection{Non-Abelian Topological Charge of Dirac nodes}
It is helpful to study another limiting case in which $\delta(q, k) = 0$ and $\epsilon'(q, k) = 0$. The effective Hamiltonian $H^s_{eff}(q, k)$ matrix can then be written as
\begin{eqnarray}
H_{eff}^s(q, k) & = & \begin{pmatrix} 0 & \Delta(q, k) \\ \Delta^*(q, k) & 0 \end{pmatrix} \nonumber  \\
\mbox{with} \ \ \Delta(q, k) &= & \begin{pmatrix} \Delta_1(q, k) & \Delta_2(q, k) \\ \Delta_2(q, k) & \Delta_1'(q, k)
\end{pmatrix}.
\end{eqnarray}
Since this Hamiltonian anti-commutes with $\begin{pmatrix} I_{2\times2} & 0 \\ 0 & -I_{2\times 2} \end{pmatrix}$, the spectrum is particle-hole symmetric, and thus the energy must be $0$ at the half-filling (CNP). With the boundary conditions given in Eqn.~\ref{Eqn:StripeMatBC}, we can readily show that $\det(\Delta(q, k))$ has the winding number of $2$ around the stripe BZ, implying that $\det(\Delta(q, k))$ vanishes at two momenta. Therefore, the spectrum must contain two nodes at zero energy. Consider the two zero modes $| \phi_1(q, k) \rangle$ and $| \phi_2(q, k) \rangle$ of a single node. The $U(1)$ gauge of these two modes can be chosen such that they are invariant under $C_2\mathcal{T}$ symmetry, i.e.~$C_2\mathcal{T} | \phi_1 \rangle = | \phi_1\rangle$ and $C_2\mathcal{T} | \phi_{2} \rangle = | \phi_2\rangle$. Due to $C_2\mathcal{T}$ symmetry, the effective Hamiltonian $H_{eff}^s$ with the matrix element of
\[ \left( \mathcal{H}_{eff}^s \right)_{ij} =  \langle \phi_i | \hat H^s_{eff}| \phi_j \rangle    \]
contains only $\sigma_1$ and $\sigma_3$ terms. This is still true even with the addition of $\epsilon(q, k)$ and $\delta(q, k)$ because the Hamiltonian should still be $C_2 \mathcal{T}$ symmetric. Therefore, introducing a small $\epsilon$ and $\delta$ terms can only shift the position of nodes without opening a gap. Furthermore, the non-zero winding number of $\det(\Delta)$ seems to suggest that the chirality of two zero energy nodes is the same, naively implying that they cannot be annihilated by meeting together. This contradicts the result of the previous subsection, in which the gapped phase is clearly robust.
\begin{figure}[htbp]
	\centering
	\subfigure[\label{Fig:StripeGapSchematic:Start}]{\includegraphics[width=0.35\columnwidth]{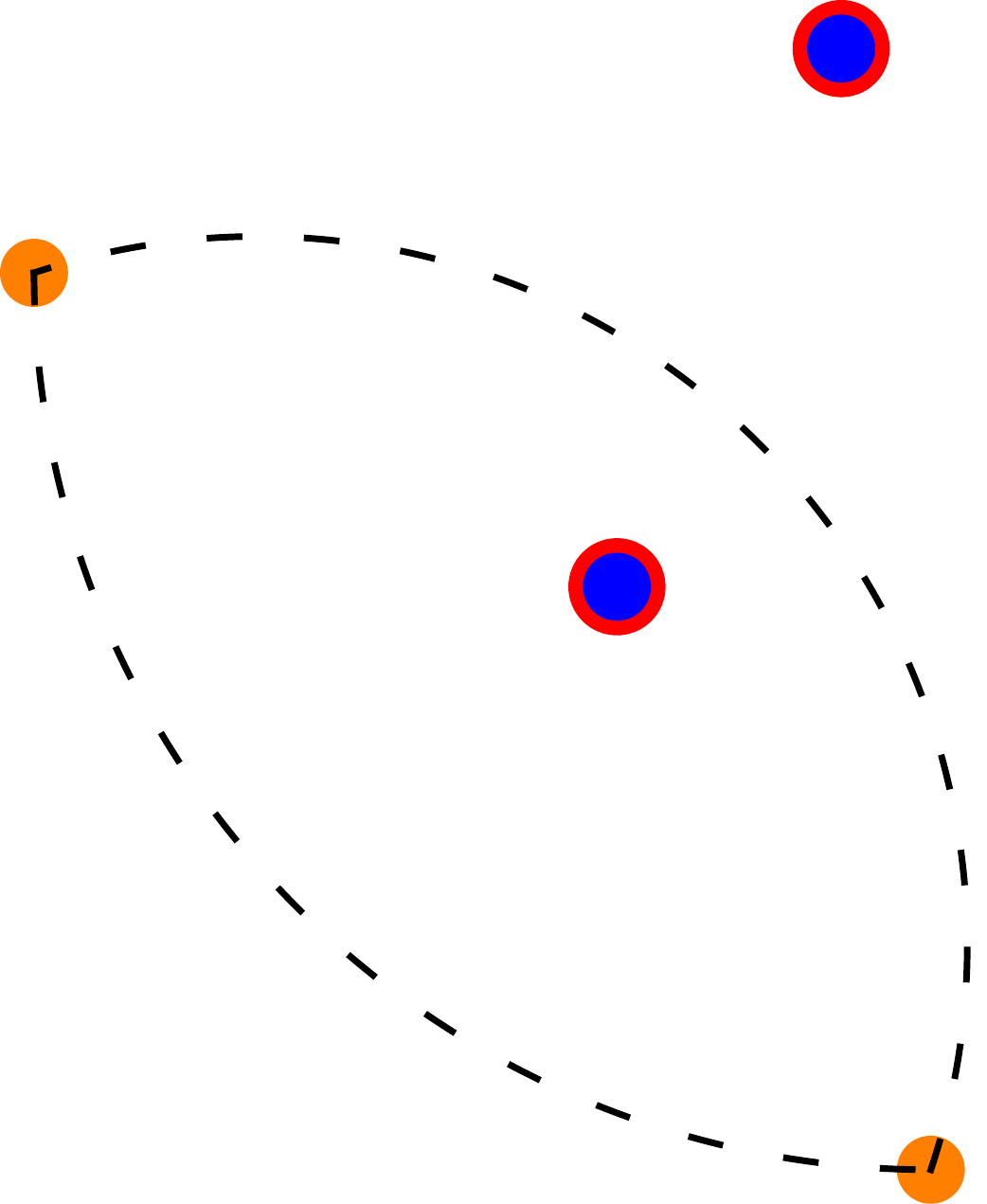}}\hspace{1cm}
	\subfigure[\label{Fig:StripeGapSchematic:End}]{\includegraphics[width=0.35\columnwidth]{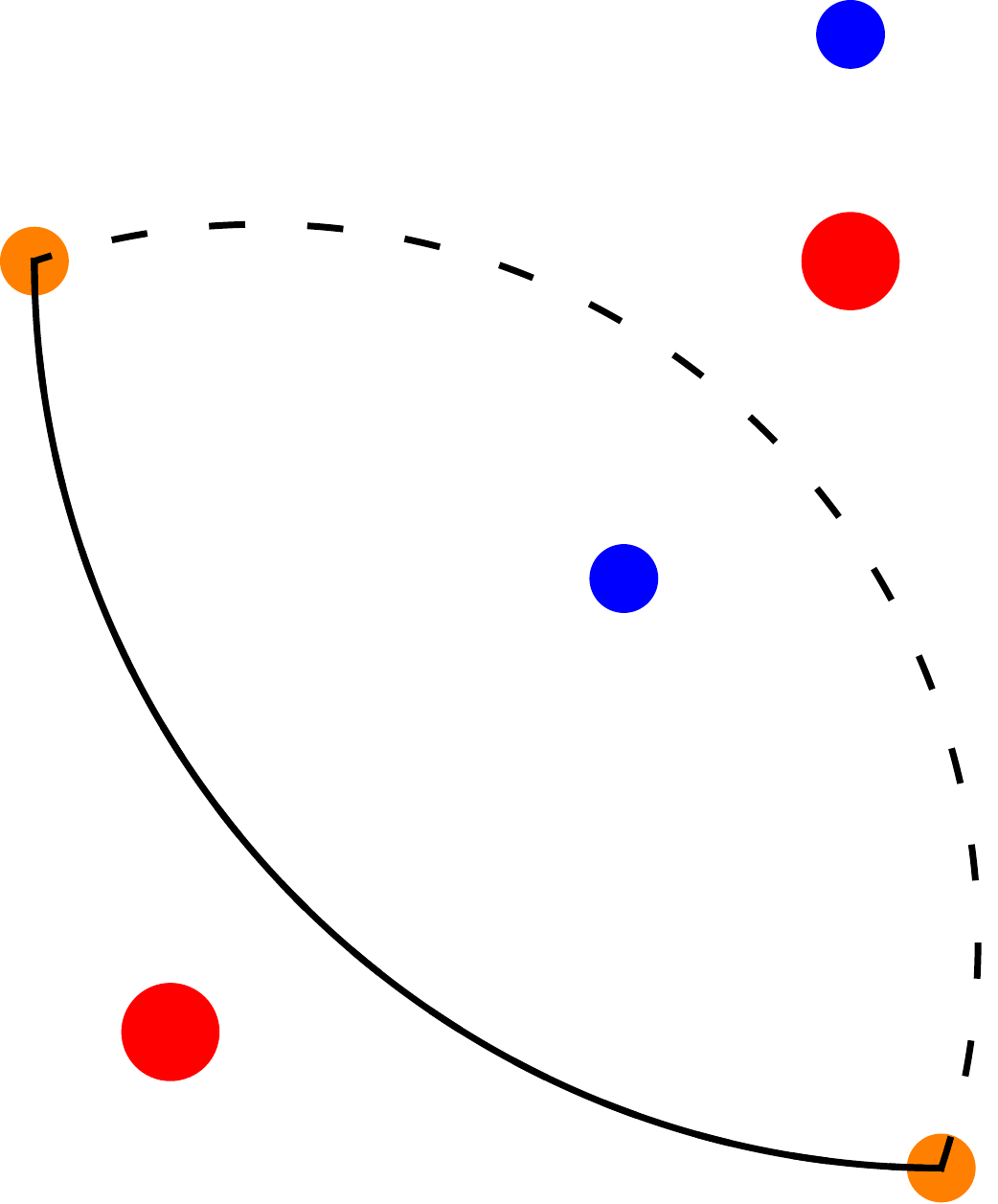}}
	\caption{Schematic plot showing how the two orange nodes around the charge neutrality point (CNP) of a spinless one valley model with the same topological charge can meet and annihilate each other in the presence of the red (blue) nodes formed by the two lower (upper) bands, respectively. The two nodes still carry the same topological charge if they move along the dashed path and can't annihilate, but they have the opposite charge and can meet, annihilate, and open a gap if they move along the solid path. (a) The red and blue nodes coincide because of the particle-hole symmetry. In this case, the two orange nodes always carry the same topological charge no matter how they move, and so can't annihilate (b) By breaking the particle-hole symmetry, the two orange nodes can move along the solid path and carry the opposite topological charge when they meet together. As a consequence, they can annihilate each other and open a gap.}
	\label{Fig:StripeGapSchematic}
\end{figure}

Before resolving this apparent contradiction, it is helpful to investigate any possible crossings between the upper two bands. We can focus on the upper two bands, because whenever there is a Dirac point between the two upper bands at the momentum $(q', k')$ with the energy of $E_0$, the lower two bands also cross at the same momentum with the energy of $-E_0$ due to the particle-hole symmetry. As a consequence, the matrix $\Delta^{\dagger} \Delta = E_0^2 I_{2\times 2}$ at $(q', k')$. Therefore, we obtain two constraints for matrix elements at $(q', k')$:
\begin{align}\label{Eqn:bands34nodes}
& |\Delta_1|  =  | \Delta_1' | \quad \mbox{and} \quad  \Delta_1^* \Delta_2 + \Delta_2^* \Delta_1' = 0.
\end{align}
Consider the term $f(q, k) = \Delta_1^* \Delta_2 + \Delta_2^* \Delta_1'$. Applying the boundary conditions listed in Eqn.~\ref{Eqn:StripeMatBC}, we obtain
\beq
f(q + \half, k) = f^*(q, k), \qquad f(q, k + 1) = - f(q, k) \ .
\eeq
Similar to the previous subsection, this boundary condition does not determine the exact winding number, but restricts the parity of the winding number to be odd. Therefore, $f(q, k)$ contains an odd number of zero points. In addition, notice that $f(q, k) = 0$ when $\Delta_2$ vanishes at the odd number of momenta because it has the winding number of $1$ inside the stripe BZ. As a consequence, the number of momentum points at which $f(q, k) = 0$ and $\Delta_2(q, k)\neq 0$ must be even, as is the number of possible crossings between the upper (lower) two bands because then the first equation in (\ref{Eqn:bands34nodes}) is automatically satisfied.

To resolve the contradiction involving nodes with equal chirality connecting the two middle bands ($E_2(q,k)$ and $E_3(q,k)$) and the possibility of a gap between the two middle bands, we follow Ref.\cite{Tomas} who described the topological properties of nodes in a multiple band system with $\mathcal{P}\mathcal{T}$ symmetry in 3D as well as $C_2\mathcal{T}$ symmetry in 2D. It is well-known that the topological charge associated with a node in a two band system with $C_2\mathcal{T}$ symmetry can be described by an integer winding number, or an element in $\mathbb{Z}$ group. However, as Ref.~\cite{Tomas} insightfully points out, this description must be modified in a system with more bands. For example, in a three-band system with $C_2\mathcal{T}$ symmetry, the topological charge of nodes should not be thought of as an integer but as a quaternion. With $N$ bands, it is an element in $\bar{P}_N$, Salingaros vee group of real Clifford algebra $C\ell_{0,N-1}$ ~Ref.~\cite{TomasSI}. For our $C_2\mathcal{T}$ symmetric period 2 stripe, $N=4$. The nodes thus anti-commute with each other if they are from the consecutive bands, and commute otherwise. The two nodes annihilate with each other if they meet and carry opposite charges, but even if their charges start out opposite, after braiding one of them with an anti-commuting node, the charge can change sign and the resulting pair can consequently annihilate.

Fig.~\ref{Fig:StripeGapSchematic} illustrates how the two nodes with the same topological charge can meet and annihilate with each other in such a four band system. For notational convenience, the bands are (still) labeled by positive integers counted from the lowest energy to the highest one. The two orange points are the nodes formed by band $2$ and $3$. We also assume that the system contains the two Dirac nodes connecting bands $1$ and $2$, labeled by red color in Fig.~\ref{Fig:StripeGapSchematic}. If the system is particle-hole symmetric, it also contains the two Dirac points at the same momentum but connecting bands $3$ and $4$, labeled by blue color. In this case, there is no path along which the orange node can change its topological charge since any closed loop contains an even number of nodes formed by neighboring bands (consistent with the winding number $2$ of the determinant found above). If the particle-hole symmetry is broken, the red and blue nodes move relative to each other. As shown in Fig.~\ref{Fig:StripeGapSchematic:End}, there exists a loop enclosing an odd number of nodes from neighboring bands. Therefore, the topological charge of the nodes connecting the two middle bands becomes opposite if they meet along the solid path. As a consequence, the system will be in the gapped phase without breaking $C_2\mathcal{T}$ symmetry.

\begin{figure}[t]
	\centering
	\includegraphics[width=\columnwidth]{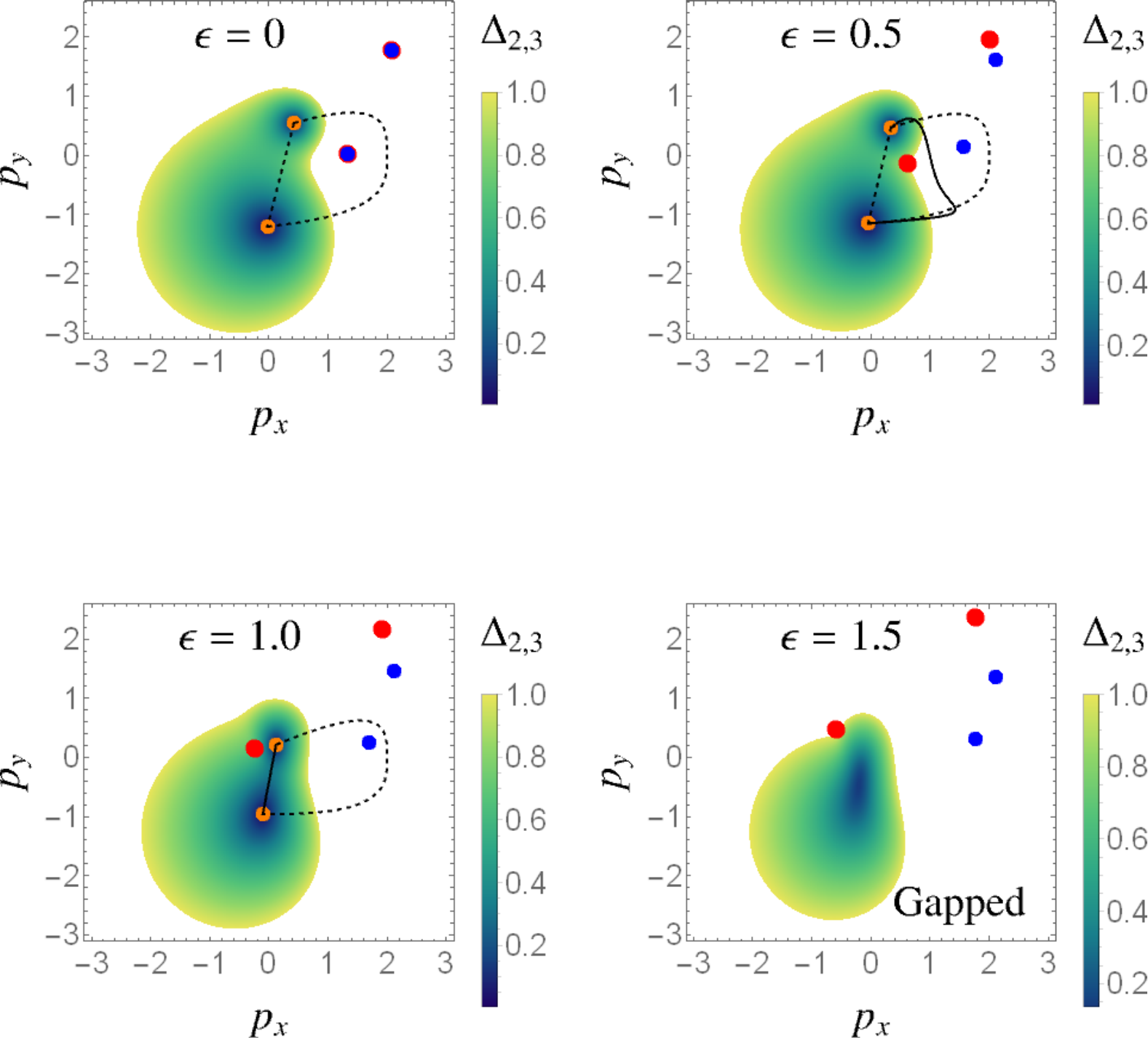}
	\caption{The annihilation of two (orange) nodes in the toy model given by Eqns.~\ref{Eqn:SelfC2TStripe},~\ref{Eqn:StripeToyModlStart} -- \ref{Eqn:StripeToyModlEnd}. The convention to label the nodes and bands is the same as the one in Fig.~\ref{Fig:StripeGapSchematic}. $\Delta_{2,3}$ gives the direct gap between band 2 and 3, and thus vanishes at the orange points. Starting with particle-hole symmetry when the two orange nodes have the same topological charge, their charges become opposite if they move along the solid curve and meet together. If they move along the dashed curves, their topological charges are still the same. The two nodes can annihilate each other only when they have opposite charges.}
	\label{Fig:StripeGapModel}
\end{figure}

\begin{figure}[htbp]
	\centering
	\includegraphics[width=0.8\columnwidth]{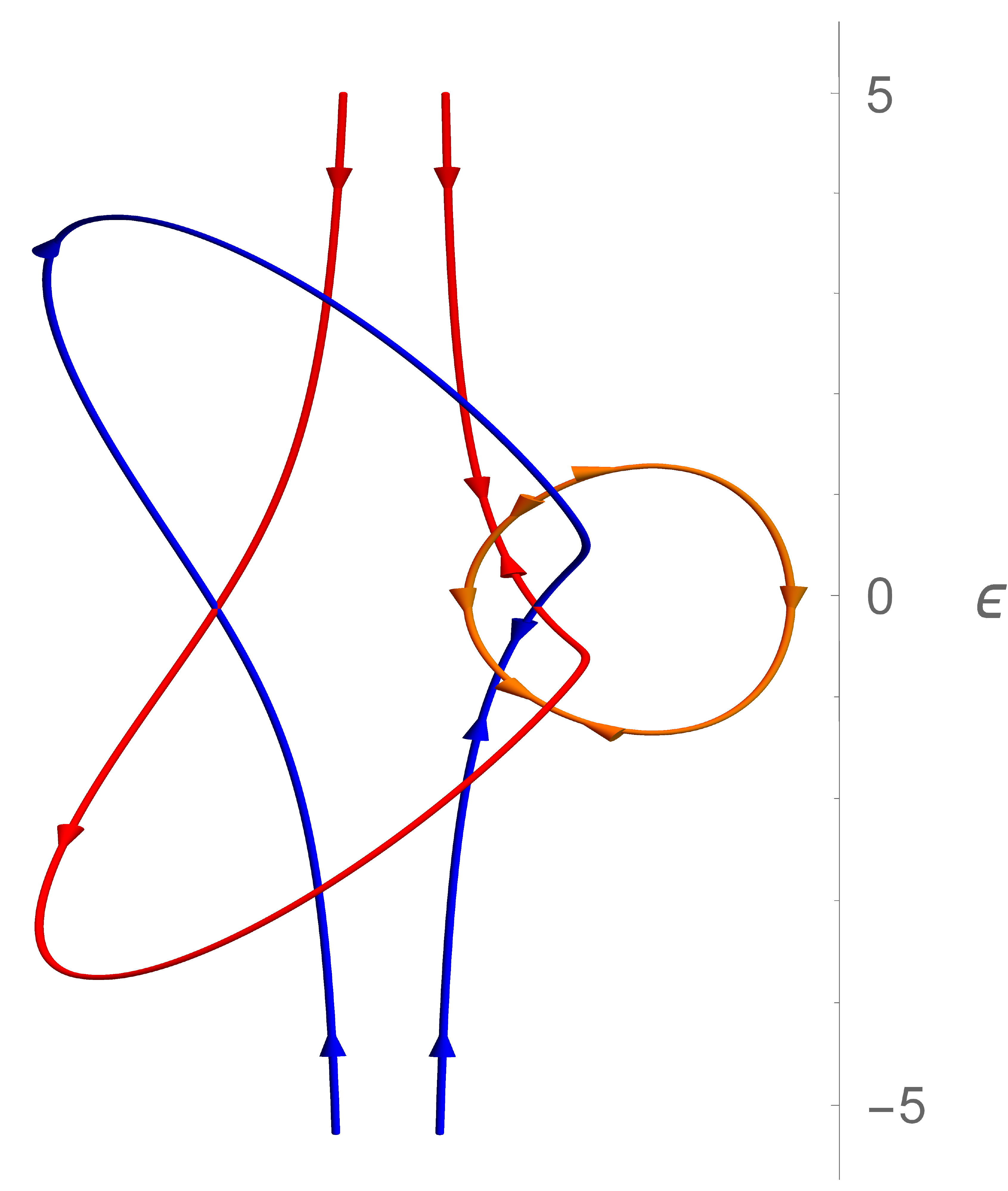}
	\caption{The ``worldlines'' of the Dirac nodes as the parameter  $\epsilon$, defined in Eqs.(\ref{Eqn:SelfC2TStripe},\ref{Eqn:StripeToyModlStart}-\ref{Eqn:StripeToyModlEnd}), varies from $-\infty$ to $\infty$. The curves are obtained using the same $C_2\mathcal{T}$ symmetric model as in the Fig.(\ref{Fig:StripeGapModel}) with matching colors for the nodes (we omit the label for the $p_x$-$p_y$ plane but it should be understood). The arrows of the colored curves follow the prescription for the orientation reversal discussed in a similar context in Ref.\cite{Tomas,TomasSI} and represent the topological charges of the corresponding nodes. Starting from $\epsilon = -\infty$, there are two blue lines representing the two nodes connecting the bands 3 and 4 (ordered in energy), with the same topological charge. The orange loops corresponds to the nodes connecting bands 2 and 3 which, as we prove in the text, must be parallel at $\eps=0$. The red lines correspond to the nodes connecting bands 1 and 2. Orange anti-commutes with the blue and the red, the latter two commute. Note the orientation reversal each time a line passes under an anti-commuting line; it forces the parallel orientation of blue and the red at large $|\eps|$.}
	\label{Fig:TopChargeRule}
\end{figure}

In order to convincingly show how the gapped state can be obtained from the gapless state containing two Dirac points with the same topological charge, we construct a four-band toy model which should accurately capture the region of the momentum space near the zero(s) of various terms, but not the periodicity in the full BZ. The matrix elements of the effective Hamiltonian (\ref{Eqn:SelfC2TStripe}) are thus set to
\begin{eqnarray}
\Delta_1 & = & \lambda e^{i\beta} (z - z_0)(z - z_1) \label{Eqn:StripeToyModlStart} \\
\Delta_2 & = & (1 - \lambda)z \\
\Delta_1' & = & \lambda e^{i \beta}  \\
\epsilon(q, k) & = & - \epsilon(q + \half, k) = \epsilon \\
\delta(q, k) & = & 0, \label{Eqn:StripeToyModlEnd}
\end{eqnarray}
where $z = p_x + i p_y$ is a complex variable, and $\beta$, $\lambda$, $z_0$, $z_1$ and $\epsilon_0$ are the parameters of this model. It is obvious that the matrix element $\Delta_2$ in this model has the winding number of $1$, and the determinant of the $\Delta$ matrix, $\det(\Delta) = \Delta_1 \Delta_1' - \Delta_2^2$ has the winding number of $2$. It is worth emphasizing that $\Delta_1^* \Delta_2 + \Delta_2^* \Delta_1'$ has the odd winding number of $1$, consistent with the analysis above. This toy model does not satisfy the boundary conditions listed in Eqn.~\ref{Eqn:StripeMatBC} and should not be extended to the whole stripe BZ. But, as mentioned, it should accurately describe the effective Hamiltonian in a region of the stripe BZ enclosing the zero points of the matrix elements.

In the calculations, we arbitrarily set $\beta = \pi/4$, $\lambda = 0.3$, $z_0 = 2 + i$, and $z_1 = 2 + 0.5i$, and vary $\epsilon$ to study the annihilation of the two nodes at the CNP. The convention for labeling the nodes and bands in Fig.~\ref{Fig:StripeGapModel} is the same as the one in Fig.~\ref{Fig:StripeGapSchematic}. At $\epsilon = 0$, when the system is particle-hole symmetric, the nodes (red points) connecting the bands $1$ and $2$, coincide with the nodes (blue points) connecting the bands $3$ and $4$. As a consequence, the two nodes (orange points) connecting the bands 2 and 3 with the same topological charge cannot annihilate each other. As shown in Fig.~\ref{Fig:StripeGapModel}, increasing $\epsilon$ leads to the particle-hole symmetry breaking. Thus the nodes connecting the bands 1 and 2 start to move relative to the nodes connecting the bands 3 and 4, and therefore allowing the two orange nodes to annihilate if they meet along the dashed path. The toy model shows that the two orange nodes annihilate each other when $\epsilon \approx 1.5$.

It is also interesting to study the final destiny of the nodes formed by bands 1 and 2, as well as bands 3 and 4, after the gap around the CNP opens. Ref.~\cite{Tomas} provides a general rule to identify the change of the topological charge when the nodes move in the momentum space as our parameter $\epsilon$ varies.  From a fixed vantage point, such change happens when a node worldline passes under an anti-commuting worldline. This orientation reversal is illustrated in Fig.~\ref{Fig:TopChargeRule}, with the arrow representing the topological charge of the nodes. If the two same-color arrows have the same orientation at fixed $\epsilon$ then the topological charges of these two nodes are the same. Otherwise, the two charges are opposite. The two nodes with the same color can annihilate each other only when they carry the opposite topological charges as they meet.

Thus, starting from $\epsilon = - \infty$, the system contains two nodes connecting bands 3 and 4 with the same topological charge as indicated by the two blue parallel arrows in Fig.~\ref{Fig:TopChargeRule}. Later, a pair of red and orange nodes are generated with opposite topological charges. Notice that both the red curve and the blue curve (which mutually commute) pass through the orange loop only once. Therefore the arrows on the blue and the red curves change their orientation only once, but the arrows on the orange curves change twice. As a consequence, the topological charge of the two orange nodes are eventually still opposite and thus can meet at $\eps\sim\pm1.5$ and annihilate. Because the charges of the blue and red nodes change only once, their ``worldlines'' cannot close into a loop. Therefore, the difference of the number of high-energy and low-energy nodes is two in any gapped phase at the CNP. This is consistent with the conclusion made at the end of the previous subsection using the analysis valid throughout the entire BZ.

\section{Discussion}
In this work, we studied the possible ground states of the TBG near the magic angle at odd integer filling, inspired by the experiments near $\nu = 3$. Using the hybrid WSs constructed within the spin and valley polarized BM model, the projected Coloumb interactions are included in the Hamiltonian to study the possible phases in the strong coupling limit. DMRG identifies different ground states as we vary the ratio of two interlayer hopping parameters $w_0$ and $w_1$ in the Bistrizer-MacDonald model. When this ratio is small ($\leq 0.7$), DMRG gives the QAH  as the lowest energy state, suggesting that the system is adiabatically connected to the chiral limit where this state is exact. On the contrary, at the larger ratio, $w_0/w_1 \geq 0.8$, DMRG identifies a state different from QAH as having lower energy. Surprisingly, this state can be well approximated as a product state of the hybrid WSs, and thus motivates our study of the competition between various phases by minimizing the energy of the DMRG inspired trial wavefunction. Our variational calculations discover three nearly degenerate states, QAH, $C_2\mathcal{T}$ nematic state, and $C_2\mathcal{T}$ period-$2$ stripe state. The tiny energy difference among them leads to strong competition between these states. Consequently, the manifold of the low energy states should include all of them even in a spin and valley polarized model, suggesting rich physics beyond the $U(4) \times U(4)$ manifold~\cite{Zalatel3}.

To further obtain the properties of these various phases, we also calculate their fermion spectrum. The QAH state is a gapped state, with the minimal gap at $\Gamma$ point, which is almost certainly further favored by the hBN alignment.
On the other hand, the $C_2\mathcal{T}$ nematic state is a gapless state, with a quadratic node or two very close Dirac nodes near $\Gamma$ point, and thus must break the $C_3$ symmetry. This state, as we discussed, has the Landau level degeneracy\cite{McCannFalko2006} of $2, 1, 1, \cdots$. We propose this $C_2\mathcal{T}$ nematic state, with spin and valley degeneracy restored, as the candidate for the gapless state observed at the charge neutrality point (CNP), because the filling factors of the Landau fan in this state are consistent with the pattern experimentally observed at the CNP\cite{Pablo1,Cory1}.

While the nodes in the $C_2\mathcal{T}$ nematic phase have been assumed to be generally protected by $C_2\mathcal{T}$ and valley $U(1)$ symmetries, we found that these nodes can be lifted by only breaking the moire translation symmetry, without breaking the $C_2\mathcal{T}$ and valley $U(1)$ symmetries, and without closing the gap to the remote bands. Our calculation shows that a gapped $C_2\mathcal{T}$ period-$2$ stripe state is nearly degenerate with the QAH and $C_2\mathcal{T}$ nematic states, and thus is a candidate state for the ground state at the filling of $\nu = 3$ without the hBN alignment. To understand how the gap is opened in $C_2\mathcal{T}$ stripe phase, we present an analysis of the topological properties of the Dirac nodes in the $C_2\mathcal{T}$ nematic state. During the transition from the $C_2\mathcal{T}$ nematic state to the $C_2\mathcal{T}$ symmetric period 2 stripe state, remarkably, the topological charge associated with these nodes should not be described by their (Abelian) winding number, but by elements of (non-Abelian) Salingaros vee group\cite{TomasSI} of real Clifford algebra $C\ell_{0,3}$. Since it is a non-Abelian group, the topological charge of these nodes depends on how they are braided with other nodes away from the CNP. Therefore, a gap at CNP can be opened even without breaking the $C_2\mathcal{T}$ and valley $U(1)$ symmetry.
We expect that this mechanism is general and applies when spin and valley degrees of freedom are fully restored, in which case a gap at odd integer filling may not necessitate translational symmetry breaking. 

Finally, the mechanism discussed makes it apparent that the gap opening in a $C_2\mathcal{T}$ symmetric, but moire translation symmetry broken, state relies on the non-Abelian topological charges of the Dirac nodes, which is effective only if the particle-hole symmetry is broken. Otherwise, the node lines providing the non-trivial braiding are glued together and the equal chirality nodes at the neutrality cannot annihilate. Therefore, if the particle hole symmetry is a good symmetry, the $C_2\mathcal{T}$ symmetric state must remain gapless even when translation symmetry is broken.
This means that it is in principle possible to be in the strong coupling limit, have weak particle-hole symmetry breaking, and end up in a state which has a gap parametrically smaller than $U_0$, the scale set by the Coulomb repulsion.

\begin{acknowledgments}
	We thank B. Andrei Bernevig, Leon Balents, Kasra Hejazi,  Nicolas Regnault, Tomo Soejima, and Michael Zaletel for discussions. We are especially grateful to Hitesh Changlani and P.~Myles Eugenio for help with the DMRG calculations.  J.~K.~is supported by Priority Academic Program Development (PAPD) of Jiangsu Higher Education Institutions and was partially supported by the National High Magnetic Field Laboratory through NSF Grant No.~DMR-1157490 and the State of Florida. O.~V.~was supported by NSF DMR-1916958. Part of this work was performed while the authors visited the Aspen Center for Physics which is supported by the National Science Foundation grant PHY-1607611, and the Kavli Institute of Theoretical Physics which is supported in part by the National Science Foundation under Grant No. NSF PHY-1748958. J.~K.~also thanks the Kavli Institute for Theoretical Sciences for hospitality during the completion of this work.
\end{acknowledgments}


\newpage

\appendix

\begin{widetext}

\begin{center}
	\textbf{\large Appendix for ``Non-Abelian Dirac node braiding and near-degeneracy of correlated phases at odd integer filling in magic angle twisted bilayer graphene''}
\end{center}
\setcounter{equation}{0}
\setcounter{figure}{0}
\setcounter{table}{0}
\makeatletter
\renewcommand{\thefigure}{S\arabic{figure}}

\section{Hybrid Wannier states}
In this section, we illustrate our approach in detail to construct the hybrid Wannier states(WS). They are the eigenstates of the position operator generalized to the periodic boundary conditions, and projected onto the narrow bands,
\begin{eqnarray}
 \hat{\mathcal{O}}  =  \hat{P}e^{-i \frac1N \bg \cdot \br } \hat{P}
\end{eqnarray}
The obtained states are maximally (exponentially) localized in one direction and extended Bloch states in the other direction~\cite{Vanderbilt,HybridWS}. The projection operator, $\hat{P}$, onto the narrow band composite can be written in terms of the (energy eigenstate) Bloch states $|\bk,m\rangle$ as $\sum_{\bk,m}|\bk,m\rangle\langle \bk,m|$, where $\bk$ resides in the first Brillouin zone (BZ), $m$ is the band index and $\hat{\bg}$ is a primitive vector of the reciprocal lattice. The Bloch state overlaps can be expressed in terms of overlaps of the periodic part of the Bloch function, $u_{\bk,m}(\br)$, as
\begin{eqnarray}
 \langle \bk, m | e^{-i\frac1N \bg \cdot \br} | \bk',m' \rangle = \sum_{\bG} \delta_{\bk', \bk + \frac1N\hat{\bg} + \bG} \sum_{\br \in uc} e^{i \bG \cdot \br} u_{\bk,m}^*(\br) u_{\bk', m'}(\br),
\end{eqnarray}
	where $\bG$ is a reciprocal lattice vector.
	The operator $\hat{\mathcal{O}}$ thus mixes only states in the first BZ which lie along the line parallel to $\hat{\bg}$. If we focus on one such line along $\bg_1$, and choose $N = N_1$ to be the number of unit cells along the direction of $\bL_1$, the operator which we wish to diagonalize is
	\begin{eqnarray}
	\hat{O}_\eta & = & \sum_{m m'} \left( \sum_{j = 0}^{N_1 - 2} | \eta \bg_2 + \frac{j}{N_1} \bg_1, m \rangle \Lambda_{mm'}(\eta, j) \langle \eta \bg_2 +  \frac{j + 1}{N_1} \bg_1,m' | \right. \nonumber \\
	& & \qquad \left. + | \eta\bg_2 + \frac{N_1 - 1}{N_1} \bg_1, m \rangle \Lambda_{mm'}(\eta, N_1 - 1) \langle \eta \bg_2, m' |
	\right),
	\end{eqnarray}
	where $0 \leq \eta < 1$ and
	\begin{eqnarray}
	\Lambda_{mm'}(\eta, j) & = & \sum_{\br \in uc} u_{\eta\bg_2 + \frac{j}{N_1} \bg_1, m}^*(\br) u_{ \eta \bg_2 + \frac{j + 1}{N_1} \bg_1, m'}(\br) ; \quad j \neq N_1 - 1 \\
	\Lambda_{mm'}(\eta, N_1 - 1) & = & \sum_{\br\in uc} e^{-i\bg_1 \cdot \br} u_{\eta\bg_2 + \frac{N_1 - 1}{N_1} \bg_1,m}^*(\br)
	u_{\eta \bg_2, m'}(\br) .
	\end{eqnarray}
	We seek a solution of the form
	\begin{eqnarray}
	& &\sum_{j = 0}^{N_1 - 1} \sum_m  \alpha_{\eta, j, m} |\eta \bg_2 + \frac{j}{N_1} \bg_1, m \rangle , \quad \text{so that} \\
	& & \hat{O}_\eta \sum_{j = 0}^{N_1 - 1} \sum_m  \alpha_{\eta, j, m} |\eta \bg_2 + \frac{j}{N_1} \bg_1, m \rangle =
	\epsilon_\eta \sum_{j = 0}^{N_1 - 1} \sum_{m}\alpha_{\eta, j, m} |\eta \bg_2 + \frac{j}{N_1} \bg_1,m \rangle  .
	\end{eqnarray}
	Clearly, this leads to the equations that
	\begin{eqnarray}
	\sum_{m'} \Lambda_{mm'}(\eta, j) \alpha_{\eta, j + 1, m'} & = & \epsilon_\eta \alpha_{\eta, j, m} \quad \textrm{with}\quad j = 0, \cdots, N_1 - 2 \nonumber\\
	\sum_{m'} \Lambda_{mm'}(\eta, N_1 - 1) \alpha_{\eta, 0, m'} & = & \epsilon_\eta \alpha_{\eta, N_1 - 1, m} .
	\end{eqnarray}
	This gives us an eigenvalue problem
	\begin{eqnarray}
	\sum_{m'} \mathcal{W}_{mm'}(\eta) \alpha_{\eta, 0, m'}  & = & \left( \epsilon_\eta \right)^{N_1} \alpha_{\eta, 0, m} \ ,
	\end{eqnarray}
	where the matrix $\mathcal{W}(\eta) = \Lambda(\eta, 0) \Lambda(\eta, 1) \cdots \Lambda(\eta, N_1 - 1)$. In the limit of $N_1 \rightarrow \infty$, $\mathcal{W}(\eta)$ is a unitary matrix. In practice, for finite $N_1$, we perform an singular value decomposition (SVD) of each $\Lambda(\eta,j) = U(\eta,j) \Sigma(\eta, j) V^\dagger(\eta,j)$ where $\Sigma(\eta,j)$ is diagonal (and very close to $1$) and $U$ and $V$ are unitary, and replace $\Sigma$ with a unit matrix.
	Once we have the eigenvectors $\alpha_{\eta, 0}$ and the eigenvalues $\epsilon_{\eta}^{N_1}$, we can construct the remaining $\alpha$'s as
	\begin{eqnarray}
	\alpha_{\eta, j} & = & \left( \epsilon_\eta \right)^j \Lambda^{-1}(\eta, j - 1) \ldots \Lambda^{-1}(\eta, 1) \Lambda^{-1}(\eta, 0) \alpha_{\eta, 0} = \left( \epsilon_{\eta} \right)^j \left( \Lambda(\eta, 0) \Lambda(\eta, 1) \ldots \Lambda(\eta, j - 1) \right)^{-1} \alpha_{\eta, 0}  \ .
	\end{eqnarray}
	Note that if $\eps_\eta^{N_1}$ is an eigenvalue, then replacement $\eps_\eta\rightarrow e^{-2\pi i\frac{n}{N_1}}\eps_\eta$ also gives an eigenvalue for an arbitrary integer $n$. Thus, we set the phase of $\epsilon_{\eta}$ to be in the range of $ -\pi/N_1 \leq \arg(\epsilon_{\eta}) < \pi/N_1$. For the two bands of interest, there are two eigenvectors $\alpha^{\pm}_0$. Our hybrid Wannier states are therefore
	\begin{eqnarray}
	| w_\pm(n,\eta \bg_2) \rangle & = & c e^{i \chi_{\eta}^{\pm}} \left( \alpha^{\pm}_{0, m}| \eta\bg_2, m \rangle + \ldots + e^{-2\pi i \frac{j n}{N_1}} \epsilon_\eta^j \left( \Lambda(\eta, 0)\Lambda(\eta, 1) \ldots \Lambda(\eta, j-1) \right)_{mm'}^{-1} \alpha^{\pm}_{0, m'} |\eta\bg_2 + \frac{j}{N_1} \bg_1, m \rangle + \right. \nonumber  \\
	& &  \left. \ldots + e^{-2\pi i \frac{(N_1 - 1) n}{N_1}} \epsilon_\eta^{N_1 - 1} \left( \Lambda(\eta, 0) \Lambda(\eta, 1) \ldots \Lambda(\eta, N_1 - 2) \right)_{mm'}^{-1} \alpha^{\pm}_{0, m'} | \eta\bg_2 + \frac{N_1 - 1}{N_1} \bg_1, m \rangle \right)  \ ,
	\label{EqnS:HybridWS}
	\end{eqnarray}
	with repeated $m,m'$ indices summed. $c$ is a positive number added for normalization. Note that under translation by $\bL_1$, $\hat{T}_{\bL_1} |w_\pm(n,\eta \bg_2) \rangle = |w_\pm(n + 1, \eta \bg_2) \rangle$, which holds if $\chi_{\eta}^{\pm}$ is $n$-independent. Under the translation by $\bL_2$, $\hat{T}_{\bL_2} |w_\pm(n, \eta \bg_2) \rangle = e^{-2\pi i \eta}|w_\pm(n,\eta \bg_2) \rangle$.
	
	\subsection{$C_2 \mathcal{T}$ Symmetry of the hybrid Wannier states}
	We first fix the phase of our Bloch states by choosing them to be eigenstates of $C_2\mathcal{T}$ with a unit eigenvalue. With this convention, it is clear that
	\beq
	\langle \fvec{r} | \fvec k, m \rangle = \langle - \fvec r | \fvec k, m \rangle^* \quad \Longrightarrow \quad u_{\fvec k, m}(\fvec r) = u^*_{\fvec k, m}(-\fvec r)  \ .
	\label{EqnS:C2TConstraint}
	\eeq
	This guarantees that $\Lambda$'s, and therefore $\mathcal{W}$, are real, and because for each valley and spin we have only two bands, it is a $2\times 2$ matrix.
	
	The only such real unitary matrix has the form $\mathcal{W}=e^{i\theta\sigma_2}$ where $\sigma_2=\left(\begin{array}{cc}0 & -i\\ i & 0\end{array}\right)$. The eigenstates of $\sigma_2$ can be chosen to be $\frac1{\sqrt{2}}(1,\pm i)^T$ with $\mathcal{W}$ eigenvalues $e^{\pm i \theta}$. We choose $\alpha_{0}^+$ in such a way that its eigenvalue winds by $2\pi$ as $\eta$ goes from $0$ to $1$, corresponding to the Chern $+1$ branch. Similarly,
	we choose $\alpha_{0}^-$ in such a way that its eigenvalue winds by $-2\pi$ as $\eta$ goes from $0$ to $1$, corresponding to the Chern $-1$ branch. Although, at each $\eta$ we may thus have either $\frac{1}{\sqrt{2}}(1,i)^T$ or $\frac{1}{\sqrt{2}}(1,-i)^T$ as $\alpha^+_0$, for either choice, complex conjugation interchanges $\alpha_0^+$ and $\alpha_0^-$, and their $\mathcal{O}_{\eta}$ eigenvalues $\epsilon_{\eta}^{\pm}$.  The action of $C_2\mathcal{T}$ on the Chern $\pm 1$ hybrid Wannier states $|w_\pm(n, \eta \bg_2)\rangle$ is therefore
	\begin{eqnarray}
	\hat{C}_2 \mathcal{T} | w_\pm(n,\eta \bg_2) \rangle & = & \hat{C}_2 \mathcal{T} c e^{i \chi_{\eta}^{\pm}} \left(\alpha^{\pm}_{0, m} | \eta\bg_2, m \rangle + \ldots +
	e^{-2\pi i \frac{j n}{N_1}} \left( \epsilon_{\eta}^{\pm} \right)^j \left(\Lambda(\eta, 0) \Lambda(\eta, 1) \ldots \Lambda(\eta, j - 1) \right)_{mm'}^{-1}
	\alpha^{\pm}_{0, m'} |\eta \bg_2 + \frac{j}{N_1} \bg_1, m \rangle \right. \nonumber\\
	& & \left. + \ldots + e^{-2\pi i \frac{(N_1 - 1)n}{N_1}} \left( \epsilon_{\eta}^{\pm} \right)^{N_1 - 1} \left( \Lambda(\eta, 0) \Lambda(\eta, 1) \ldots \Lambda(\eta, N_1 - 2) \right)_{mm'}^{-1} \alpha^{\pm}_{0, m'} |\eta \bg_2 + \frac{N_1 - 1}{N_1} \bg_1, m \rangle \right)  \nonumber \\
	& = & c e^{-i \chi_{\eta}^{\pm}} \left( \alpha^{\mp}_{0, m} | \eta \bg_2, m \rangle + \ldots + e^{2\pi i \frac{j n}{N_1}} \left( \epsilon_{\eta}^{\mp} \right)^j \left( \Lambda(\eta, 0) \Lambda(\eta, 1) \ldots \Lambda(\eta, j - 1) \right)_{m m'}^{-1}
	\alpha^{\mp}_{0, m'} |\eta \bg_2 + \frac{j}{N_1} \bg_1,m \rangle + \right. \nonumber \\
	& & \left. \ldots + e^{2\pi i \frac{(N_1 - 1)n}{N_1}} \left( \epsilon^{\mp}_{\eta} \right)^{N_1 - 1} \left( \Lambda(\eta, 0) \Lambda(\eta, 1) \ldots \Lambda(\eta, N_1 - 2) \right)_{mm'}^{-1} \alpha^{\mp}_{0, m'} |\eta \bg_2 + \frac{N_1 - 1}{N_1} \bg_1, m \rangle \right).
	\end{eqnarray}
	Therefore, our hybrid WSs satisfy the constraint
	\begin{eqnarray}
	\hat{C}_2T|w_\pm(n,\eta \bg_2) \rangle & = & |w_\mp(-n, \eta \bg_2)\rangle.
	\end{eqnarray}
	as long as $\chi_{\eta}^+ = -\chi_{\eta}^-$. The $C_2\mathcal{T}$ symmetry is therefore implemented ``on-site'' in the hybrid Wannier state basis, where $n$ and $k\equiv \eta \bg_2$ are the generalized ``sites''.
	
	\subsection{Continuity of the hybrid WSs}
	Before proceeding to the discussion of other symmetries, we should address the continuity of he hybrid WSs. We require that $| w_{\pm}(n, \eta \bg_2) \rangle$ should be continuous in terms of $\eta$ and $| w_{\pm}(n, (1 + \eta)\bg_2) \rangle = | w_{\pm}(n\pm 1, \eta \bg_2) \rangle$. For this purpose, we consider how to fix the phase factor $\chi^+_{\eta} = - \chi_{\eta}^-$.
	
	First, notice that the phase of the Bloch states are almost fixed by $C_2 \mathcal{T}$ except their signs:
	\[ C_2 \mathcal{T} | \fvec k, m \rangle =  | \fvec k, m \rangle \quad \Longleftrightarrow \quad  C_2 \mathcal{T} \left( -| \fvec k, m \rangle \right) =  -| \fvec k, m \rangle  \ ,  \]
	To further remove this sign freedom, we apply the constraints that
	\begin{equation}
	\Re \left( \sum_{\fvec r \in uc} u^*_{\eta \bg_2, m}(\fvec r) u_{(\eta + \delta \eta) \bg_2, m}(\fvec r) \right) > 0  \ .
	\label{EqnS:BlochPhase}
	\end{equation}
	With this convention set for Bloch states, it is obvious that the hybrid WSs defined in Eqn.~\ref{EqnS:HybridWS} are continuous in terms of $\eta$ as long as $\chi_{\eta}^{\pm}$ are smooth functions of $\eta$.
	
	It is also interesting to investigate $| w_{\pm}(n, (1 + \eta)\bg_2 \rangle$. In the most general form, the Bloch states
	\beq
	| (1 + \eta) \bg_2 + \frac{j}{N_1} \bg_1, m \rangle = e^{i \theta_{\eta, j, m}} | \eta \bg_2, \frac{j}{N_1} \bg_1, m \rangle .
	\eeq
	Applying $C_2 \mathcal{T}$ on both sides, we found that $e^{i \theta_{\eta, j, m}} = e^{-i \theta_{\eta, j, m}}$, ie.~$\theta_{\eta, j, m} = 0$ or $\pi$. Since the phase of Bloch states with $j = 0$ are fixed by Eqn.~\ref{EqnS:BlochPhase}, $\theta_{\eta, j =0, m}$ should be a smooth function of $\eta$. As a consequence, $\theta_{\eta, j = 0, m}$ is independent of $\eta$. Numerically, we found that $\theta_{\eta, j = 0, m}$ always vanishes for both bands.
	
	Now, consider $| w_{\pm}(n, \eta \bg_2) \rangle$ defined in Eqn.~\ref{EqnS:HybridWS}. As the hybrid WSs carry the Chern indices $\pm$, the corresponding eigenvalues of the projected position operator $\mathcal{O}$ have the properties of  $\epsilon_{1 + \eta}^{\pm} = \epsilon_{\eta}^{\pm} e^{\mp i 2\pi/N_1}$. Furthermore, since
	\begin{align}
	& |(1 + \eta) \bg_2 + \frac{j}{N_1} \bg_1, m \rangle  =  |  \eta \bg_2 + \frac{j}{N_1} \bg_1, m \rangle  \nonumber \\
	\Longrightarrow \quad  & u_{(1 + \eta) \bg_2 + \frac{j}{N_1} \bg_1, m}(\br) =  e^{ - i \bg_2 \cdot \br}  u_{\eta \bg_2 + \frac{j}{N_1} \bg_1, m}(\br) \quad \text{and} \quad  \Lambda_{m m'}(1 + \eta, j) =  \Lambda_{m m'}(\eta, j) \nonumber \\
	& | w_\pm(n, (1 + \eta) \bg_2) \rangle  =  e^{i \chi_{1 + \eta}^{\pm}} \left( \alpha^{\pm}_{0, m}   | \eta \bg_2, m \rangle + \ldots + \right. \nonumber \\
	& \qquad \qquad e^{-2\pi i \frac{j (n \pm 1)}{N_1}}    \left( \epsilon_\eta^{\pm} \right)^j \left( \Lambda(\eta, 0)\Lambda(\eta, 1) \ldots \Lambda(\eta, j-1) \right)_{mm'}^{-1} \alpha^{\pm}_{0, m'} |\eta\bg_2 + \frac{j}{N_1} \bg_1, m \rangle +   \ldots + \nonumber  \\
	&  \qquad   \left.  e^{-2\pi i \frac{(N-1)(n \pm 1)}{N_1}}   \left( \epsilon_\eta^{\pm} \right)^{N_1 - 1} \left( \Lambda(\eta, 0) \Lambda(\eta, 1) \ldots \Lambda(\eta, N_1 - 2) \right)_{mm'}^{-1} \alpha^{\pm}_{0, m'} | \eta\bg_2 + \frac{N_1 - 1}{N_1} \bg_1, m \rangle \right) \nonumber \\
	\Longrightarrow \quad   & |w_{\pm}(n, (1 + \eta) \bg_2) \rangle = e^{i (\chi_{1+\eta}^{\pm} - \chi_{\eta}^{\pm})} | w_{\pm}(n \pm 1, \eta \bg_2) \rangle
	\end{align}
	Now, it is easy to see that we can set $\chi_{\eta}^{\pm} = 0$ so that the hybrid WSs $| w_{\pm}(n, \eta \bg_2) \rangle$ is a smooth function of $\eta$, and satisfies
	\[   |w_{\pm}(n, (1 + \eta) \bg_2) \rangle = | w_{\pm}(n \pm 1, \eta \bg_2) \rangle \quad \mbox{and} \quad  C_2 \mathcal{T} | w_{\pm}(n, \eta \bg_2) \rangle = | w_{\mp}(-n, \eta \bg_2) \rangle  \ . \]
	Assuming the Bloch states are properly normalized, $\langle \bg, m| \bg', m' \rangle = \delta_{\bg \bg'} \delta_{m m'}$, we obtained
	\begin{align}
	\langle w_{\alpha}(n, \eta \bg_2) | w_{\alpha'}(n' \eta' \bg_2) \rangle \propto \delta_{\alpha\alpha'} \delta_{n n'} \delta_{\eta \eta'} \quad \mbox{and} \quad
	\langle w_{\alpha}(n, \eta \bg_2) | w_{\alpha}(n, \eta \bg_2) \rangle = N_1 c^2
	\end{align}
	We choose that $c = (N_1)^{-1/2}$ for normalization.

	\subsection{$C_2''$ Symmetry of the Hybrid WSs}
	We also wish to find how the $C''_2$ transformation acts on the hybrid Wannier states. To this end we note that for the Bloch states,
	\begin{eqnarray}
	\hat{C}''_2 | \eta \bg_2 + \frac{j}N \bg_1, m\rangle & \propto & |(1 - \eta)\bg_2 + (\frac{j}N - \eta)\bg_1, m \rangle  \ .
	\end{eqnarray}
	Let $\eta = \frac{h}N$, where $h$ is an integer. Then, because Bloch states at wavevectors related by a reciprocal lattice vector are identical, and
	\begin{align}
	& C_2'' (x \fvec L_1 + y \fvec L_2) C_2'' = x (\fvec L_1 - \fvec L_2) + y \fvec L_2 \quad \Longrightarrow \quad C_2'' e^{-i \frac1{N_1} \bg_1 \cdot \br} C_2'' = e^{-i \frac1{N_1} \bg_1 \cdot \br}     \\
	\Longrightarrow \quad &  \hat{C}''_2 \hat{O}_{\eta} \hat{C}''_2 = \hat{O}_{1 - \eta} \quad \xRightarrow{\big( C_2'' \big)^2 = 1} \quad  \hat{O}_{1 - \eta} \hat{C}''_2 | w_\pm(n,\eta \bg_2) \rangle = \epsilon_{\eta}^{\pm} e^{-i 2\pi n/N_1} \hat{C}''_2 | w_\pm(n, \eta \bg_2) \rangle   \ .
	\end{align}
	Note that the state at $1-\eta$, obtained from the state at $\eta$ has the same eigenvalue $\eps_\eta$.
	But, because the phase of the eigenvalues of $\mathcal{W}$ winds by $\pm 2\pi$ as $\eta$ changes from $0$ to $1$, we must have $\left( \epsilon_{\eta}^{\pm} \right)^{N_1} = \left( \epsilon_{1 - \eta}^{\mp} \right)^{N_1}$. Therefore, up to a phase factor, the $\pm1$ Chern index of the hybrid Wannier states are interchanged under $C''_2$:
	\[ e^{i\phi_{\eta}^{\pm}(n)} | w_{\mp}(n,(1 - \eta) \bg_2) \rangle =   \hat{C}''_2 |w_{\pm}(n, \eta \bg_2)   \rangle   \ .  \]
	Consider the case when $n = 0$, we found
	\beq
	C_2'' | w_+(0, \eta \bg_2) \rangle = e^{i \phi_{\eta}^+(0)} | w_-(0, (1 - \eta) \bg_2) \rangle \quad \Longrightarrow \quad  C_2'' | w_-(0, \eta \bg_2) \rangle =  e^{- i \phi_{\eta}^+(0)} | w_+(0, (1 - \eta) \bg_2) \rangle
	\eeq
	The second formula is derived by applying $C_2 \mathcal{T}$ to both sides of the first formula. Thus, $\phi_{\eta}^+(0) = - \phi_{\eta}^-(0)$. By applying $C_2''$ to both sides of the first formula, we obtain
	\[ | w_+(0, \eta \bg_2) \rangle = e^{i \phi_{\eta}^+(0)} C_2'' | w_-(0, (1 - \eta) \bg_2) \rangle \quad \Longrightarrow \quad \phi_{1-\eta}^-(0) = - \phi_{\eta}^+(0) \quad \Longrightarrow \quad \phi_{\eta}^{\pm}(0) = -\phi_{\eta}^{\mp}(0) =  \phi_{1-\eta}^{\pm}(0) \ .  \]
	Since the hybrid WSs are smooth with respect to $\eta$, $\phi^{\pm}_{\eta}(0)$ are also smooth functions of $\eta$. We redefining the phase factor $\chi_{\eta}^{\pm} \rightarrow \chi^{\pm}_{\eta} - \phi^{\pm}_{\eta}(0)/2$. Therefore,
	\begin{align}
	& C_2'' | w_{\pm}(0, \eta \bg_2) \rangle = | w_{\mp}(0, (1 - \eta)\bg_2) \rangle \ , &  C_2 \mathcal{T} | w_{\pm}(n, \eta \bg_2) \rangle = | w_{\mp}(-n, \eta \bg_2) \rangle \nonumber \\
	& | w_{\pm}(n, (1 + \eta) \bg_2) \rangle = | w_{\pm}(n \pm 1, \eta \bg_2) \rangle \ .
	\end{align}
	Because $\hat{C}''_2 \hat{T}_{n \bL_1} = \hat{T}_{n(\bL_1 - \bL_2)} \hat{C}''_2$, we have
	\begin{eqnarray}
	\hat{C}''_2 | w_\pm(n, \eta \bg_2) \rangle & = & \hat{C}''_2 \hat{T}_{n\bL_1} | w_\pm(0, \eta \bg_2) \rangle=
	\hat{T}_{- n \bL_2} \hat{T}_{n \bL_1} \hat{C}''_2| w_\pm(0, \eta \bg_2) \rangle = \hat{T}_{-n \bL_2} \hat{T}_{n \bL_1} | w_\mp(0,(1-\eta) \bg_2) \rangle \nonumber\\
	& = & \hat{T}_{-n \bL_2} | w_\mp(n,(1-\eta) \bg_2) \rangle = e^{-2\pi i n\eta} |w_{\mp}(n,(1-\eta) \bg_2) \rangle  \ .
	\end{eqnarray}
	This means that $C_2''$ is also implemented ``on-site''.
	
	\subsection{Kinetic energy}
	\begin{figure}[htbp]
		\centering
		\subfigure[\label{FigS:HKin:N1}]{\includegraphics[width=0.4\columnwidth]{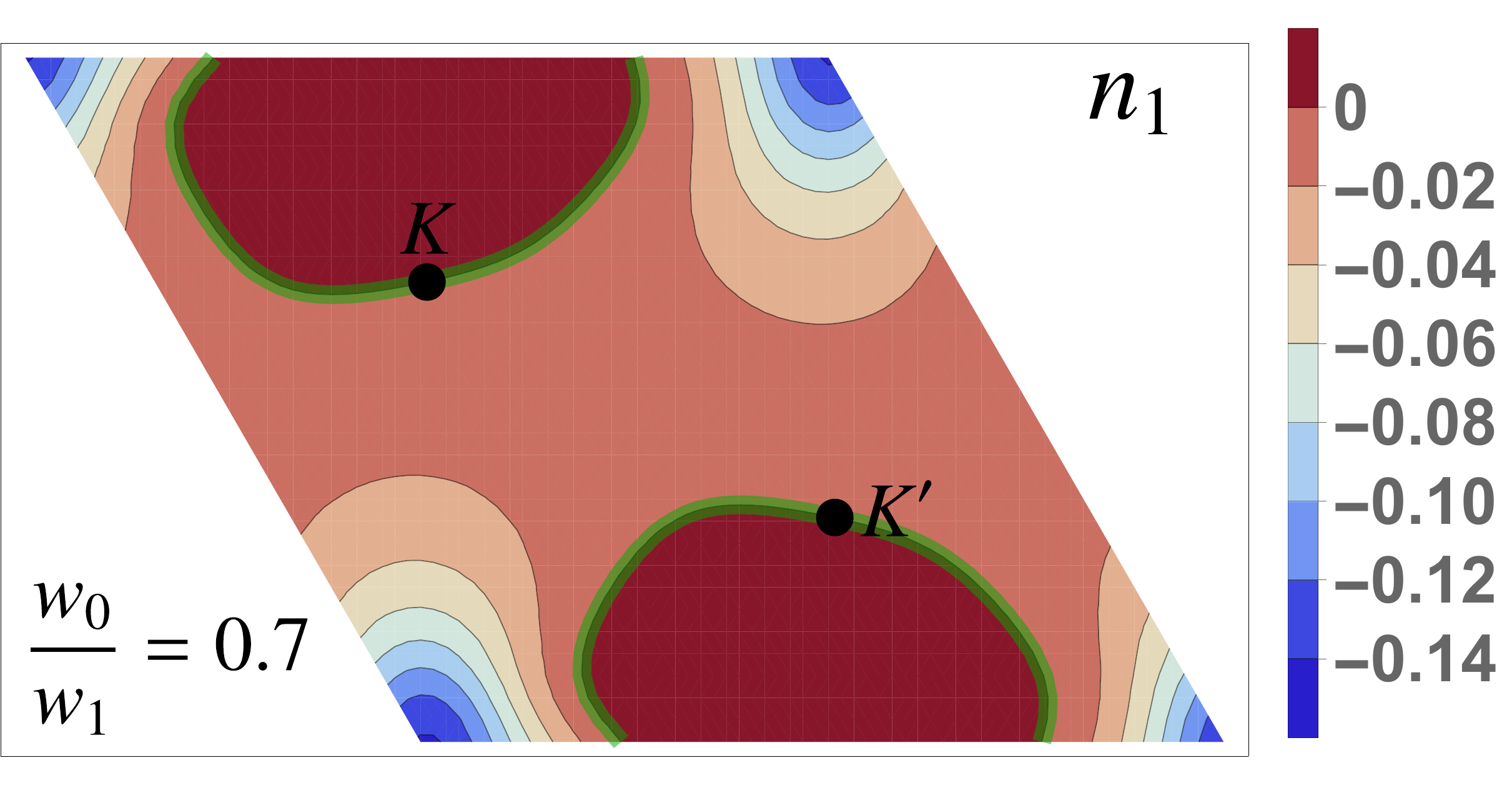}}
		\subfigure[\label{FigS:HKin:N2}]{\includegraphics[width=0.4\columnwidth]{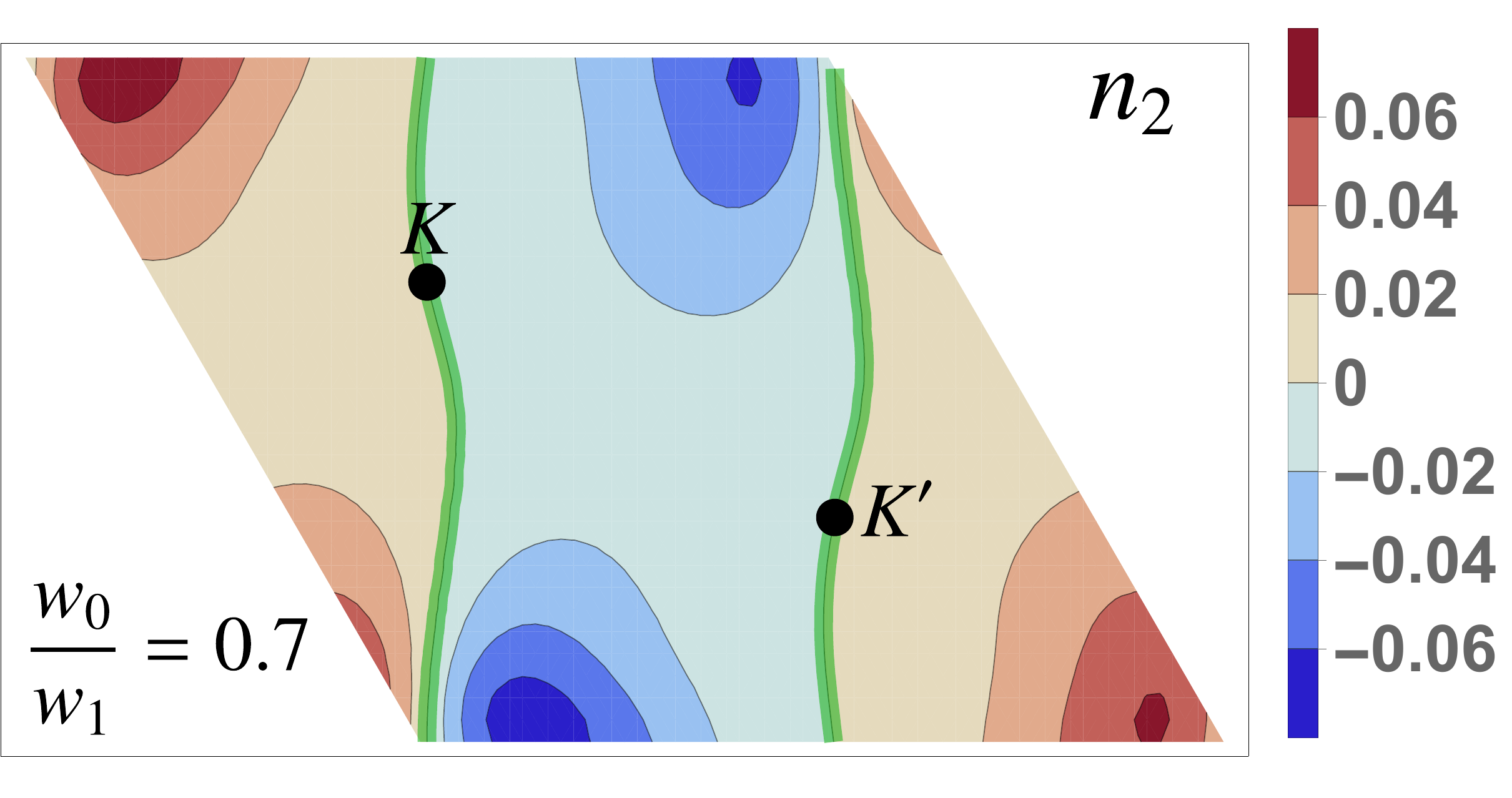}}
		\caption{The kinetic energy vector fields ( Eqn.~\ref{EqnS:HkinGauge} ) obtained from (locally) smooth gauge using hybrid Wannier states (left) $n_1(q,k)$ and (right) $n_2(q,k)$. The bold green lines correspond to the points where $n_{1,2}(q,k) = 0$. The intersection of the green lines results in the Dirac nodes. As shown in Fig.~\ref{Fig:Dirac}, the chirality of the two Dirac points\cite{VafekVishwanath2014} is the same, seemingly contradicting the fermion doubling theorem. Note, however, that one of the assumptions of the theorem does not hold, namely, the vector fields $n_{1,2}(q,k)$ need to be smooth and periodic. As this figure shows, $n_2(q,k)$ is periodic but is not globally smooth, instead, it has a branch-cut discontinuity at $k=0$ and $k=1$.}
		\label{FigS:HKin}
	\end{figure}

	\begin{eqnarray}
	H_{kin} & = &\sum_{n,n'} \sum_k \sum_{\alpha, \alpha' = \pm}  \langle w_{\alpha}(n, k \bg_2) | H | w_{\alpha'}(n', k \bg_2) \rangle d^\dagger_{\alpha, n, k} d_{\alpha', n', k} = t_{\alpha \alpha'}(n - n', k)  d^\dagger_{\alpha, n, k} d_{\alpha', n', k} \ .
	\end{eqnarray}
	It is obvious that $t_{\alpha \alpha'}(n, k) = t_{\alpha' \alpha}^*(-n, k)$. Due to $C_2\mathcal{T}$, we have
	\begin{eqnarray}
	t_{++}(n-n', k) & = &\langle w_+(n, k \bg_2) | H | w_+(n', k \bg_2) \rangle  =  \langle w_-(-n, k \bg_2) | H | w_-(-n', k \bg_2) \rangle^*  \nonumber \\
	& = & \langle w_-(-n', k \bg_2) | H | w_-(-n, k \bg_2) \rangle = t_{--}(n - n', k)  \label{EqnS:C2TEkin}
	\end{eqnarray}
	Moreover, due to $C''_2$
	\begin{eqnarray}
	t_{+-}(n-n',k) & = & \langle w_+(n, k\bg_2) | \hat{C}''_2 H \hat{C}''_2 | w_-(n', k\bg_2) \rangle \nonumber\\
	& = & e^{2\pi i k(n - n')} \langle w_-(n, (1 - k) \bg_2) | H | w_+(n', (1 - k) \bg_2) \rangle = e^{2\pi i k(n-n')} t_{-+}(n - n', 1 - k)\ . \label{EqnS:C2''Ekin}
	\end{eqnarray}
	Fourier transforming gives the hybridized band operators, $d_{\pm, n, \eta} = N^{-1/2} \sum_{q = 0}^{N-1} e^{2\pi i n q } b_{\pm, q, k}$, in terms of which the kinetic energy has the form
	\begin{eqnarray}
	\label{EqnS:HkinGauge}
	H_{kin} & = & \sum_{k, q} \begin{pmatrix} b_{+,q,k} \\ b_{-,q,k} \end{pmatrix}^{\dagger}
	\begin{pmatrix} n_0(q, k) + n_3(q, k) & n_1(q,k) - i n_2(q,k) \\  n_1(q,k) + i n_2(q,k) & n_0(q,k) - n_3(q, k) \end{pmatrix}
	\begin{pmatrix}  b_{+,q,k} \\ b_{-,q,k} \end{pmatrix} \nonumber \\
	& = & \sum_{\alpha \alpha' = \pm} \sum_{q, k} t_{\alpha \alpha'}(q, k) b^{\dagger}_{\alpha, q, k}  b_{\alpha', q, k} \  , \\
	\mbox{where}  \quad t_{\alpha \alpha'}(q, k) & = & \sum_{\delta n} t_{\alpha \alpha'}(\delta n, k) e^{2\pi i q \delta n }
	\end{eqnarray}
	With Eqn.~\ref{EqnS:C2TEkin}, it is obvious that $t_{++}(q, k) = t_{--}(q, k)$, ie.~$n_3(q, k) = 0$. Furthermore, with Eqn.~\ref{EqnS:C2''Ekin} by $C_2''$ symmetry (\cite{Senthil1,SenthilTop}),
	\begin{align}
	t_{+-}(q, k) & =  \sum_{\delta n} t_{+-}(\delta n, k) e^{2\pi i q \delta n} = \sum_{\delta n} t_{-+}(\delta n, 1 - k) e^{2\pi i k \delta n} e^{2\pi i q\delta n} = t_{-+}(q + k, 1- k)  \nonumber \\
	\Longrightarrow \quad n_1(q,k) & =  n_1(q + k, 1 - k) \ , \qquad n_2(q, k)  = -n_2(q + k, 1 - k)  \ .
	\end{align}
	This in turn implies that, if the two Dirac nodes are present, then the winding numbers of at the two Dirac nodes are the same. The hybrid Wannier states thus provide means to construct a locally smooth gauge with on-site representation of the $C_2T$ and $C''_2$ symmetries.
	However, the gauge is not globally smooth, in that there are branch-cuts at $k=0$ and $k=1$. Otherwise, there would be an additional pair of Dirac nodes at the location of the branch-cuts, both nodes with the same chirality, cancelling the overall chirality in the Brillouin zone.

\section{Gate-Screened Coulomb Interaction}
	
In this section, we calculate the metallic gate-screened Coulomb interaction in the bilayer system, with two graphene layers separated by the distance $d_{\perp}$, located in the middle of two gates. Assuming the distance between two gates is $\xi$, due to the gate screening effect, the screened Coulomb interaction is
\begin{align}
 V_{intra}(\br) & = \frac{e^2}{4\pi \epsilon_{hBN}} \sum_{n = -\infty}^{\infty} \frac{(-)^n}{\sqrt{r^2 + (n \xi + ((-)^n - 1) d_{\perp}/2)^2}} \nonumber \\
 & =  \frac{e^2}{4\pi \epsilon_{hBN}} \sum_{n = -\infty}^{\infty} \left( \frac1{\sqrt{r^2 + (2 n \xi)^2}} - \frac1{\sqrt{r^2 + ((2n +1)\xi - d_{\perp})^2}} \right) \\
 V_{inter}(\br) & = \frac{e^2}{4\pi \epsilon_{hBN}} \sum_{n = -\infty}^{\infty} \frac{(-)^n}{\sqrt{r^2 + (n \xi + ((-)^n + 1) d_{\perp}/2)^2}} \nonumber \\
 & =  \frac{e^2}{4\pi \epsilon_{BN}} \sum_{n = -\infty}^{\infty} \left( \frac1{\sqrt{r^2 + (2 n \xi + d_{\perp})^2}} - \frac1{\sqrt{r^2 + ((2n +1)\xi)^2}} \right)
\end{align}
To calculate the Fourier transform of $V(\br)$, notice that
\begin{align}
 & \int \rmd^2 \fvec r\ \frac{e^{i \fvec k \cdot \fvec r}}{\sqrt{r^2 + r_0^2}} = \frac1{\pi} \int \rmd^2 \fvec r\ \rmd z\ \frac{e^{i \fvec k \cdot \fvec r}}{r^2 + z^2 + r_0^2} = \frac1{\pi} \int \rmd^3 \fvec r \frac{e^{i \fvec k_{\parallel}\cdot \fvec r}}{r^2 + r_0^2} = 2 \int \rmd r\ \rmd \cos\theta\ \frac{r^2 e^{i k r \cos\theta}}{r^2 + r_0^2} \nonumber \\
 = & \frac{2}{i k} \int \rmd r\ \frac{r}{r^2 + r_0^2} \left( e^{i k r} - e^{-i k r} \right) = \frac{2}{i k} \int_{- \infty}^{\infty} \rmd r \ \frac{r e^{i k r}}{r^2 + r_0^2} = \frac{2\pi}k e^{-k r_0} \nonumber \\
 \Longrightarrow  \quad & V_{intra}(\bq) =  \frac{e^2}{4 \pi \epsilon} \frac{2\pi}q \frac{ (e^{q d_{\perp}} - e^{-q \xi}) (e^{-q d_{\perp}} - e^{-q \xi}) }{1 - e^{- 2 q \xi}} \\
 \Longrightarrow \quad & V_{inter}(\bq) =  \frac{e^2}{4 \pi \epsilon} \frac{2\pi}q \frac{ e^{q d_{\perp}} (e^{-q d_{\perp}} - e^{-q \xi})^2 }{1 - e^{- 2 q \xi}}
\end{align}
	When $q d_{\perp} \ll 1$, it is clear that
	\[ V_{intra}(\bq) \approx V_{inter}(\bq) \approx \frac{e^2}{4 \pi \epsilon} \frac{2\pi}q \tanh\left( \frac{q \xi}2 \right) \]
		
\section{Energetics at $w_0/w_1 = 0.3$}
	\begin{figure}[h]
	\centering
	\includegraphics[width=0.5\columnwidth]{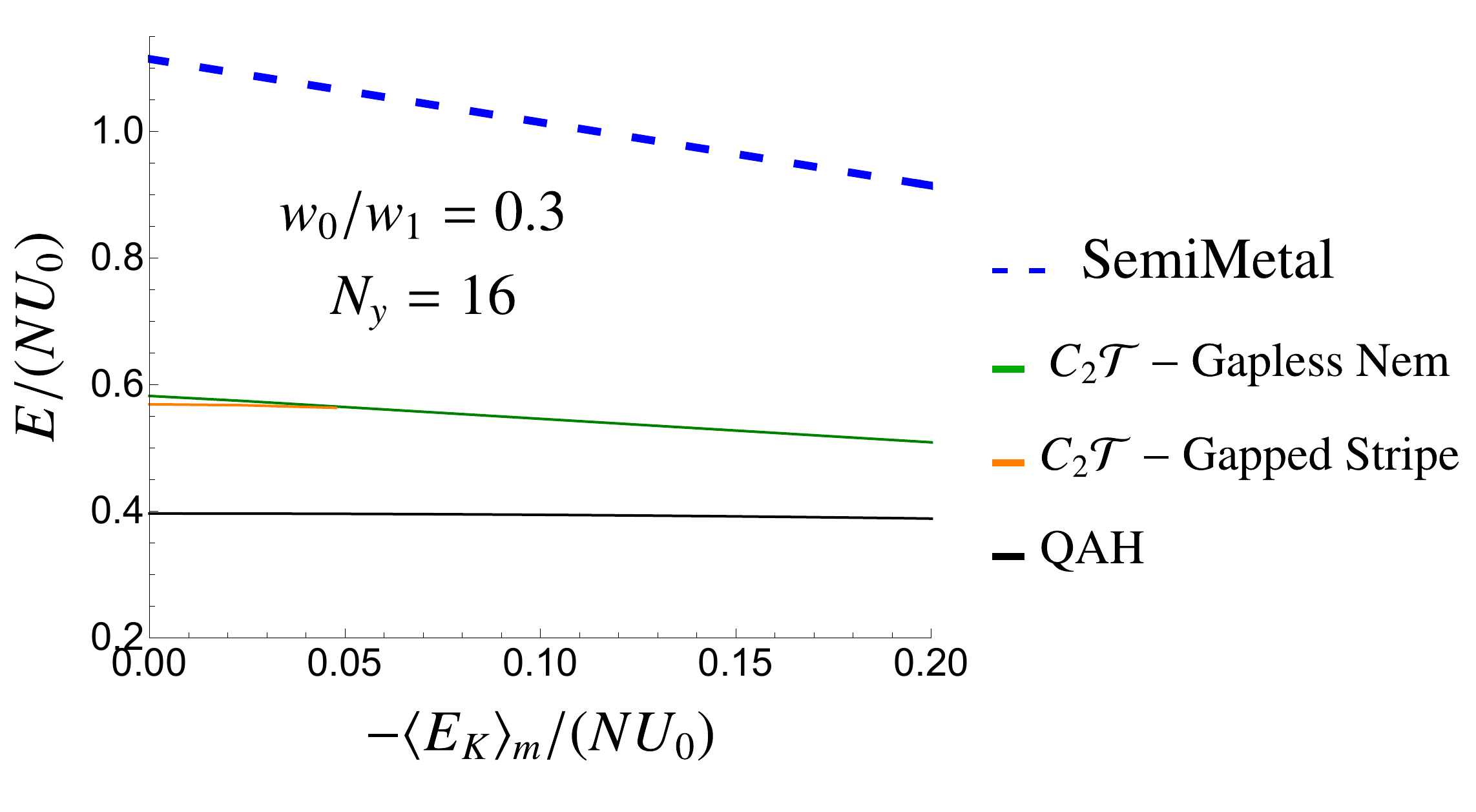}
	\caption{The energy of various states with the trial function in Eqn.~\ref{Eqn:TrialGS} at $w_0/w_1=0.3$. The energies are normalized by $U_0 = e^2/(4\pi \epsilon L_m)$. The figure includes the energies of four different states: $C_2 \mathcal{T}$ broken state, $C_2\mathcal{T}$ nematic state, $C_2 \mathcal{T}$ period-2 stripe state, and the semi-metal state obtained by minimizing the kinetic energy only. The QAH state is clearly the ground state, with $0.17 U_0 \approx 3$meV below the two $C_2\mathcal{T}$ symmetric states, suggesting that the system with $w_0/w_1 = 0.3$ is close to the chiral limit.}
	\label{FigS:DMRGEne030}
\end{figure}

\section{Parameterization of $C_2\mathcal{T}$ Stripe Phase}
In this section, we discuss how to parameterize the $C_2\mathcal{T}$ symmetric period-2 stripe phase. As mentioned in the main text, we consider the states that can be written in the product form so that the Wick's theorem can be applied, ie.
\begin{align}
| \Psi^s \rangle & = \prod_{\substack{k \in [0, 1)\\ q \in [0, 1/2)}}  \chi_1^{\dagger}(q, k) \chi_2^{\dagger}(q, k) | \emptyset \rangle  \label{EqnS:C2TStripeState}\\
\mbox{with}  \quad \chi_i^{\dagger}(q, k) & =  u_i(q, k) b^{\dagger}_{+, q, k}  + u_i(q + \half, k) b^{\dagger}_{+, q + \half, k} + v_i(q, k) b^{\dagger}_{-, q, k}  +  v_i(q + \half, k) b^{\dagger}_{-, q + \half, k}   \quad i = 1\ , \ 2 \ .
\end{align}
Since the many-body state is $C_2\mathcal{T}$ symmetric, the effective Hamiltonian $H_{eff}^s$ in Eqn.~\ref{Eqn:StripeEffHam} for the stripe phase is also $C_2\mathcal{T}$ symmetric. The two vectors
\begin{equation}
 \varphi_i(q, k) = \left( u_i(q, k) \ , \ u_i(q + \half, k) \ , \ v_i(q, k) \ , \ v_i(q + \half, k) \right)^T   \label{EqnS:C2TStripeVector}
\end{equation}
for $i = 1$ and $2$ can be chosen to be the eigenstates of $H_{eff}^s$ and thus also be $C_2\mathcal{T}$ symmetric. This leads to the constraints that $v_i(q, k) = u_i^*(q, k)$ and $v_i(q + \half, k) = u_i^*(q + \half, k)$. Therefore, we can write the vectors as
\[\varphi_i(q, k) = \begin{pmatrix}
\psi_i(q, k) \\ \psi_i^*(q, k)
\end{pmatrix}   \quad \mbox{with} \quad \psi_i(q, k) = \begin{pmatrix}
u_i(q, k) \\ u_i(q + \half, k)
\end{pmatrix}  \ . \]
The normalization gives $\psi^{\dagger}_i(q, k) \psi_i(q, k) = 1/2$. Furthermore, the two vectors $\varphi_1(q, k)$ and $\varphi_2(q, k)$ are orthogonal to each other, leading to another constraint that $\psi_2^{\dagger}(q, k) \psi_1(q, k) + c.c. = 0$. Therefore, $\psi_2^{\dagger}(q, k) \psi_1(q, k)$ is purely imaginary. We write $\psi_1 = \frac1{\sqrt{2}} | \hat n, \uparrow \rangle$, meaning it is the eigenstate of the operator $\hat n \cdot \fvec \sigma$ with the eigenvalue of $1/\sqrt{2}$. Notice that the overall phase of $\psi_1$ is also important, because the state $\varphi_1$ is changed with an additional phase added to $\psi_1$. We also write $\psi_2(q, k) = \frac1{\sqrt{2}} \left( i \alpha | \hat n \uparrow \rangle + \beta | \hat n \downarrow \rangle \right)$ with $|\alpha|^2 + |\beta|^2 = 1$. It is obvious that $\alpha$ must be real to ensure the orthogonality condition. Thus, $\psi_2 = e^{i \frac{\pi}2 \hat n' \cdot \fvec \sigma}  \psi_1 = i \hat n' \cdot \fvec \sigma  \psi_1$ with $\hat n'$ being an arbitrary three dimensional unit vector.

However, this configuration for $\varphi_1$ and $\varphi_2$ still contains redundancy, since any real orthogonal transformation that mixes these two vectors leaves the two dimensional subspace unchanged. This transformation can be written as
\[ \psi_1' = \cos\frac{\omega}2 \psi_1 - \sin\frac{\omega}2 \psi_2 =  \cos\frac{\omega}2 \psi_1 - i \sin\frac{\omega}2 \hat n' \cdot \fvec \sigma \psi_1 = e^{- i \frac{\omega}2 \hat n' \cdot \fvec \sigma} \psi_1  \quad \psi_2' = i \hat n' \cdot \fvec \sigma \psi_1' \ .  \]
Thus we can rotate the spinor $\psi_1$ and $\psi_2$ around the $\hat n'$ unit vector, and obtain an equivalent many-body state. Since $\psi_2$ is acquired by rotating $\psi_1$ by $\pi$ around the same unit vector $n'$, we can always choose the spinor $\psi_1 = \frac1{\sqrt{2}} | \hat n \uparrow \rangle$ so that $\hat n$ has the same azimuthal angle as $\hat n'$. Therefore, we can parameterize these two unit directions as
\[ \hat n = \left( \sin(\theta_1 + \theta_2) \cos\phi_1 \ , \ \sin(\theta_1 + \theta_2) \sin\phi_1 \  , \ \cos(\theta_1 + \theta_2) \right)  \quad \mbox{and} \quad \hat n' =  \left( \sin\theta_1 \cos\phi_1 \ , \ \sin\theta_1 \sin\phi_1 \  , \ \cos\theta_1  \right). \]
This gives the spinors $\psi_1$ and $\psi_2$ as
\begin{align}
\psi_1 = \frac1{\sqrt{2}} e^{i \phi_2} \begin{pmatrix} \cos\frac{\theta_1 + \theta_2}2 \\ \sin\frac{\theta_1 + \theta_2}2  e^{i\phi_1} \end{pmatrix}
 \qquad \psi_2 = i \hat n' \cdot \sigma \psi_1 = \frac{i}{\sqrt{2}} e^{i \phi_2} \begin{pmatrix} \cos\frac{\theta_1 - \theta_2}2 \\ \sin\frac{\theta_1 - \theta_2}2  e^{i\phi_1} \end{pmatrix}.
\end{align}
Therefore, the state in Eqn.~\ref{EqnS:C2TStripeState} can be described by $4$ parameters living on $S^2 \times S^2$ at every momentum. The projector can be written as
\begin{align}
\varphi_1 \varphi_1^{\dagger} + \varphi_2 \varphi_2^{\dagger} = \half \begin{pmatrix}
1 + \cos\theta_1 \cos\theta_2 &  e^{-i \phi_1} \sin\theta_1 \cos\theta_2 & -e^{2 i \phi_2} \sin\theta_1 \sin\theta_2  & e^{i (\phi_1 + 2 \phi_2)}  \cos\theta_1 \sin\theta_2 \\
e^{i \phi_1} \cos\theta_2 \sin\theta_1 & 1 - \cos\theta_1 \cos\theta_2 &  e^{i (\phi_1 + 2 \phi_2)}  \cos\theta_1 \sin\theta_2 & e^{2i(\phi_1 + \phi_2)} \sin\theta_1 \sin\theta_2 \\
- e^{-2 i \phi_2} \sin\theta_1 \sin\theta_2 & e^{-i (\phi1 + 2 \phi_2)} \cos\theta_1 \sin\theta_2 & 1 + \cos\theta_1 \cos\theta_2 & e^{i \phi_1} \sin\theta_1 \cos\theta_2 \\
e^{-i (\phi_1 + 2 \phi_2)}  \cos\theta_1 \sin\theta_2 & e^{-2i(\phi_1 + \phi_2)} \sin\theta_1 \sin\theta_2 & e^{- i \phi_1} \cos\theta_2 \sin\theta_1 & 1 - \cos\theta_1 \cos\theta_2
\end{pmatrix}.
\end{align}

\section{Self-Consistent Equation}
In this section, we present the detailed derivation of the self-consistent equations \ref{Eqn:SelfConsistent} and \ref{Eqn:SelfC2TStripe}, as well as the expression of the operator $\mathcal{F}$ in Eqns.~\ref{Eqn:SelfConsistentO} and \ref{Eqn:SelfC2TStripeO}. First, we consider the state with translation symmetry. For a state given by Eqn.~\ref{Eqn:GeneralTrial}, the fermion correlation function is
\begin{align}
\langle c^{\dagger}(\br) c(\br') \rangle = \sum_{q, k} \sum_{\alpha, \beta} g^*_{\alpha, q, k}(\br)  g_{\beta, q, k}(\br') M_{\alpha \beta}(q, k)   \quad \mbox{with} \quad M(q, k) = \begin{pmatrix} |u(q, k)|^2 & u^*(q, k) v(q, k) \\ v^*(q, k) u(q, k) & |v(q, k)|^2 \end{pmatrix} \ ,  \label{EqnS:FermionCorr}
\end{align}
where $g_{\alpha, q, k}(\br) = \langle \br | \phi_{\alpha}(q, k)\rangle$ is the wavefunction of the Chern Bloch state at the momentum of $(q, k)$ with the Chern number $\alpha$. Since the trial state is a product state, by Wick's theorem
\begin{eqnarray}
\langle \hat V_{int} \rangle & = & \frac12  \int \rmd\br \rmd \br' \  V(\br, \br') \langle c^{\dagger}(\br) c^{\dagger}(\br') c(\br') c(\br) \rangle \nonumber \\
& = & \frac12  \int \rmd\br~\rmd \br' \  V(\br, \br')  \left[ \langle c^{\dagger}(\br) c(\br) \rangle  \langle c^{\dagger}(\br') c(\br') \rangle - \langle c^{\dagger}(\br) c(\br') \rangle \langle c^{\dagger}(\br') c(\br) \rangle  \right]  \label{EqnS:IntFormula}  \\
\langle \hat H_{kin} \rangle & = & \int\rmd \br~\rmd\br'\ t(\br, \br') \langle c^{\dagger}(\br) c(\br') \rangle   \label{EqnS:KinFormula} \\
E & = & \langle \hat H_{kin} \rangle + \langle \hat V_{int} \rangle \label{EqnS:EneFormula}
\end{eqnarray}
Notice that $u(q, k)$ and $v(q, k)$ show only in the matrix $M_{\alpha\beta}(q, k)$. After some calculations, we obtain
\begin{eqnarray}
\begin{pmatrix} \frac{\delta E}{\delta u^*(q, k)} \\ \frac{\delta E}{\delta v^*(q, k)} \end{pmatrix} & = & H_{eff}(q, k) \begin{pmatrix}  u(q, k) \\ v(q, k)  \end{pmatrix} \\
\left( H_{eff}(q, k) \right)_{\alpha \beta} & = & \int \rmd\br~\rmd \br' \  V(\br, \br') \left[ \langle c^{\dagger}(\br) c(\br) \rangle g^*_{\alpha, q, k}(\br') g_{\beta, q, k}(\br') - \langle c^{\dagger}(\br') c(\br) \rangle g^*_{\alpha, q, k}(\br) g_{\beta, q, k}(\br')   \right] \nonumber \\
& & + \int \rmd\br~\rmd\br'\ t(\br, \br') g^*_{\alpha, q, k}(\br) g_{\beta, q, k}(\br')
\end{eqnarray}
Note that we have used the relation $V(\br, \br') = V(\br', \br)$.

Therefore, $\left( H_{eff}(q, k) \right)_{\alpha \beta}$ can be written as  $\langle \phi_{\alpha}(q, k) | \mathcal{F} | \phi_{\beta}(q, k) \rangle$ with
\begin{eqnarray}
\mathcal{F}(\br_1, \br_2) = \langle \br_1 | \mathcal{F} | \br_2 \rangle = \delta(\br_1 - \br_2) \int \rmd\br\ V(\br, \br_1) \langle c^{\dagger}(\br) c(\br) \rangle - V(\br_1, \br_2) \langle c^{\dagger}(\br_2) c(\br_1) \rangle + t(\br_1, \br_2) \  , \label{EqnS:OprF}
\end{eqnarray}
where the fermion correlation is given by Eqn.~\ref{EqnS:FermionCorr}. It is clear that the operator $\mathcal{F}$ is independent of the momentum $(q, k)$, and thus has the winding number of $0$. Since $V(\br_1, \br_2) = V(\br_2, \br_1)$ and $t(\br_1, \br_2) = t(\br_2, \br_1)$, $\mathcal{F}$ is a hermitian operator.

In addition, if the state is $C_2\mathcal{T}$ symmetric,
\begin{align}
& \langle c^{\dagger}(\fvec r) c(\fvec r) \rangle = \langle c^{\dagger}(-\fvec r) c(-\fvec r) \rangle \quad \mbox{and} \quad   \langle c^{\dagger}(\fvec r_1) c(\fvec r_2) \rangle = \langle c^{\dagger}(-\fvec r_1) c(-\fvec r_2) \rangle^* \quad \Longrightarrow \quad   \mathcal{F}(\fvec r_1, \fvec r_2) =  \mathcal{F}^*(-\fvec r_1, -\fvec r_2)
\end{align}
Combined with hermiticity of the operator $\mathcal{F}$, we conclude $\mathcal{F}(\fvec r_1, \fvec r_2) = \mathcal{F}(-\fvec r_2, - \fvec r_1)$.

For the period-$2$ stripe state described by Eqn.~\ref{EqnS:C2TStripeState}, we can follow the same approach to find the expression of $\mathcal{F}$. For this purpose, define the wavefunction to be
\[ g_{q, k}(\br) = \left( \langle \br | \phi_+(q, k) \rangle \ , \  \langle \br | \phi_+( q + \half, k) \rangle \ , \ \langle \br | \phi_-(q, k) \rangle \ , \ \langle \br | \phi_-(q + \half, k) \rangle   \right)   \]
As a consequence, the fermion correlation function can be expressed as
\begin{eqnarray}
\langle c^{\dagger}(\br) c(\br') \rangle = \sum_{\substack{k \in [0, 1)\\ q \in [0, 1/2)}} \sum_{i = 1}^2 \sum_{\alpha, \beta = 1}^4 g_{\alpha, q, k}^*(\br) g_{\beta, q, k}(\br') \left(  M_i(q, k) \right)_{\alpha\beta} \quad \mbox{with} \quad M^i(q, k) = \left( \varphi^i(q, k) \right)^* \left( \varphi^i(q, k) \right)^T \ ,  \label{EqnS:C2TStripeFermionCorr}
\end{eqnarray}
where $\varphi^i(q, k)$ is defined in Eqn.~\ref{EqnS:C2TStripeVector}. The energy can be calculated by applying the Wick's theorem, and we obtain the same expression as Eqns.~\ref{EqnS:IntFormula} -- \ref{EqnS:EneFormula}. As a consequence,
\begin{eqnarray}
& & \begin{pmatrix} \frac{\delta E}{\delta u^*(q, k)} & \frac{\delta E}{\delta u^*(q + \half, k)} & \frac{\delta E}{\delta v^*(q, k)} & \frac{\delta E}{\delta v^*(q + \half, k)} \end{pmatrix}^T  =  H^s_{eff}(q, k) \begin{pmatrix}  u(q, k) & u(q + \half, k) & v(q, k) & v(q + \half, k)  \end{pmatrix}^T \\
& & \left( H^s_{eff}(q, k) \right)_{\alpha \beta}  =  \int \rmd\br~\rmd \br' \  V(\br, \br') \left[ \langle c^{\dagger}(\br) c(\br) \rangle g^*_{\alpha, q, k}(\br') g_{\beta, q, k}(\br') - \langle c^{\dagger}(\br') c(\br) \rangle g^*_{\alpha, q, k}(\br) g_{\beta, q, k}(\br')   \right] \nonumber \\
& & \qquad + \int \rmd\br~\rmd\br'\ t(\br - \br') g^*_{\alpha, q, k}(\br) g_{\beta, q, k}(\br') \label{EqnS:C2TStripeMatEle}
\end{eqnarray}
Thus, $\left( H^s_{eff}(q, k) \right)_{\alpha \beta} = \langle \eta_{\alpha}(q, k) | \mathcal{F}^s | \eta_{\beta}(q, k) \rangle$, and  the operator $\mathcal{F}^s$ for the stripe state has the same expression as Eqn.~\ref{EqnS:OprF}, but with the fermion correlation given in Eqn.~\ref{EqnS:C2TStripeFermionCorr}. Notice that $\mathcal{F}^s$ is also hermitian. Similar to the nematic phase, if the stripe state is also $C_2 \mathcal{T}$ symmetric, $\mathcal{F}^s(\br_1, \br_2) =  \mathcal{F}^s(-\br_2, -\br_1)$. In the next section, this property will be used to derive a series of periodic properties of the effective Hamiltonian.

\section{Properties of the Effective Hamiltonian of the $C_2\mathcal{T}$ Stripe State}
In this section, we discuss the properties of the matrix elements of $H_{eff}^s$ given in Eqn.~\ref{Eqn:SelfC2TStripe} and \ref{EqnS:C2TStripeMatEle}. As illustrated in the previous section, the matrix element can be expressed as Eqn.~\ref{Eqn:SelfC2TStripeO}. With the gauge of Chern Bloch states chosen in Eqn.~\ref{Eqn:chernBlochk},
\begin{align}
 \epsilon(q, k + 1) & = \epsilon(q, k)  & \epsilon(q + 1, k) & = \epsilon(q, k)  & \delta(q + \half, k)  &= \delta^*(q, k) & \delta(q, k + 1) & = - \delta(q, k) \nonumber \\
 \Delta_1(q, k + 1) & = e^{4\pi i q} \Delta_1(q, k) & \Delta_1(q + 1, k) & = \Delta_1(q, k) & \Delta_2(q + \half, k) & = \Delta_2(q, k) & \Delta_2(q, k + 1) & = -e^{4\pi i q} \Delta_2(q, k)  \label{EqnS:C2TStripeBC}
\end{align}
Notice that the definition of the operator $\mathcal{F}^s$ is needed to prove $\Delta_2(q + \half, k) = \Delta_2(q, k)$:
\begin{eqnarray}
\Delta_2(q + \half, k) & = & \langle \phi_+(q + \half, k) | \mathcal{F}^s | \phi_-(q, k) \rangle = \int \rmd\br_1~\rmd\br_2\ g_{+, q + \half, k}^*(\br_1) \mathcal{F}^s(\br_1, \br_2) g_{-,q,k}(\br_2) \nonumber \\
& = & \int \rmd\br_1~\rmd\br_2\ g_{-, q+\half, k}(-\br_1) \mathcal{F}^s(-\br_2, -\br_1) g^*_{+,q,k}(-\br_2) \nonumber \\
& = &  \Delta_2(q, k)
\end{eqnarray}
It is obvious that $\Delta_2$ has the winding number of $1$ around the stripe BZ.

Next, we consider how the double degeneracy between the two low (high) energy bands is lifted by both $\delta(q, k)$ and $\Delta_1(q, k)$. Based on Eqns.~\ref{Eqn:StripeGapStates} -- \ref{Eqn:PerturbH1}, the first order perturbation gives
\begin{align}
H_{12}^+(q, k) = \Delta_1 \cos^2\frac{\theta}2 + \delta \sin\theta \frac{\Delta_2}{|\Delta_2|} + \Delta_1'^* \sin^2 \frac{\theta}2 \left( \frac{\Delta_2}{|\Delta_2|} \right)^2  \label{EqnS:C2TStripeH12Plus}\\
H_{12}^-(q, k) = \Delta_1'^* \cos^2\frac{\theta}2 - \delta \sin\theta \frac{\Delta_2^*}{|\Delta_2|} + \Delta_1 \sin^2 \frac{\theta}2 \left( \frac{\Delta_2^*}{|\Delta_2|} \right)^2  \label{EqnS:C2TStripeH12Minus}
\end{align}
Applying the boundary conditions listed in Eqn.~\ref{EqnS:C2TStripeBC}, we obtain
\begin{align}
H_{12}^+(q, k + 1) & = e^{i 4\pi q} H_{12}^+(q, k)  &  H_{12}^+(q + \half, k) & = \left( H_{12}^+(q, k) \right)^* \left( \frac{\Delta_2}{|\Delta_2|} \right)^2  \nonumber \\
H_{12}^-(q, k + 1) & = e^{-i 4\pi q} H_{12}^-(q, k)  &  H_{12}^-(q + \half, k) & = \left( H_{12}^-(q, k) \right)^* \left( \frac{\Delta_2^*}{|\Delta_2|} \right)^2
\end{align}
Now, we prove that the winding number of $H_{12}^{\pm}$ around the strip BZ must be even. Consider the closed contour $\mathcal{C}=(0, 0) -(\half, 0) - (\half, 1) - (0, 1) - (0, 0)$, and define the change of phase as
\[ \delta\theta_1 = \frac1{i} \int_0^{\half} \rmd q \frac{\partial_q H_{12}^+(q, 0)}{H_{12}^+(q, 0)}  \ , \ \delta\theta_2 = \frac1{i} \int_0^1 \rmd k \frac{\partial_k H_{12}^+(\half, k)}{H_{12}^+(\half, k)}  \ , \ \delta\theta_3 = \frac1{i} \int^{\half}_0 \rmd q \frac{\partial_q H_{12}^+(q, 1)}{H_{12}^+(q, 1)}  \ , \ \delta\theta_4 = \frac1{i} \int_1^0 \rmd k \frac{\partial_k H_{12}^+(0, k)}{H_{12}^+(0, k)} \ .  \]
Applying the formula $H_{12}^+(q, k + 1) = e^{4\pi i q}H_{12}^+(q, k)$, it is easy to prove that $\delta\theta_1 + \delta \theta_3 = -2\pi$. For notational convenience, introduce $\phi_2$ as the phase of $\Delta_2$. Applying the boundary condition $H_{12}^+(q + \half, k) = \left( H_{12}^+(q, k) \right)^* e^{2i\phi_2(q, k)}$, we obtain
\begin{align}
\delta\theta_2 & = \frac1i \int_0^1 \rmd k \frac{\partial_k H_{12}^+(\half, k)}{H_{12}^+(\half, k)} = \frac1i \int_0^1 \rmd k \frac{\partial_k \left( \big( H_{12}^+(0, k) \big)^* e^{2i\phi_2(0, k)} \right) }{\big( H_{12}^+(0, k) \big)^* e^{2i\phi_2(0, k)} } = \frac1i \int_0^1 \rmd k \left( \frac{\partial_k  \big( H_{12}^+(0, k) \big)^*  }{\big( H_{12}^+(0, k) \big)^* } + 2 i \partial_k \phi_2(0, k)  \right) \nonumber \\
& = - \frac1i \int_0^1 \rmd k \frac{\partial_k H_{12}^+(0, k)   }{ H_{12}^+(0, k)  } + 2  \int_0^1 \rmd k\ \partial_k \phi_2(0, k) = \delta\theta_4 +  2  \int_0^1 \rmd k\ \partial_k \phi_2(0, k)
\end{align}
Since $\Delta_2(0, 1) = -\Delta_2(0, 0)$, $\int_0^1 \rmd k\ \partial_k \phi_2(0, k) = (2 n + 1)\pi$. In addition, the boundary condition $H_{12}^+(0, 1) = H_{12}^+(0, 0)$ gives $\delta\theta_4 = 2 m \pi$. Therefore, the total change of the phase is
\beq
\delta\theta_1 + \delta\theta_2 + \delta\theta_3 + \delta\theta_4 = -2\pi + 2\delta\theta_4 + 2 \int_0^1 \rmd k\ \partial_k \phi_2(0, k) = 2\pi \left( -1 + 2m + (2n + 1) \right) = 4\pi (n - m ) \ .  \label{EqnS:WindingNum}
\eeq
Therefore, the winding number must be even. Similar derivation can be done for $H_{12}^-$, and we obtain the same conclusion.

As mentioned in the text, the formula in Eqns.~\ref{EqnS:C2TStripeH12Plus} and \ref{EqnS:C2TStripeH12Minus} are ill-defined at a particular momentum $(q', k')$ at which $\Delta_2(q', k') = 0$ and $\epsilon'(q', k') < 0$. Obviously, Eqns.~\ref{EqnS:C2TStripeH12Plus} and \ref{EqnS:C2TStripeH12Minus} are well defined in the region enclosed by the closed contours $\mathcal{C}$ and $\gamma_{q', k'}$. The winding numbers on the contour $\mathcal{C}$ are still given by Eqn.~\ref{EqnS:WindingNum}. On the small contour $\gamma_{q', k'}$, Eqns.~\ref{EqnS:C2TStripeH12Plus} and \ref{EqnS:C2TStripeH12Minus} are dominated by their last terms. Since $\theta(q', k') = \pi$, the winding numbers of $H_{12}^{\pm}$ on the contour $\gamma_{q', k'}$ are $\pm 2$ respectively. Thus, we conclude that the total winding numbers on the combined contour, $\mathcal{C}$ and $\gamma_{q', k'}$ are still even. As a consequence, the numbers of Dirac points of $H^{\pm}$ are still even in the stripe BZ.

\end{widetext}

\end{document}